\documentstyle [12pt] {report}
\def\be{\begin{equation}}
\def\ee{\end{equation}}
\date{   }
\begin{document}
\parindent=5mm
\pagestyle{empty}
\begin{titlepage}

\title {QUANTUM CORRECTIONS TO BARYON PROPERTIES\\
        IN CHIRAL SOLITON MODELS}
\author {F. Meier and H. Walliser\\
         Fachbereich Physik, Siegen University,\\
         57068 Siegen, Germany\\}
\maketitle
\vfill
\begin{abstract}

Chiral lagrangians as effective field theories of $QCD$ are
successfully applied to meson physics at low energies in the framework
of chiral perturbation theory. Because of their nonlinear structure
these lagrangians allow for static soliton solutions which
may be interpreted as baryons. Their semiclassical quantization, which
provides the leading order in an $1/N_C$ expansion ($N_C$ is the number
of colors) turned out to be insufficient in many cases to obtain good
agreement with empirical baryon observables. However with $N_C=3$,
large corrections are expected in the next-to-leading order which is
carried by pionic fluctuations around the soliton background. The
calculation of these corrections requires renormalization to $1$-loop of
the underlying field theory. We present a procedure to calculate the $1$-loop
contributions for a variety of baryonic observables. In contrast to
chiral perturbation theory, terms with an arbitrary number of gradients
may in principle contribute and the restriction to low chiral orders
can only be justified by the investigation of the scale
independence of the results. The results generally give the right sign
and magnitude to reduce the discrepancy between theory and experiment
with one exception: the axial quantities. These suffer from the fact
that the underlying current algebra mixes different $N_C$ orders, which
suggests a large and positive next-to-next-to-leading order
contribution, which is probably sufficient to close the gap to experiment.

\end{abstract}
\end{titlepage}
\pagestyle{plain}
\tableofcontents
\chapter{Introduction}
\section{Preliminaries}

During the past 15 years, effective field theories have been
proven to be valuable tools in different branches of physics. In hadron
physics, the need for effective field theories is brought about by the intractability of
$QCD$ at low energies around  $1 GeV$. Of course, the
distinction between $QCD$ on the one hand and effective field theories on the other is
somewhat artificial at the conceptual level since $QCD$ itself is
probably an effective field theory of something more fundamental. However, for practical
purposes, it is clear that the examples of effective field
theories we are considering, namely chiral lagrangians formulated in terms of
pseudoscalar fields,
are much more restricted in their domain of validity and richness
of phenomena they can describe than $QCD$.

The most prominent, i. e. energetically lowest lying, degrees of
freedom of such hadron effective field theories are pions (and kaons)
thought to be the
goldstone modes of the spontaneously broken chiral symmetry of $QCD$.
What makes these mesonic effective field theo\-ries tractable in a concise fashion is the
concept of chiral
perturbation theo\-ry ($ChPT$), pioneered by Weinberg  \cite{wei}-\cite{wei3} and for the
first time exploited by Gasser and Leutwyler
\cite{galeu} \cite{galeu2}, which has been successfully applied to a
variety of meson- and meson-baryon phenomena (for the latter case, see
e.g. \cite{gassa}).
$ChPT$ allows to calculate processes in a way that starting from a
meson lagrangian ${\cal L}$ systematically
includes all relevant terms up to some power of external momenta, which
correspond to the number of derivatives (chiral order) contained in the vertices
generated from 
${\cal L}$, and the quark masses. The only theoretical input are
universally accepted principles like chiral symmetry (and a certain
pattern of symmetry breaking), Lorentz covariance etc. . The
essential attraction of $ChPT$ stems from the fact that calculations are no longer model
dependent in the traditional sense. Unless some of the underlying
general principles  could be proven wrong, and as long as higher
degrees of freedom (e.g. vector mesons) can safely assumed to be
frozen, one only has to decide on the power in external momenta to
which the calculation will be carried, and in this way
the error introduced is under control.

The price to pay for this conceptual stringency rests with the fact
that even at small orders, there is a vast number of independent terms 
in the effective lagrangian,
each comes along with its own low energy constant ($LEC$) to be fixed
from experimental information. Since experimental knowledge, as well as
the practical ability to include terms in a calculation are both
limited, so is the applicability of $ChPT$.

Baryons have been introduced into the overall framework of effective field theories in
several different ways: One possibility, directly related to $ChPT$, is to treat
them as external fields coupled in the most general chiral invariant
manner to pions. A similar momentum (and quark mass) power counting
like the one outlined above may then be employed to calculate the
'dressing' of the originally pointlike particle by meson loops,
provided the baryon is treated as a heavy field ($HBChPT)$
\cite{jenkma}. Unfortunately, in this form, the
proliferation of terms allowed from chiral symmetry alone already at the
lowest nontrivial chiral orders is even more worrisome than in the
purely mesonic case, and consequently the applicability
of $HBChPT$ is restricted even further compared to meson $ChPT$.
is lost.

Therefore, the idea to create the baryons as solitons from the meson
fields themselves without explicit fermionic degrees of freedom has
been considered an especially attractive alternative, which has the
advantage that one might directly build upon the effective lagrangian
established in $ChPT$ in the mesonic sector. It is this
aspect that defines the scope of our paper, namely, to investigate the
extent to which the concept of baryons as solitons in effective meson
theories can be linked to the methods and results of purely mesonic
$ChPT$. Evidently, the consideration of loop corrections will be the
crucial link in this connection.

In order to provide the reader with some motivation of
what follows in the course of our report, we shall in this
introduction recall the state of affairs in topological soliton models,
outline the problems one has to face in trying to go beyond this status
quo, and shed some light on the relationship between (baryon) $ChPT$
and the soliton model calculation we are going to consider.   

\section{The Skyrme model}
The Skyrme model \cite{thrsk}, invented about 35 years ago, yet
reconsidered only within the timeframe mentioned in the beginning, emerges from 
the realization that
nonlinear effective meson theories allow for topologically nontrivial
static solutions (solitons) whose conserved topological index is then 
tentatively interpreted  as the baryon number. Spin and isospin of the
soliton are supplied by quantizing the zero modes of the problem.
The reasons which left this model dormant for about 20 years after its
inception (except for it being somehow orthogonal to the mainstream
developments in medium energy physics) were essentially twofold: Firstly, in the 
absence of a suitable
expansion parameter there was no compelling reason why zero modes
should be preferred over other fluctuations in the soliton's
quantization. Secondly, there was no argument to constrain the soliton
to have half integer spin. The solution to both these problems is of
course well
known: The difficulty with half integer spin could be removed upon
consideration of the Wess-Zumino-Witten anomaly, whereas  
the work of t'Hooft 
\cite{tho}, \cite{tho2} and Witten 
\cite{ewi} on the expansion of $QCD$ in powers of the inverse number of
colours $1/N_C$, besides making the concept of solitonic baryons more
palatable as a low energy approximation to $QCD$, showed that the
pion-baryon coupling is suppressed in $1/N_C$ (pions being
identified with non-zero mode fluctuations of the soliton), and in this guise
saved the zero mode quantization as a first approximation for baryon
properties.  

Since we will be concerned with corrections to this approximation, it
is at this point worthwhile to recall
some results given in reference \cite{ewi}. There, it was
shown that the $QCD$ coupling constant scales like $N_C^{-1/2}$ if
one assumes a smooth large $N_C$ limit for the $1$-loop gluon self
energy graph which would give rise to the worst divergences in $N_C$.
Essentially from that result it follows that the pion baryon
coupling must be of ${\cal O}(N_C^{1/2})$, whereas the baryon mass is
of ${\cal O}(N_C)$. 
Provided one accepts the conclusions of \cite{ewi}, one can calculate
many observables to leading ${\cal O}(N_C)$, and  
since the picture of meson physics predicting baryon physics seemed
rather appealing, the cited references consequently
triggered a plethora of papers on the Skyrme model and its
generalizations, examples of which are  \cite{adnawi}-\cite{sch93}. 
The leading contributions in $1/N_C$ to static baryon properties,
formfactors and pion
nucleon scattering were calculated.
What marred these calculations and what made the 'model' character of
the Skyrme model was the difficulty to establish a firmly posed
lagrangian from which to start, since the $1/N_C$ expansion gives no
argument as to the general form of ${\cal L}$. This meant that terms in
${\cal L}$ were mostly chosen for computational convenience  and
carried adjustable parameters. The resulting possibility to 'tailor' the
lagrangian in order to obtain a best value for some quantity under
consideration makes a clear judgement of the merits and defects of the
Skyrmion approach to baryon physics difficult, because most
calculations were limited
to one or a few baryon observables for which the model used could be
fine tuned. Since we want to provide the reader
with a motivation for our project, we shall now investigate in some
detail the status of several quantities within the Skyrme model.

In the following, we present a list of quantities which have been
calculated in $SU(2)$ Skyrme-type models. Although even this fairly
long list is not exhaustive, it may serve as a yardstick
against which
the quality of the model can reliably be measured, and which clearly
illustrates the problems mentioned above. Therefore, in table
\ref{introres} we present entries obtained from the standard Skyrme
model with one free parameter $e$, in one case adjusted
to give the best possible overall
picture of baryon properties and in a second case 
chosen in order to cure already at leading  ${\cal O}(N_C)$ what
was perceived to be a major shortcoming,
namely the too large soliton mass.  It should be evident, that
in the latter case, optimization of one value can only be achieved at
the expense of all others. Eventually, we have included in table
\ref{introres} ranges of the results for the different quantities as can
be found in the skyrmion literature.
\begin{table}[h]
\begin{center} \parbox{10cm}{\caption{\label{introres}
Tree contribution to various
quantities for
the standard Skyrme model with
various values of the Skyrme parameter $e$. Contained is one column that
lists typical ranges for these quantities as found in the skyrmion literature.}}
\begin{tabular}{|c|c|c|c|c|}
\hline
&  & & Skyrmion & \\
&$e=4.25$ & $e=7.3$& Literature & exp.\\
\hline
$M [MeV]$ & 1648 & 941 & 1300-2100 & 939 \\
\hline
$\sigma$ $[MeV]$ & 42 & 11 & 30-60 & 45 $\pm$ 7 \\
\hline
$<\!r^2\!>^S$ $[fm^2]$ & .98 & .56 & $\simeq$ 1 & 1.6 $\pm$ .3 \\
\hline
$g_A$ & .90 & .32 & .8-1.1 & 1.26 \\
\hline
$<\!r^2\!>_A$ $[fm^2]$ & .09 &.06 & $\simeq .1$ & .42 ${+.18 \atop -.08}$ \\
\hline
$<\!r^2\!>_E^S$ $[fm^2]$ & .61 & .47 & .5-.8 & .59 \\
\hline
$\mu^V$ & 1.6 & .47 & 1.5-2.0 & 2.35 \\
\hline
$\mu^S$ & .18 & .29 & .15-.25 & .44 \\
\hline
$<\!r^2\!>_M^V$ $[fm^2]$ & .70 & .40 & .5-.7 & .73 \\
\hline
$\alpha$ $[10^{-4} fm^3]$ & 17.0 & 2.3 & 14-22 & 9.5 $\pm$ 5 \\
\hline
\end{tabular}
\end{center}
\end{table}

Some examples from this list are
\begin{itemize}
\item the {\bf nucleon mass} as a prominent failure
of the Skyrme model, since it comes out
too large by a factor of almost  $2$
regardless of the detailed lagrangian used. An early effort to
reproduce the experimental value in leading ${\cal O}(N_C)$
\cite{adnawi} required the reduction of $f_{\pi}$ to one half of its
experimental value.

\item the 
$\mbox{\boldmath$ \sigma$} $ {\bf term}, where the picture is less
clear. Depending on the model used, it is over or underestimated.
However, this quantity, extracted from the isospin even
part of the $\pi$N scattering amplitude has also seen a
considerable reduction of its experimental value according to the
latest $\sigma$-term update \cite{gals91}. The comparably large error bar is
partly due to inconsistencies in the available data and partly due to
some model dependence in the data analysis.  

\item the {\bf axial vector constant} which is a weak point of Skyrme type
models, too. It is an excellent example for the statement that
adjusting the model to reproduce one quantity exactly is very much
of a mixed blessing with respect to all others. 

\item  the {\bf electromagnetic formfactors}, where it has long been recognized
that magnetic moments generally come out too small, most notably the 
isoscalar one, which is wrong by a factor of $2$, but also the isovector moment
for which our parameter gives $1.6$ nuclear magnetons in contrast to
the experimental finding of $2.35$. Interestingly, the defective
$\mu^S$ has been
impossible to repair even with some weird choice of parameters.

\item the {\bf static electric polarizability}, which is generally too
large if one tries to obtain an overall good fit to the data for other
properties as well. However, there has been some
confusion in the literature concerning the exact status of this
quantity in soliton models which we will have to discuss in more depth
later on (section 3.5). 
\end{itemize}

From the preceeding  discussion, two points should be obvious: There is
a definite need for improvements in the numerical values of several
static quantities if the soliton picture is to proceed from a
qualitative to a more quantitative
description of the baryon, and such improvements can only come from the
consideration of higher orders in $1/N_C$. Because the next to leading
order contribution in $1/N_C$ is carried by pionic fluctuations, this
involves calculation of pion loops, with all associated problems of
regularization and renormalization of what is, strictly speaking, a 
nonrenormalizable theory.
The second point is that it would be highly desirable to reduce the
amount of arbitrariness in the model, preferably by establishing a
tighter connection to meson physics. 

On the other hand, in view of the
obvious problems of Skyrme-type models, one might ask
why a deeper investigation of the soliton picture is at all worthwhile,
since $HBChPT$ seems to provide a systematic way to calculate
baryon properties. This point of view, with which we definitely
disagree, nevertheless deserves a closer look.
\section{Baryon $ChPT$ vs. soliton picture: a comparison}
As mentioned above, $ChPT$, for mesons as well as baryons, is an
expansion in powers of external momenta
and quark masses (plus inverse baryon mass in case of $HBChPT$) 
without any reference to $N_C$ as an expansion
parameter. Nevertheless, the
$N_C$ counting rules outlined above apply generally, not just in
solitonic theories, and one could look into their consequences for
theories coupling external baryons to pions . This in fact has 
been accomplished \cite{mano}, \cite{doma}. The picture that emerges 
is the
following: Suppose one starts with a meson-baryon lagrangian of the
form ${\cal L}={\cal L}_M+{\cal L}_{MB}$ where the subscripts
denote the purely mesonic part and the meson
baryon interactions, and wherein the baryon has been introduced as an
external field and provided ${\cal L}_M$ supports a stable soliton, then the
leading ${\cal O}(N_C)$ of this solitonic theory
corresponds to the set of
{\it all} multi loop graphs without a {\it closed} meson loop generated from
${\cal L}$ and evaluated in the limit of a static baryon. (As a
shorthand, and in order to maintain the distinction between solitonic
theory and $HBChPT$, we shall furtheron call $ChPT$ loops the ones
generated by
${\cal L}$.)
Although, from refs \cite{mano} \cite{doma},
this connection is not strictly proven, it nevertheless seems very
plausible. 
With the same comment, one might go one step further and state 
that the $1$-loop level of the soliton model corresponds to the set of 
multi-loop graphs deduced from ${\cal L}$ containing
one closed pion loop and
evaluated in heavy baryon approximation. 

The calculational scheme provided by (baryon) $ChPT$ now ties
the number of loops to which one carries the
calculation to the specific form of ${\cal L}$: The more $ChPT$ loops, the more
complicated becomes ${\cal L}$. Stated differently, once one has chosen
${\cal L}$, one has a limit on the number of $ChPT$ loops to which the
calculation might be consistently extended. For an infinite number of
loops, ${\cal L}$ would have to contain an infinite number of terms.

Two consequences follow from these remarks: Provided the
picture about the connection of $HBChPT$ and soliton models
outlined in \cite{mano} \cite{doma} is correct, the leading $N_C$ part
of the $ChPT$ 1-loop contribution must be contained in the tree level
of a soliton model. We shall demonstrate this in the course of our
report on several occasions. The other consequence is that a soliton
model must be accompanied by an additional assumption not inherent in
$ChPT$ to justify working with a lagrangian that consists of a finite
number of terms despite summing an infinite number of $HBChPT$ loops
already at leading ${\cal O}(N_C)$.

The practicality of a project involving the calculation of loop
corrections to baryon observables in soliton models therefore hinges
upon the answers to the following questions:
\begin{itemize}
\item
Can we justify the choice of a particular
chiral lagrangian used to construct the soliton, which of necessity must be
truncated to some finite chiral order, by arguments other than
small-derivative-ones? 
\item
Once this general form of lagrangian then has
been established, can we make
contact to meson phenomenology upon using experimental $LECs$ within the
theory, and have good values for all baryon properties?
\end{itemize}

In the course of this report we shall try to answer these questions.
In order to maintain the tightest possible connection to the meson
sector, we will show that it is necessary to use a lagrangian
that is different from the ones found in the skyrmion literature.
Before entering the calculation of loop corrections, we have to
recalculate the entire set of tree level quantities from this
lagrangian, mainly because all tree quantities
should be calculated using the same lagrangian and the same parameter combination, and because
we found several earlier tree level calculations to be in need for improvements,
specifically those for the magnetic polarizability and the electromagnetic ratio of
the $\Delta$ photodecay amplitudes.
Additionally, we calculated the neutron proton split of 
the electric
polarizability, which had not been derived in soliton models before.

For these
reasons and also in order to clarify the relationship between $1$-loop
calculations in the meson and the soliton sector, large parts of the paper
are of review character.

This report is organized as follows: Chapter 2 provides the general
framework for the calculations, namely a brief review of $ChPT$ (section
2.1), the quantization of the soliton and its renormalization in the
presence of those external fields related to the baryon quantities of
interest (section 2.1) as well as the set up of the lagrangian (section
2.2). Chapter 3 then presents the investigations for specific baryon
observables, in detail mass (section 3.1), scalar properties (section
3.2), axial properties (section 3.3), electromagnetic formfactors
(section 3.4) and polarizabilities (section 3.5) and
electromagnetic properties of the $\Delta$ resonance (section 3.6).
Results are
summarized and discussed in chapter 4. In this chapter we also draw some
conclusions concerning several problems which occured during the
process of the calculations to finally return to the questions posed in
the beginning. 
 
\chapter{General framework}

As pointed out in the introduction,
effective field theories are designed to describe more fundamental
theories (which may well be again effective) at low energies where only
part of the degrees of freedom are important. In the standard model
these effective degrees of freedom are assumed to be mesons constrained
by the requirement of global chiral symmetry which is spontaneously
broken into a non chirally invariant vacuum state. Restriction to the
lowest-lying meson states which may be interpreted as Goldstone modes
of the broken symmetry leads to the lagrangian
\be \label{chila}
{\cal{L}}_{eff}(U) ={\cal{L}}^{(2)} + {\cal{L}}^{(4)} + \ldots
{\cal{L}}^{(N)} + \ldots
\ee
where the $ChO \, N$ counts the number of gradients
contained in each term ${\cal{L}}^{(N)}_i$ of
\be
{\cal{L}}^{(N)} = \sum_i \ell^{(N)}_i {\cal{L}}^{(N)}_i
\ee
which comes with an $LEC$ to be fixed from experimental information.
For convenience $U$ is here an $SU(2)$ matrix
\be
U = e^{{i\mbox{\boldmath$ \tau \pi$}}/f_{\pi}} \in SU(2)
\ee
which depends nonlinearly on the pion field. The restriction to pionic
degrees of freedom is of course meaningful only below energies where the next
higher resonances become important. For example, starting from a theory
which includes explicit vector meson degrees of freedom we could
replace the vector through pseudoscalar degrees of freedom by virtue of
a gradient expansion in
powers of the inverse vector meson mass $m_V$. This would formally
generate no terms not
already present in (\ref{chila}), yet make sense numerically only if
the gradients involved were smaller than $m_V$.

A loop expansion of a theory like (\ref{chila}) necessarily brings about the
need for renormalization. Provided the
regularization procedure used respects chiral symmetry, then
(\ref{chila}) is renormalizable in the sense that no counterterm can be
produced which is not present already. However, this statement is
rather academical because for practical reasons it is impossible to
treat ${\cal{L}}_{eff}$ as a whole, instead it must be truncated to
some finite number of terms preferably at some finite chiral order. The
question is of course whether this truncation can be justified.

For the answer, we have to distinguish the vacuum sector from the
soliton sector. Generally, Weinberg \cite{wei} realised that matrix elements
of (\ref{chila}) behave like some power $p$ of external momenta carried by the
source fields. In a graph containing $n_N$ vertices with $N$
derivatives this power is related to the number of loops $n_L$
\be
p = 2 n_L + 2 + \sum_N n_N(N-2).
\ee
One loop graphs from the $N\ell\sigma$ model ${\cal{L}}^{(2)}$ always
come along with a power $ p=4$ irrespective of the number of their
vertices and the divergencies can always be absorbed into the tree
graph coefficients of ${\cal{L}}^{(4)}$. Proceeding to higher chiral
orders this situation changes fundamentally. For example already one
loop graphs from ${\cal L}^{(4)}$ produce all powers $p=4+2n_4$
depending on the number of vertices $n_4$, and consequently their
renormalization affects the coefficients of tree graphs to all chiral
orders in (\ref{chila}).

The characteristic feature of the meson sector now is that external
momenta can be made small by construction i.e. by designing the
experiment such that this requirement is fulfilled. The truncation of
$\cal{L}$ in (\ref{chila}) is then justified, with $ChPT$ being the result. The
crucial difference in the soliton sector is that the soliton itself
constitutes "external" fields, which cannot be made weak by assumption.
In fact, gradients of the soliton profiles are typically of the order
of 700 MeV which is dangerously close to the scale of $m_V$. Therefore,
in the soliton sector, we can neither disregard the counterterms of any
$p$ generated by a $1$-loop graph nor can we, at first glance, get rid
of multi-loop  graphs.

For the latter problem there is a way out. Following Witten \cite{ewi}, higher
loops are suppressed by additional powers of the inverse number of
colours $N_C$. Thus, for a first correction to the leading tree
approximation in the soliton sector, we consider the $1$-loop
contribution to be satisfactory. In contrast to this, the problem of
counterterms to all chiral orders may only be solved by an ad hoc
assumption, namely we must assume that the renormalized $LECs$ of
higher chiral orders are small. Whether this assumption is justified
will be investigated during the course of this report.

\section {Quantum corrections versus $1$-loop corrections in the
soliton sector}

As discussed above, in the meson sector, the calculation of quantum
corrections, next-to-leading order in $N_C$, corresponds to evaluating
the set of $1$-loop graphs generated by the chiral lagrangian. This is
different in the soliton sector. There, the existence of a nontrivial
static solution, the hedgehog
\be \label{he}
U_0 = e^{i \mbox{\boldmath$ \tau {\hat r}$} F(r)}
\ee
means that the classical "vacuum" field configuration is no longer
invariant with respect to the rotational and translational symmetries
of the lagrangian. Instead, one has a set of degenerate "vacua"
connected to each other by these global symmetry transformations.

A fluctuation which parametrizes such a transformation doesn't remain
small in the course of its time evolution and therefore has to be
treated to all orders. Technically, this amounts to introducing
collective coordinates \cite{raja}. It was one of the central assumptions in
the early paper on baryon properties by Adkins, Nappi and Witten
\cite{adnawi} as well as in Skyrme's original work \cite{thrsk}
that the quantization of these collective modes already supplies a
reasonable approximation. In this sense we should distinguish in the
soliton sector the quantum corrections due to collective degrees of
freedom, which have been taken into account from the earliest attempts
on, from quantum corrections caused by $1$-loop diagrams in which
interest emerged much later \cite{ch87}, \cite{zawime}-\cite{mou}.

\subsection{Collective coordinates}

As a prelude to the calculation of $1$-loop corrections, we shall
briefly recapitulate the collective coordinate method used in the
soliton sector \cite{adnawi}. We introduce the position $\mbox{\boldmath$  X$}(t)$ of the
soliton centre and an $SU(2)$ matrix $A(t)$ which parametrizes the
soliton's isorotation through three time dependent Euler angles
\be 
U(\mbox{\boldmath$  x$}, t) = A(t) U_0(\mbox{\boldmath$  x$} -
\mbox{\boldmath$  X$}(t)) A^{\dagger}(t).
\ee
Due to the peculiar structure of the hedgehog solution (\ref{he}) a rotation
in isospace is equivalent to a rotation in coordinate space. With the
angular velocities $A^{\dagger} \dot A = \frac{i}{2} \mbox{\boldmath$  \tau   \Omega$}^R$
one obtains
\be
\dot U = A \left( \frac{i}{2} [\mbox{\boldmath$  \tau \Omega$}^R, U_0]
-  \dot{\mbox{\boldmath$  X$}}
\mbox{\boldmath$ \nabla$} U_0\right) A^{\dagger}
\ee
and the lagrangian
\be \label{slom}
L = - M_0 + \frac{1}{2} M_0 \dot{\mbox{\boldmath$ X$}}^2 + \frac{1}{2} \Theta (\mbox{\boldmath$ 
\Omega$}^R)^2 
\ee
in terms of the soliton mass $M_0$ and the moment of inertia $\Theta$
to be specified later. For the derivation of (\ref{slom}) slow motion of the
soliton has been assumed. With the angular and linear momenta
\begin{eqnarray}
R_a & = & - \frac{\partial L}{\partial \Omega^R_a} \nonumber \\
L_a & = & D_{ab}R_b, \quad D_{ab}  = \frac{1}{2} \langle \tau_a A 
\tau_b A^{\dagger} \rangle \\
P_a & = & \frac{\partial L}{\partial \dot X_a} \nonumber
\end{eqnarray}
( $\langle \quad \rangle $ denotes the trace in isospace ) following the canonical quantization procedure
\be
[R_a, R_b] = - i \varepsilon_{abc} R_c, \quad [L_a, L_b] = i
\varepsilon_{abc} L_c, \quad [P_a, X_b] = - i{\delta}_{ab}
\ee
the hamiltonian is obtained as
\be
H = M_0 + \frac{\mbox{\boldmath$ P$}^2}{2M_0} + \frac{\mbox{\boldmath$ R$}^2}{2 \Theta}
\label{ndel}
\ee
and for the construction of eigenstates with definite spin $(S,S_3)$
and isospin $(T,T_3)$ quantum numbers
\be \label{sta}
< \mbox{\boldmath$  x$}, A | \psi^{T=S}_{T_3, S_3} (\mbox{\boldmath$  P$}, t) > =
\sqrt{\frac{2T+1}{8\pi^2}} (-)^{T-S_3}D^{T=S}_{T_3,-S_3}(A) e^{i(\mbox{\boldmath$  P$}
\mbox{\boldmath$ X$} - E t)}
\ee
$SU(2)$ $D$-functions may be employed. The calculation of matrix
elements with the states (\ref{sta}) for the various currents of interest
will be explained in the corresponding sections.

\subsection{Continuum contributions: The phaseshift formula}

In order to quantize the continuum modes of the chiral lagrangian we
consider the path integral \cite{cheli}
\be
W = N \int d[U] e^{-iZ[U]},\quad Z[U] = \int d^4 x {\cal{L}}
\ee
where $Z$ is the effective action and $N$ a normalization factor.
Fluctuations are now introduced through the following ansatz \cite{galeu}
\be
\label{ansat}
U = A \xi e^{i\mbox{\boldmath$ \tau  \eta$}(\mbox{\boldmath$  x $}- \mbox{\boldmath$  X$})/f_\pi} \xi A^\dagger,
\quad \xi = \sqrt{U_0}.
\ee
$A$ and $\mbox{\boldmath$  X$}$ denote the collective coordinates introduced in the
previous section. The next step then is to expand $\cal{L}$ to
quadratic order in the fluctuations. In contrast to these fluctuations
which come only with a factor $1/f_\pi = {\cal{O}} (N_C^{-1/2})$, time
derivatives on collective coordinates count as ${\cal{O}}(N^{-1}_c)$ and
may be neglected because first of all we are interested in the leading
loop corrections only. Using this adiabatic approximation the path
integral reads
\be
W = N \int d[A]d[\mbox{\boldmath$  X$}]d[\mbox{\boldmath$ \eta$}] e^{-i \int d^4 x[{\cal{L}}(U_0, A, \mbox{\boldmath$ 
X$}) + \frac{1}{2} \eta_a \Omega_{ab} \eta_b]} \; .
\ee
For the decomposition of the measure $d[U] = d[A] d[\mbox{\boldmath$  X$}]d[\mbox{\boldmath$ \eta $}]$
, orthogonality of the fluctuations on the collective modes has to be
presumed in order to guarantee independence of the integration
variables and also to avoid double counting. For that purpose, the zero
modes have to be excluded from the space of allowed proper
fluctuations. Formally, integrating over $\mbox{\boldmath$ \eta$}$, the generating
functional 
\be
Z = \int d^4 x {\cal{L}} (U_0, A, \mbox{\boldmath$  X$}) + \frac{1}{2} \int d^4 x
\langle \ell n \Omega \rangle
\ee
contains a trace log of the operator governing the time evolution of
the fluctuations .
The task now is to evaluate this trace log and, in
particular, to isolate the divergencies residing in it.

Here, one encounters the principal technical difficulty of a soliton
sector calculation: The presence of the static soliton implies that
axial, vector and (pseudo) scalar sources are contained in the
equations of motion (e.o.m) for the fluctuations even in the absence
of true external fields. These sources lead to a metric appearing in
the kinetic part of the e.o.m
\be
\Omega_{ab} = -\partial_t n^2_{ab} \partial_t - h^2_{ab} \, .
\label{metr}
\ee
Although the fact that the norm $n^2_{ab}$ is time independent
simplifies matters somewhat, so far, we are not aware that anybody has
been able to perform a
heat kernel expansion for this kind of operator, which is the method
used in \cite{galeu} to find the residues of the poles in the trace log of
$\Omega$. 

Consequently, one has to resort to a different procedure. The
alternative is the numerical determination of the divergencies from an
exact diagonalization of the e.o.m for the fluctuations, which
according to their time dependence $\sim e^{-i\omega t}$ may be written as 
\be \label{fleom}
h^2_{ab} \eta_b = \omega^2 n^2_{ab} \eta_b \, .
\ee
For convenience fluctuations normalized with respect to a flat metric
may be introduced
\be
\tilde h^2_{ab} \tilde \eta_b = \omega^2 \tilde \eta_b\, ,
 \quad \tilde h^2_{ab}
= n^{-1}_{ac} h^2_{cd} n^{-1}_{db}\, , \quad \tilde \eta_a = n_{ab} \eta_b\, , 
\ee
the spectrum, or, more precisely, the eigenvalues and phaseshifts, are
not affected by this transformation. If $\tilde h^2$ is time
independent and does not contain time derivatives, one has 
\be \label{umf}
\frac{i}{2} \int d^4 x \langle \ell n \Omega \rangle = - \frac{T}{2}
\int d^3\!r\,\, \langle \tilde h \rangle \, ,
\ee
where $T$ limits the time integral. For a one-dimensional potential
scattering problem, it can be shown \cite{dahane} that the continuum part of the
scattering operator's trace may be accounted for through a phaseshift
integral. Using the hedgehog ansatz in the adiabatic approximation the
e.o.m for the fluctuations (\ref{fleom}) may be decomposed into partial waves
\cite{waeck} for which the one-dimensional phase shift formula applies. The
total trace of the scattering operator is then obtained by replacing
$\delta$ by the sum over phaseshifts of all angular momenta
\begin{eqnarray}  
\frac{1}{2} \int d^3\!r\,\, \langle \tilde h - h_0 \rangle & = & \frac{1}{2\pi}
\int^\infty_{m_{\pi}} d \omega \; \omega \; \frac{d}{d\omega} \delta
(\omega) \nonumber \\
& = & \frac{1}{2\pi} \int^\infty_o dp \sqrt{p^2 + m^2_\pi} \frac{d}{dp}
\delta (p) \;.
\label{shift}
\end{eqnarray}
Here, $h^2_0 = - \Delta + m^2_\pi$ corresponds to the free Klein-Gordon
equation in the absence of the soliton fields. In the last step the
$\omega$ integration,  $\omega = \sqrt{p^2 + m^2_\pi}$, has been converted
to a momentum integration. From (\ref{shift}) it is clear that divergencies
are related to the high momentum behaviour of the phaseshifts \cite{mou}
\be
\label{mouas}
\delta (p) \stackrel{p \to \infty}{\longrightarrow} a_0 p^3 + a_1 p +
\frac{a_2}{p} + {\cal{O}} (p^{-3}) \; .
\ee
The explicitly denoted terms give rise to at least logarithmically
divergent expressions. The strategy is now to subtract the worrisome
terms in the phase shift integral and add them separately. Using
dimensional regularization
\begin{eqnarray}
\int \frac{d^{d-1} p}{(2 \pi)^{d-1}} (p^2 + m^2_\pi)^{1/2} & = & m^4_\pi
\left(\lambda + \frac{1}{32\pi^2} (-\frac{1}{2} + \ell n \frac{m^2_\pi}
{\mu^2}) \right) \nonumber \\
\int \frac{d^{d-1} p}{(2 \pi)^{d-1}} (p^2 + m^2_\pi)^{-1/2} & = & 4 m^2_\pi
\left(\lambda + \frac{1}{32\pi^2} \ell n \frac{m^2_\pi}{\mu^2}) 
\right) \nonumber \\
\int \frac{d^{d-1} p}{(2 \pi)^{d-1}} (p^2 + m^2_\pi)^{-3/2} & = & - 8
\left(\lambda + \frac{1}{32\pi^2} (1 + \ell n \frac{m^2_\pi}{\mu^2})
\right) \; ,
\end{eqnarray}
which involves a scale $\mu$ to render the arguments in the logarithms
dimensionless, the poles as $d \to 4$ which reside in
\be \label{lamb}
\lambda = \frac{\mu^{d-4}}{16 \pi^2} \left( \frac{1}{d-4} - \frac{1}{2}
(\Gamma'(1) + \ell n (4 \pi) + 1) \right)
\ee
may be isolated
\be \label{cas1}
\frac{1}{2} \int d^3\!r\,\, \langle \tilde h - h_0 \rangle = E_{cas}(\mu) +
\lambda ( 3 \pi m ^4_\pi a_0 - 4 \pi m^2_\pi a_1 + 8 \pi a_2 ) \; .
\ee
The Casimir energy
\begin{eqnarray} 
E_{cas}(\mu) & = & \frac{1}{2 \pi} \left\lgroup \int_0^{\infty}
\frac{dp}{\sqrt{p^2 + m^2_\pi}} [-p(\delta(p) - a_0p^3 - a_1p) + a_2]  -
m_\pi \delta(0)  \right. \nonumber\\
 & + & \left. \frac{m^4_\pi a_0}{16} (\frac{1}{2} + 3 \ell n
\frac{m^2_\pi }{\mu^2})
 - \frac{m^2_\pi a_1}{4} \ell n \frac{m^2_\pi}{\mu^2} + \frac{a_2}{2}
(1 + \ell n \frac{m^2_\pi}{\mu^2}) a_2  \right\rgroup  \;
\label{cas2}
\end{eqnarray}
is a finite but scale-dependent expression. For the derivation of
eqs.(\ref{cas1}, \ref{cas2}) an integration by parts has been performed and
$m^2_\pi$
has been added to the denominator of the $a_2$ term in order to avoid
infrared problems. The coefficients $a_0, a_1$ and $a_2$ contain in
general all $ChOs$ as is noticed from the fact that $\tilde h =
\sqrt{h^2_0 + \tilde w} $ with some potential $\tilde w$.
Therefore the trace (\ref{cas1}) contains not only all $ChOs$ via the
soliton's stability condition but also an infinite numer of explicit
gradients. The part proportional to the divergent $\lambda$ may
formally be expanded
\be \label{exp}
3 \pi m^4_\pi a_0 - 4\pi m^2_\pi a_1 + 8 \pi a_2 = \sum_{i, N}
\gamma^{(N)}_i \int d^3\!r\,\, {\cal{L}}^{(N)}_i, \quad N \geq 4
\ee
into the complete set of terms ${\cal{L}}_i^{(N)}$ contained in the
chiral lagrangian (\ref{chila}). The coefficient $\gamma^{(N)}_i$ for a
definite term of $ChO \, N$ is in general complicated and scale
dependent through all the renormalized $LECs$ of $ChO$ smaller than
$N$. Again, it is only the $1$-loop contribution to the $N\ell \sigma$
model which corresponds to the lowest non-vanishing $ChO \, 4$ in (\ref{exp}),
where the situation becomes simple. For that particular case
$(a_0^{(0)} =0) \quad a_1^{(2)}$ and $a_2^{(4)}$ are analytically known
\cite{galeu} \cite{mouk} and the coefficients $\gamma^{(4)}_i$ are simple numerical
factors (section 2.2).

According to (\ref{exp}) the divergencies in (\ref{cas1}) may finally
be absorbed
into a redefinition of the lagrangians $LECs$. The total soliton mass
(tree $+$ $1$-loop) is obtained as
\begin{eqnarray}
M (\mu) & = & - \sum_{i,N} (\ell^{(N)}_i - 
\gamma_i^{(N)} \lambda) \int d^3\!r\,\, {\cal{L}}^{(N)}_i + E_{cas} (\mu)
\nonumber \\
& = & M_0 (\ell^{(N)r}_i) + E_{cas}(\mu) \; ,
\label{solmat}
\end{eqnarray}
where $E_{cas}$ represents the Casimir energy (\ref{cas2}) and
\be
M_0(\ell_i^{(N)r}) = - \sum_{i, N} \ell^{(N)r}_i \int d^3\!r\,\,
{\cal{L}}^{(N)}_i
\ee
is the tree contribution to the soliton mass calculated with the
renormalized $LECs$
\be \label{lec}
\ell^{(N)r}_i = \ell^{(N)}_i - \gamma^{(N)}_i \lambda \; .
\ee
It is this relation which for $N=4$ establishes the close connection to
$ChPT$ in the meson sector with the conventions adopted by Gasser and
Leutwyler. With $\lambda$ defined in (\ref{lamb}) the renormalized $LECs$
$\ell^{(4)r}_i$ (\ref{lec}) just coincide with those defined in
\cite{galeu}  and
auspiciously we may take over their analysis (section 2.2).

According to (\ref{lec}) the renormalized $LECs$ for the higher $ChOs$, in
particular their scale dependence, are fixed in principle, however for
practical purposes we do not know the $\gamma^{(N)}_i$ for the
individual higher $ChO$ terms. Even if we knew the value of the
corresponding renormalized $LEC$ at some scale we could not, according
to (\ref{lec}) determine its scale dependence without knowledge of the
$\gamma^{(N)}_i$. For a detailed examination of the scale dependence of
the soliton's mass see section 3.1.2.

As mentioned, the phaseshift formula (\ref{shift}) takes care of the continuum
only; if there exist any bound states at energies $\omega_c$ the zero
point energy $\frac{1}{2} \sum_c \omega_c$ has to be added to the
Casimir energy. In the absence of external fields, the spectrum of
$h^2$ contains no true bound states, instead there exist zero modes due
to the rotational and translational symmetries discussed in section
2.1 . Because of $\omega_c =0$ the zero modes do not explicitly
contribute. However, in the presence of external fields, which violate
rotational or translational symmetries,
zero modes are shifted to finite energies. Their treatment is described
in section 2.1.4.

\subsection{External fields}

The above formulae allow to calculate the static soliton's energy
including quantum corrections due to pion loops. However, the aim of
this investigation is to calculate other baryon properties on the same
footing, too. Such quantities may be defined as the linear change of
the baryon's energy in the presence of an external stimulus.

To be specific, we study the coupling of an external field $j$ with
strength $\varepsilon$ to the lagrangian (\ref{chila})
\be \label{chilaex}
{\cal{L}}(\varepsilon) = {\cal{L}}(\ell_i, U) -  \varepsilon j \cdot J(\ell_i, U) \; ,
\ee
where $J(\ell_i, U)$ denotes the corresponding current density. The
external field has to be chosen suitably so as to give the desired
quantity, e.g. for the calculation of the magnetic moment, $J$
represents the electromagnetic current density and $j \cdot J$
corresponds to the magnetic moment density. Matrix elements of $j \cdot
J$ are then obtained as a derivative of the soliton mass (tree $+$ $1$
loop) in the presence of the external field with respect to its
strength 
$$
\langle N| \int d^3\!r\,\, j J | N \rangle = \left. \frac{\partial
M(\varepsilon)}{\partial \varepsilon} \right|_{\varepsilon =0}
$$
\be
M(\varepsilon) = M_0(\varepsilon,\ell_i^r)+E_{cas}(\varepsilon,\mu)\;.
\label{exter}
\ee
Here, $| N \rangle$ represents the one baryon state (\ref{sta}) .

Considering static baryon properties, the external field can be
specialised to a simple, in many cases space-time-independent form, and
can generally be assumed to be weak. Consequently, we shall proceed as
follows: 
\begin{itemize}
\item[(i)]
The external field is chosen such that the quantity of interest
is obtained from (\ref{exter}) in tree approximation.

\item[(ii)] 
The static soliton is computed from (\ref{chilaex}) for some small
$\varepsilon$. It turns out that for all quantities considered here,
the hedgehog still solves the classical e.o.m in the presence of the
external field.

\item[(iii)] 
The operators $h^2(\varepsilon)$ and $n^2(\varepsilon)$ are obtained 
by expanding ${\cal{L}}(\varepsilon)$ to quadratic
order in the fluctuations. The stability condition for the hedgehog (ii)
ensures that the term linear in the fluctuations is absent.

\item[(iv)] 
The e.o.m for the fluctuations $h^2(\varepsilon) \eta 
= \omega^2(\varepsilon) n^2(\varepsilon) \eta(\varepsilon) $ 
in the presence of the
external field is solved for the scattering phaseshift up to
sufficiently large linear and angular momenta, respectively.

\item[(v)] 
The Casimir energy $E_{cas}(\mu , \varepsilon)$ is computed
according to (\ref{cas2}) and the soliton mass in tree $+$ $1$-loop is
obtained in the presence of the external field.

\item[(vi)] 
Repeating the whole procedure for several small values of
$\varepsilon$ , the derivative $\left. \partial M(\varepsilon) /
\partial \varepsilon \right|_{\varepsilon =0}$
 may finally be computed, which
equals the desired quantity including quantum corrections.

\end{itemize}
We again stress that in this method the integral has to be restricted
to the continuum fluctuations orthogonal on the zero modes. Whereas
this orthogonality holds automatically in the absence of external
fields, it generally has to be imposed in their presence (c.f. next
section) .

A second caveat concerns eq. (\ref{umf}) . If the external field
creates a time-derivative interaction as is the case for the electric
polarizabilities (section 3.5)
\be
\label{cave1}
\Omega = -\partial_t n^2 \partial_t -2 i\varepsilon w \partial_t - h^2 \,
\ee
where $w$ (and $\tilde w=n^{-1}w n^{-1}$) is time independent, we have
\begin{eqnarray}
\label{cave2}
\frac{i}{2} \int d^4\!x \langle \ell n \Omega \rangle &=& 
i\frac{T}{2}\int dp_0\int d^3\!r\,\, \langle \ell n (p_0^2-2 \varepsilon p_0
\tilde w-\tilde h^2) \rangle \nonumber \\
 &=& - \frac{T}{2}\int d^3\!r\,\, \langle \sqrt{\tilde h^2+
\varepsilon^2 \tilde w^2} \rangle \;
\end{eqnarray}
i.e. there is no term linear in $\varepsilon$ and the leading
contribution is of ${\cal{O}}(\varepsilon^2)$.

The specific choice of external field, as well as other details, will be
given separately for each quantity in question within the corresponding
subsection of the next chapter.  

\subsection{Treatment of zero modes in the presence of external
fields} 

Due to translational and rotational invariance the system of e.o.m for
the adiabatic fluctuations possesses 6 zero mode solution $z_a^c$ with
$\omega_c=0$
\be
h_{ab}^2z_b^c=0 ,\quad c=1...6
\ee
which enter the phaseshift formula for the Casimir energy only
indirectly via Levinson's theorem $\delta(0)=6\pi$. Because of
$\omega_c=0$ , there is no direct bound state contribution as e.g. in the
case of the kink's breathing mode \cite{raja}.

In the presence of external fields, the situation changes, since
rotational and/or translational symmetries are in general violated. 
This is always the case if the external field coupled to the lagrangian
is not rotationally or translationally invariant. If such a field is
switched on weakly with strength $\varepsilon$, the zero
modes of the corresponding broken symmetry are shifted to finite 
energies $\omega_c \sim \sqrt{\varepsilon}$. This is understood
immediately, because the external field generates an additional term in
the e.o.m for the fluctuations $h^2(\varepsilon)=h^2+\varepsilon w$
which in perturbation theory leads to $\omega_c^2=\varepsilon \int d^3\!r\,\,
z_a^c v_{ab} z_b^c $ if the matrixelement does not vanish because of
rotational or translational invariance. For the case of the axial
vector coupling constant $g_A$ and the corresponding external axial
field which violates the former but not the latter symmetry (c.f.
section 3.3 ) we calculated the energies of the rotational zero modes
numerically for different field strengths $\varepsilon$. The result,
plotted in fig. \ref{emze} illustrates the behaviour $\omega_c \sim
\sqrt{\varepsilon}$ for the zero modes which, considered as true bound
states with contribution $\frac{1}{2}\sum_c \omega_c $ to the Casimir
energy would lead to a desaster: The corresponding baryon property 
$\left. \partial M(\varepsilon) /
\partial \varepsilon \right|_{\varepsilon =0}$
would aquire infinite quantum corrections.
\vspace*{7.5cm}
\begin{figure}[h]
\begin{center} \parbox{8.2cm}{\caption{ \label{emze} Energy of the rotational
zero modes as 
function of the strength $\varepsilon$ of a constant external axial field.
Because the external field violates rotational invariance, the zero modes are
shifted to finite energies proportional to the square root of the external
field strength $\varepsilon$.}}
\end{center}
\end{figure}

The solution to this problem is obtained by removing the zero modes,
which are not confined by a restoring force, from the space of allowed
small amplitude fluctuations. In fact, their contribution has already
been taken into account in tree approximation by the introduction of
collective coordinates. To remove the redundant variables in order to
avoid double counting we must impose constraints \cite{dir} (see also section
3.5 for details)
\be \label{constr}
\int d^3\!r\,\, z_a^c n_{ab}^2 \eta_b=0 ,\quad \int d^3\!r\,\, z_a^c n_{ab}^2
z_b^d=\delta_{cd} 
\ee
which have to be added with multipliers $\lambda_c$ to the lagrangian.
Only then the Legendre transformation to the hamiltonian will be well
behaved. Without external fields, the constraints (\ref{constr}) are
automatically fulfilled and the calculation of the soliton's Casimir
energy remains untouched.

However, in the presence of symmetry breaking external fields, the
constraints become active and contribute to the e.o.m
\be \label{fleom2}
h^2_{ab}(\varepsilon) \eta_b(\varepsilon)-\lambda_c(\varepsilon)
n_{ab}^2 z_b^c = \omega^2 n^2_{ab}(\varepsilon) \eta_b(\varepsilon) \, .
\ee
The dependence on $\varepsilon$ is indicated wherever necessary; for
the constraint itself this dependence is irrelevant, since
${\cal{O}}(\varepsilon)$ is sufficient to calculate the baryon property
 $\left. \partial M(\varepsilon) /
\partial \varepsilon \right|_{\varepsilon =0}$ of interest and $\lambda_c$ is already of ${\cal{O}}(\varepsilon)$ :
\be
\lambda_c =\int d^3\!r\,\, z_a^c[h^2_{ab}(\varepsilon) - \omega^2
n^2_{ab}(\varepsilon)] \eta_b(\varepsilon) \sim \varepsilon \,
\ee
In fact, we are going to demonstrate that apart from removing the zero
mode  bound states the effect of the constraints on the scattering
states is ${\cal{O}}(\varepsilon^2)$ and may therefore safely be
discarded.      

For this purpose, we first solve the e.o.m (\ref{fleom2}) without
constraints 
\be
h^2_{ab}(\varepsilon)\bar \eta_b(\varepsilon) = \omega^2 
n^2_{ab}(\varepsilon) \bar \eta_b(\varepsilon) \,
\ee
and then treat the constraints as perturbation calculating the DWBA
matrix element which is related to the phaseshift change
\be
{\scriptstyle \triangle} \delta(p)= p \left\{ \int d^3\!r\,\,
\bar \eta_a n_{ab}^2(\varepsilon) z_b^c \right\}
\left\{ \int d^3\!r\,\, z_e^c[h^2_{ed}(\varepsilon) - \omega^2
n^2_{ed}(\varepsilon)] \bar \eta_d \right\} \,  .
\ee
Both integrals vanish linearly with $\varepsilon$ and the phaseshift
change is clearly ${\cal{O}}(\varepsilon^2)$.

Thus the situation becomes quite advantageous in this respect: For the
phaseshift calculation the constraints may be ignored and in the
Casimir energy they just remove the unwanted contribution 
$\frac{1}{2}\sum_c \omega_c$. The
phaseshift formula (\ref{cas1}, \ref{cas2}) as it stands remains correct
in the presence of external fields which violate rotational and/or
translational symmetry.

\section{Chiral lagrangian in the soliton sector}
We now turn to the discussion of the lagrangian to be used in the
soliton sector. Since for consistency reasons we have to solve the
static e.o.m of the same lagrangian that generates the loop graphs, it
is clear that such an object has to contain at least $ChO \; \, 4$
terms - an N$\ell \sigma$ model doesn't support a stable soliton. The
question then is, what chiral order, if any, would be sufficient?

Recalling the remarks of section 2.1 we have to face the
facts that counterterms will contain all chiral orders and, even more
disturbingly, that an expansion in powers of external momenta must
fail because of the size of gradients on the soliton field.

Consequently, it would be meaningless to simply count gradients in
Weinberg-Gasser-Leutwyler fashion. Instead, either the renormalized
$LECs$ beyond some chiral order must be small or a cancellation between
different terms must occur in order to assure negligible contributions.

The smallness of a renormalized $LEC$ is of course to some extent
dependent on the choice of scale. But this choice is restricted in
several ways: A pragmatic restriction stems from the requirement of
a small symmetric $4$th order term in ${\cal L}$ so as not to destroy
the soliton. Another, more fundamental point is that the renormalized
$LECs$ must be adjusted to be predominantly of ${\cal O}(N_C)$ since
otherwise the $1/N_C$ expansion must fail. The lagrangian up to
$ChO \; \; 4$ given by Gasser and Leutwyler \cite{galeu} reads
\begin{eqnarray}
{\cal L}^{GL}&=& -\frac{f^2}{4}<\alpha_{\mu}\alpha^{\mu}>
+\frac{f^2m^2}{4}<U+U^{\dagger}> \nonumber \\
&&+ \frac{\ell_1^r}{4}<\alpha_{\mu}\alpha^{\mu}>^2
+\frac{\ell_2^r}{4} <\alpha_{\mu} \alpha_{\nu}>^2 \nonumber \\
&&- \frac{\ell_4^r}{4} m^2 <\alpha_{\mu}\alpha^{\mu}(U+U^{\dagger})>
+\frac{\ell_3^r+\ell_4^r}{16}m^4<U+U^{\dagger}>^2 \nonumber \\
&&+ \ell_5^r <F_{\mu\nu}^L U F^{\mu\nu R} U^{\dagger}>
\nonumber \\
&&+ i\frac{\ell_6^r}{2}<F_{\mu\nu}^L D^{\mu} U D^{\nu} U^{\dagger}
+F_{\mu\nu}^R D^{\mu} U^{\dagger} D^{\nu} U> 
\;,
\label{galla}
\end{eqnarray}
where
$$
\alpha_{\mu}=U^{\dagger} D_{\mu} U \, , \quad D_{\mu}=\partial_{\mu}
+i[v_{\mu},\cdot]-i\{a_{\mu},\cdot \} 
$$
$$
F_{\mu\nu}^{L,R}=\partial_{\mu}(v_{\nu} \pm
a_{\nu})-\partial_{\nu}(v_{\mu} \pm a_{\mu})-i[(v_{\mu} \pm
a_{\mu}),(v_{\nu} \pm a_{\nu})] 
$$
and $v_{\mu} \; (a_{\mu})$ are external vector (axialvector) fields.
Renormalized $LECs$ and the respective $\gamma_i$ are given in table
\ref{lect}. Using these numbers (\ref{galla}) allows for a soliton
solution; however, with the value of $\ell_2^r$ corresponding to a
Skyrme parameter $e = 7.24$ this soliton is unphysically small,
compare table \ref{introres}.
Therefore one has to conclude that something is missing. To quantify     
'something' and to get a better understanding of higher $ChO$
contributions at least at tree level, we investigate in the following a
pionic N$\ell \sigma$ model coupled to several other mesonic degrees of
freedom, namely scalars, isovector vector and isoscalar vector mesons.

Neither is this list of resonances exhaustive nor is their inclusion
into a chiral lagrangian unambiguous.
Concerning these objections, at least the quality of the vector meson
dominance (VMD) assumption on which our way to incorporate vector mesons is
based can be judged using experimental values for $ChO \, \, 4$ $LECs$ 
as a reference.

\subsection{Contributions from scalar mesons}
To estimate the effects of scalar mesons in a purely pseudoscalar
model, we start from an N$\ell \sigma$ model in which a scalar $\sigma$
meson has been introduced as a dilaton \cite{schwewe}, \cite{schec} 
\cite{gojajosche}. 
This model contains two
parameters, the glueball condensate $C_G \simeq (300 MeV)^4$ \cite{shivaza}
and the
glueball mass $m_G \simeq 1200 MeV$ leading to a scalar meson mass 
$m_{\sigma}=1209 MeV$.

In order to obtain a purely pionic lagrangian up to $ChO \; \; 6$, we
expand the scalar $\sigma$ field in powers of $1/C_G$. Only the
first two terms in this expansion are of relevance, higher ones would
affect the result only at $ChO > 6$. Inserting the approximate $\sigma$
field back into ${\cal L}$, we arrive at
\begin{eqnarray}
\label{locsig}
{\cal L} &=& \frac{f_{\pi}^2}{2}\left\lbrace 
\mbox{\boldmath$ \alpha$}_{\mu} \mbox{\boldmath$ \alpha$}^{\mu}
-2m_{\pi}^2 (1-u)\right\rbrace \nonumber \\
&+& \frac{f_{\pi}^4}{8C_G}\left\lbrace \mbox{\boldmath$ \alpha$}_{\mu}
\mbox{\boldmath$ \alpha $}^{\mu}
-3m_{\pi}^2 (1-u)\right\rbrace^2  \\
&+& \frac{2}{m_G^2 C_G} \partial_{\mu} \sigma^{(2)}\partial^{\mu}
\sigma^{(2)} -\frac{4}{3C_G^2}(\sigma^{(2)})^3-\frac{3f_{\pi}^2
m_{\pi}^2}{2C_G^2}(\sigma^{(2)})^2 (2-u)
\;,\nonumber
\end{eqnarray}
where we used 
\begin{eqnarray}
\sigma^{(2)} &=& \frac{f_{\pi}^2}{4}\left\lbrace \mbox{\boldmath$
\alpha$}_{\mu} \mbox{\boldmath$ \alpha $}^{\mu}
-3m_{\pi}^2 (1-u)\right\rbrace \nonumber \\
\alpha_{\mu} &=& i\mbox{\boldmath$ \tau \cdot \alpha$}_{\mu} \, , \quad
u=\frac{1}{4}<U+U^{\dagger}> 
\;.
\end{eqnarray}
Numerically, we have calculated the chiral angle $F$ from the
original lagrangian \cite{schwewe} and inserted into (\ref{locsig}) (note
that for the present purpose (\ref{locsig}) is understood not to
contain external fields). The 
contributions to mass, $\pi N$ $\sigma$-term and $g_A$ are then compared
in table \ref{locsva}. Convergence in this case is excellent. (The 
Skyrmeterm, which is not affected from $\sigma$ meson
exchange was omitted in this comparison.)
\begin{table}[t]
\begin{center}
\parbox{10.2cm}{\caption{\label{locsva} 
Comparison of exact and approximate scalar meson contributions.
Each of the three intermediate columns contains the sum of approximate pieces
up to 
and including the indicated $ChO$.}} 
\begin{tabular}{|c|c|c|c|c|}
\hline
& $ChO=2$ & $ChO \le 4$ & $ChO \le 6$ & exact \\
\hline
Mass $[MeV]$ & 818.5 & 691.7 & 712.8 & 716.9 \\ 
\hline
$\sigma_{\pi N}$ $[MeV]$ & 45.8 & 28.9 & 34.0 & 34.4 \\
\hline
$g_A$ & .473 & .384 & .40 & .404 \\
\hline
\end{tabular}
\end{center} 
\end{table}

Likewise $LECs$ could be read off and confronted with experimental
values; however, we cannot expect agreement since we deliberately
chose only to include the dilaton $\sigma$ meson; therefore we skip
this here.
\subsection{Contributions from vector mesons}
Proceeding to calculate in analogous manner the effects of vector
mesons, we use as a starting point the so called minimal model \cite{adna}
\cite{meiss}
including symmetry breakers \cite{symmbr} and coupled to external sources.
This is not the most general implementation of $\omega$ and $\varrho$
mesons as far as anomalous pieces of $\cal L$ are concerned
\cite{meiss} \cite{kasche}
, but the
differences can be shown \cite{wall} to manifest themselves only at $ChO \ge
8$.

Expansion of the $\varrho$ and $\omega$ fields in powers of the
inverse vector meson mass and subsequent reinsertion into $\cal L$
yields (again up to $ChO \; \; 6$)
\begin{eqnarray}
{\cal L} &=&  \frac{f_{\pi}^2}{2}\left\lbrace \mbox{\boldmath$
\alpha$}_{\mu} \mbox{\boldmath$ \alpha $}^{\mu}
-2m_{\pi}^2 (1-u)\right\rbrace \nonumber \\
&-& \frac{1}{8}<{\varrho_{\mu\nu}^{(1)}}^2>
-\frac{m_{\pi}^2}{2m_{\varrho}^2}<(D^{\mu}\varrho_{\mu\nu}^{(1)}-ig[V^{\mu},
\varrho_{\mu\nu}^{(1)}]) r^{\nu(1)}> - \frac{N_C}{2}v_{\mu}^0 B^{\mu}
\nonumber \\
&-& \frac{1}{4m_{\varrho}^2} <(D^{\mu}\varrho_{\mu\nu}^{(1)}-ig[V^{\mu},
\varrho_{\mu\nu}^{(1)}])^2> -\frac{m_{\pi}^4}{4m_{\varrho}^2}<r_{\mu}^{(1)} r^{\mu
(1)}> \nonumber \\
&-& \frac{1}{2}v_{\mu\nu}^0 \omega^{\mu\nu (3)} +2(\frac{N_C
g_{\omega}}{2m_{\omega}})^2B_{\mu}B^{\mu} 
\;.
\label{locro}
\end{eqnarray}
Here, the following abbreviations have been used: 
\begin{eqnarray}
\varrho_{\mu}^{(1)}&=& V_{\mu}-\frac{1}{g}v_{\mu} \nonumber \\
V_{\mu} &=& \frac{i}{2g}(\xi^{\dagger}D_{\mu} \xi-D_{\mu}
\xi \xi^{\dagger}) \nonumber \\
\omega_{\mu}^{(3)} &=& -\frac{N_C g_{\omega}}{m_ {\omega}^2}B_{\mu} \nonumber
\\ 
r_{\mu}^{(1)} &=& i \frac{d}{g} 
\xi^{\dagger} ([U,
\alpha_{\mu}]+[\alpha_{\mu}, U^{\dagger}]) \xi
\;.
\end{eqnarray} 
 $B_{\mu}$ is Skyrme's baryon current. The strength $d$ of the standard
symmetry breaker for vector mesons is related to the $SU(3)$ meson mass
differences 
$d=(m_{K^{*}}^{2}-m_{\varrho}^2)/8(m_K^2-m_{\pi}^2)$. Moreover $v_{\mu}^0$ means an
isoscalar vector field and the field strength tensors are defined in the
usual way
$$
f_{\mu\nu}=\partial_{\mu} f_{\nu}-\partial_{\nu}
f_{\mu}-i(g)[f_{\mu},f_{\nu}] 
$$
where the factor $g$ is in place if $f$ denotes a vector meson while it
is absent in case of external fields.
The coupling to external fields was considered since the local
approximation can result in nonminimal couplings.

As an aside, one notes that $\varrho_0$ as well as $\omega_i$ are
induced through the external vector field. Such components are
important for certain properties (e.g.
polarizabilities, see section 3.5) and have so far been missed
in the calculations.

Without external fields, numbers are again computed by inserting the exact
chiral angle obtained from the lagrangian according to  \cite{meiss} \cite{symmbr} for
$g=2.9,\; g_{\omega} = 2.2$ into (\ref{locro}). The comparison in table
\ref{resro} then shows that the $ChO \; 6$ contribution due to the
$\varrho$ is indeed much smaller (by a factor of $4$) than the one of 
$ChO \; 4$; nevertheless the approximation does not improve - the
exact value is, including $ChO \; 6$ underestimated by the same
amount that it was overestimated using only $ChO \; 4$. For the
$\omega$ the first approximation differs from the exact result by a
factor of $2$.   
\begin{table}[h]
\begin{center} \parbox{10.5cm}{\caption{\label{resro} 
Comparison of exact and approximate contributions to the soliton mass (in
$MeV$) for $\varrho$
and $\omega$ mesons.
The two intermediate columns contains the sum of approximate pieces
up to 
and including the indicated $ChO$. They are further divided so as to display
the contributions to symmetric ($\sim m_{\pi}^0$) and symmetry breaking
($\sim m_{\pi}^{(2n)}, \;\; n>0$) pieces individually. Note that, although the 
original Lagrangian
contains terms with $n=1$ at most, its approximate may comprise terms of arbitrary
$n$.}} 
\begin{tabular}{|c|c|c|c|}
\hline
& $ChO \le 4$ & $ChO \le 6$ & exact \\
& $m_{\pi}^0 \quad\quad m_{\pi}^2$ & $m_{\pi}^0 \quad\quad m_{\pi}^2 
\quad\quad m_{\pi}^4$ &  $m_{\pi}^0 \quad\quad m_{\pi}^2$ \\
\hline
$\varrho$ & 494.6 $\quad$ - $\;\;$ & 377.6 $\quad$ 5.9 $\quad$ $-$.36 & 439.0 
$\quad \;$ 5.66 \\ 
\hline
$\omega$ & - $\quad\quad$ -& 475.3 $\quad\;\;$ - $\quad\quad\;$ - $\;$ & 
235.85 $\quad\quad$ - $\;\;$\\
\hline                           
\end{tabular}
\end{center} \end{table}

Since $\omega$ can be expressed exactly in terms of the
baryon source, it is easy to go beyond this approximation. We have
\begin{eqnarray}
\omega_0 (r)&=& -N_C g_{\omega} \int d^3\!r^{\prime}\, \int
\frac{d^3\!q\,}{(2\pi) ^3} \frac{e^{i\mbox{\boldmath$ q(r-r^{\prime})$}}}{\mbox{\boldmath$ q$}^2+m_{\omega}^2}
B_0(r^{\prime}) \nonumber \\
&=& -\frac{N_C g_{\omega}}{m_{\omega}^2}
\sum_n \left( \frac{\partial_r^2}{m_{\omega}^2}\right)^n B_0(r)
\end{eqnarray}
for the (only nonzero) static field $\omega_0$. If $\partial^2_r B_0$
and $m_{\omega}^2$ are of the same magnitude, the approximation will not
converge. This indeed seems to be the case.

From the above reasoning, it is obvious that the $ChO \; 4$
lagrangian accounts for all effects from scalar mesons and to a large
extent for $\varrho$ mesons, too. This last statement is also obvious
from the comparison of VMD predicted versus experimentally known $LECs$,
table \ref{lect}. (The $ \gamma_i $ used in this table are taken from 
\cite{galeu},
$ LECs $ from \cite{edar}).                          
\begin{table}[h]
\begin{center} \parbox{12.5cm}{\caption{\label{lect}
$ LECs $ at scales of $ \mu=m_{\eta} $ and $ \mu=m_{\varrho} $, respectively, 
compared to VMD prediction. 
}} 
\begin{tabular}{|c|c|c|c|c|}
\hline
& $\gamma_i$ & $\ell_i^r(\mu=m_{\eta}) \cdot 10^3 $ & 
$\ell_i^r(\mu=m_{\varrho}) \cdot 10^3$ & VMD \\
\hline
$\ell_1^r$ & 1/3 & $-$3.94 $\pm$1.3 & $-$4.65 $\pm$1.3 & - \\
\hline
$\ell_2^r$ & 2/3 & 6.35 $\pm$1.5 & 4.92 $\pm$1.5 & $1/(16g^2)=$7.3 \\
\hline
$\ell_3^r$ & $-$1/2 & $-$0.22 $\pm$3.8 & $-$0.85 $\pm$3.8 & - \\
\hline
$\ell_4^r$ & 2 & 9.74 $\pm$5.7 & 5.46 $\pm$5.7 & - \\
\hline
$\ell_5^r$ & $-$1/6 & $-$5.88 $\pm$0.7 & $-$5.22 $\pm$0.7 & $-1/(16g^2)=-$7.3 \\
\hline
$\ell_6^r$ & $-$1/3 & $-$14.5 $\pm$1.2 & $-$13.8 $\pm$1.2 &  $-1/(8g^2)=-$14.6 \\
\hline
\end{tabular}
\end{center} \end{table}
\vspace*{6.5cm} 
\begin{figure}[h]
\begin{center} \parbox{8.2cm}{\caption{ \label{endens} Comparison 
of energy densities of purely pionic
models with the three
parameter sets $A$, $B$ and $C$ (solid lines) and the exact density
from  the VMD
lagrangian with $g=2.9,\;\; g_{\omega}=2.2$ (dashed line).}}
\end{center}
\end{figure}
\vfill 
\newpage
What is missing in (\ref{galla}) essentially is due
to the $\omega$ meson. How then can we account for this?
To answer, we compared the energy density of a VMD minimal model with
$g=2.9 \;\; g_{\omega}=2.2$ \cite{meiss} to its local approximate
((\ref{locro}) without symmetry breakers and $6$th order contributions
from $\varrho$).
The chiral angle was in this case computed from the
approximate lagrangian, and the result is plotted in fig.
\ref{endens}. 
(For a converging local approximation, this
computational difference shouldn't matter; however, since we suspected
lack of convergence for $\omega$, and since we would use a chiral
angle from a purely pionic model in our calculations of $1$-loop
corrections, we had to do the
comparison in this way.) 
From fig. \ref{endens} it is obvious that much better agreement may be
obtained by calculating the chiral angle from an approximate lagrangian
with parameters reduced to $g_{\omega} \simeq 1.0$ and $e = 4.5$: 
the missing higher chiral orders renormalize
the original parameters to effective values. Eventually, a
computationally convenient model with $e=4.25$ and $g_{\omega}=0$ is
not worse than the original $e=5.8,   \;\;\; g_{\omega}=2.2$
combination.

These considerations result in the lagrangian  
\begin{eqnarray}
{\cal L}/f_{\pi}^2 &=&  \left. \frac{1}{2} \mbox{\boldmath$
\alpha$}_{\mu} \mbox{\boldmath$ \alpha $}^{\mu}
+ m_{\pi}^2(u-1) \right. \nonumber \\
 &-& \left. \frac{1}{4} c^a_4\left [ (\mbox{\boldmath$
\alpha$}_{\mu} \mbox{\boldmath$ \alpha$}^{\mu})^2-(\mbox{\boldmath$
\alpha$}_{\mu}\mbox{\boldmath$ \alpha$}_{\nu})^2 \right ] +\frac{1}{2}c_4^s(\mbox{\boldmath$
\alpha$}_{\mu}\mbox{\boldmath$ \alpha$}^{\mu})^2 \right. \nonumber \\
&+& \left. c_4^k (\mbox{\boldmath$ \alpha$}_{\mu}\mbox{\boldmath$\alpha$}^{\mu}) (u-1) +c_4^m
(u-1)^2 \right. \nonumber
\\
&+& \left. c_4^e <F_{\mu\nu}^L U F^{\mu\nu R} U^{\dagger}>
+ i\frac{c_4^f}{2}<F_{\mu\nu}^L D^{\mu} U D^{\nu} U^{\dagger}
+F_{\mu\nu}^R D^{\mu} U^{\dagger} D^{\nu} U>
\right.
\nonumber \\
&-& \left. \frac{N_C}{2f_{\pi}^2} v_{\mu}^0
B^{\mu}- 2\pi^4 c_6 B_{\mu} B^{\mu} -\frac{1}{4m_{\omega}^2} v_{\mu\nu}
B^{\mu\nu} \right.
\label{stanlag}
\end{eqnarray}
\begin{eqnarray}
c_4^a &=&\frac{4\ell_2^r}{f_{\pi}^2}=\frac{1}{f_{\pi}^2 e^2} \quad c_4^s
=2\frac{\ell_1^r+\ell_2^r}{f_{\pi}^2} \quad
c_4^k =m^2\frac{\ell_4^r}{f_{\pi}^2} \quad c_4^m
=m^4 \frac{\ell_3^r+\ell_4^r}{f_{\pi}^2}
\nonumber \\
c_4^e&=&\frac{\ell_5^r}{f_{\pi}^2}=-\frac{c_4^a}{4}
\quad \; \; c_4^f=\frac{\ell_6^r}{f_{\pi}^2}=-\frac{c_4^a}{8} \quad \;\; 
c_6=\frac{1}{4\pi^4 f_{\pi}^2} (\frac{N_C g_{\omega}}{m_{\omega}})^2 \;.
\nonumber 
\end{eqnarray}
Note that only the photon $v_{\mu}$ can couple
nonminimally to the baryon current.
For comparison purposes, in the following evaluations, we therefore use
(\ref{stanlag}) with the parameter combinations mentioned above, which will henceforth be
referred to as models $$A \equiv (e=4.25,g_{\omega}=0) \;, $$
$$B \equiv (e=4.5,g_{\omega}=1.0) \;, $$
$$C \equiv (e=5.8,g_{\omega}=2.2) \;. $$ 
These parameters are used to determine
$c^a_4$ and $c_6$ as well as the constants $c_4^e,\; c_4^f$ regulating the
contributions of nonminimal terms, whereas $c^k_4$ and $c^m_4$, both scalar
meson induced, are kept at their
experimental values for $\mu =m_{\varrho}$. One notes that the constant
in front of the nonminimal term produced by the $\omega$  depends solely on the 
vector meson mass and is
therefore not affected by the use of 'effective' coupling constants.
Unless otherwise stated,
$c^s_4$ is set to zero in accordance with experiment. Finally, the finite
renormalization of $f\to f_{\pi}$, $m \to m_{\pi}$ is expressed through the
formulae
\be
f_{\pi}^2=f^2+2\ell_4^r m^2, \qquad f_{\pi}^2 m_{\pi}^2=f^2
m^2+2(\ell_3^r+\ell_4^r)m^4
\;.
\ee

To summarise, we have investigated the local approximations on models
containing explicit scalar and vector meson degrees of freedom in order to
better understand the role of higher ($ChO > 4$)  chiral orders in the chiral 
lagrangian,
which, in contrast to standard $ChPT$ cannot be dismissed out of hand in the 
soliton sector. We concluded that terms of higher chiral
orders due to scalar mesons and isovector vector mesons are essentially unimportant, but
that there is no a priori justification to drop those induced
by the isoscalar vector
meson. Since we could not take those terms into account properly, we used, as
an alternate means to handle this problem, 'effective' coupling constants
in the terms generated by vector mesons 
These coupling constants differ from the ones in the 
original vector meson lagrangian; comparison of exact and approximate energy
densities led to parameter combination $B$, whereas $A$ and $C$ would serve as
reference sets, with $C$ using paramters unchanged relative to the exact
lagrangian and $A$ being particularly  convenient to work with.

\section{One-loop corrections for lagrangian with explicit vector and
scalar mesons}

Although we use a purely pseudoscalar lagrangian for the calculation of
$1$-loop corrections in this report, we briefly discuss the inclusion of
other meson species like vector and scalar mesons. Such models \cite{adna} 
\cite{meikawiwe}-\cite{schwewe}  have been quite sucessful in some respects
(high-energy behaviour of pion-nucleon phaseshifts, formfactors). 

The most general lagrangian of this kind would be of the form
\be \label{chilax}
{\cal{L}}_{x} ={\cal{L}}_{eff}(U) + \sum_x{\cal{L}}_{int}(U,x) +
\sum_x{\cal{L}}_{res}(x)
\;, 
\ee
where  $x=\varrho , \omega ,\sigma \dots $ generically denotes all
possible resonance degrees of freedom and ${\cal L}_{eff}$ is of the
form (\ref{chila}). It is clear that such an object would be even less
manageable than the purely pseudoscalar lagrangian and has to be
accompanied by some simplifying assumptions. The first one is obviously
the restriction to low lying resonances, i.e. vector and (two kinds of)
scalar mesons. The second one uses the fact that upon expanding the
resonances in terms of pseudoscalars, ${\cal L}_x$ itself is of the
form (\ref{chila}). Calculating $1$-loop processes, we have
$\ell_i^{(N)} \to \ell_i^{(N)r}(\mu)$ as previously, but for $N \ge 4$,
$\ell_i^{(N)r}$ may be decomposed into a resonant and a direct part
\be
\ell_i^{(N)r}(\mu)=\ell_{i \,\, res}^{(N)
}+\ell_{i \,\, direct}^{(N)r}(\mu)
\;,
\ee
where $\ell_{i \,\, res}^{(N)}$ depends solely on resonance mass and
coupling constant, but carries no scale dependence. The assumption then
is
that there exists a scale such that $\ell_{i \,\, direct}^{(N)r}(\mu) \simeq
0$, which is an alternative way to spell the concept of vector (and
scalar) meson dominance. This postulate is considerably weaker than
its analogue in the purely pseudoscalar case, namely to assume the
existence of a scale where $\ell_i^{(N)}(\mu) \simeq 0$ for all
higher chiral order terms which cannot be accomodated within our
formalism. 
This constitutes the main conceptual advantage of a calculation
involving resonances explicitly.
The total soliton mass in tree $+$ $1$-loop (\ref{solmat}) generalizes to
\be
M (\mu)  =  M_0 +  \sum_{x} E^x_{cas}(\mu)
\label{solmax}
\;.
\ee
The classical soliton mass $M_0$ depends on parameters introduced by
the additional mesons, e.g. $g_{\varrho},g_{\omega}\dots$ .
All mesons contribute to the Casimir
energy via their phaseshift $\delta^x(p)$ which is a sum over all
channel eigen phases. The number of channels may be considerable, e.g.
9 channels for the $\varrho$-meson, which makes an accurate determination
of the phaseshifts technically difficult. A nice feature is however
that in the high momentum region these phaseshifts behave well, i.e.
in the individual partial wave they tend to zero in contrast to those
of purely pseudoscalar models. Consequently the determination of the
asymptotical constants
\be
\delta^x (p) \stackrel{p \to \infty}{\longrightarrow} a^x_1 p +
\frac{a^x_2}{p} + {\cal{O}} (p^{-3})
\ee
should be less critical compared to the pure pseudoscalar case which
involves an additional term $a_0 p^3$ (compare eq.(\ref{mouas})). With these
preparations the contributions of the individual meson species $x$ to
the Casimir energy may be evaluated with the phaseshift formula 
(\ref{cas2})
\begin{eqnarray} 
E_{cas}(\mu) & = & \frac{1}{2 \pi} \left\lgroup \int_0^{\infty}
\frac{dp}{\sqrt{p^2 + m^2_x}} [-p(\delta^x(p) - a^x_1p) + a^x_2]  -
m_x \delta^x(0)  \right. \nonumber\\
 & & \left. 
 - \frac{m^2_x a^x_1}{4} \ell n \frac{m^2_x}{\mu^2}  + \frac{a^x_2}{2}
(1 + \ell n \frac{m^2_x}{\mu^2})   \right\rgroup  \; .
\label{cas2x}
\end{eqnarray}
Because the meson mass $m_x$ is smallest for pions it is expected that
their contribution dominates the Casimir energy, the contributions of
other mesons being suppressed by their larger mass.

There remains the question concerning the chiral scale $\mu$. For a 
$\pi \varrho \omega$ model which in addition includes both sorts of scalar mesons the
$LECs$ in $ChO \,\, 4$ compare well \cite{edar} with those obtained in
$ChPT$  at
$\mu \simeq m_{\varrho}$ (table \ref{lect}). Therefore, such a choice should be
quite reasonable. Anyhow the Casimir energy will not react very
sensitively on small changes in the chiral scale because it enters only
logarithmically.

Below, we list merits and problems of using
lagrangians with explicit vector- and scalar-mesons.
\begin{flushleft}
Advantages:
\end{flushleft}
\begin{itemize}
\item The $LECs$ of higher chiral  order terms are fixed form the 
assumption of vector (scalar) meson dominance and not simply neglected
as in the pseudoscalar case. 
\item the high momentum phaseshifts behave well
\item all parameters are fixed in the meson sector, these models
possess no free parameters.
\end{itemize}
Disadvantages:
\begin{itemize}
\item the inclusion of additional mesons is not unique as far as the
anomaly is concerned
\item next higher resonances (e.g. axial vectors etc.) are treated only
approximately
\item the tree calculations have to be updated (induced components,
e.g. polarizabilities)
\item a huge coupled channel problem has to be solved with sufficiently
high precision.
\end{itemize}

In the end we decided to use a pseudoscalar lagrangian only. We are
confident that, concerning $1$-loop corrections, the results for a
lagrangian with explicit $\varrho$ and $\omega$ -mesons would come 
close to those of our model $B$. In particular the problems with the
axial quantities discussed in section 3.3 are not expected to be cured
by the introduction of vector mesons.

\chapter{Baryon properties}
In the following sections, we are going to display the specifics of the 
way to calculate
various baryon properties of interest.

For the
presentation, we shall adopt the following pattern:
For each quantitiy under investigation, we will first give some basic
definitions and experimental findings. In
parallel, we explain the choice of external field so as to admit a hedgehog
solution.                                                                   
We then show a recalculation of the tree
level value comprising all relevant terms of the lagrangian (\ref{stanlag}). In
case of the electric polarizability, doubts had been voiced regarding the correct way
to calculate it; we will discuss this issue and conclude that the standard way is
indeed correct.

Proceeding to the $1$-loop calculation, we derive the e.o.m for the
fluctuations and discuss problems which arise in several
instances. They concern the proper meaning of 
vacuum subtraction ($\sigma$-term, e.m. formfactors) and the treatment of
apparently non local terms (electric polarizability).

For a more convenient comparison, numerical results will then be discussed in
a separate section. 
As always, there is an exception from the rule; in the
present case this involves the axial coupling constant, where we give results
immediately since we are forced to
investigate in detail the implications of current algebra for the $1/N_C$
expansion of this quantity.

The full list of calculated quantities involves
\begin{itemize}
\item the Baryon mass (section 3.1),
\item the $\pi N$ $\sigma$ term and  scalar radius (section 3.2),
\item the axial coupling constant $g_A$ and the corresponding axial radius
(section 3.3),
\item the electromagnetic formfactors (more specifically, the isovector 
magnetic
moment and corresponding radius and the isoscalar electric radius)
(section 3.4),
\item the electric polarizability of the nucleon (section 3.5).
\item the electromagnetic properties of the $\delta$ isobar
(section 3.6).
\end{itemize}
With respect to the polarizability, we opined that its neutron proton split
deserves a derivation, which has
therefore been included in section 3.5.3 . 

Although the list of calculated
quantities is fairly long, it is by no means exhaustive, since we are limited
to quantities where the correction is brought about by 'adiabatic loops', i.e.,
loop graphs, in which the pion can be treated in adiabatic approximation.
A calculation involving nonadiabatic fluctuations would also have to deal with 
(adiabatic) 2-loop graphs  which appear at the same level in the  $1/N_C$ 
counting. Such a project is well beyond our present abilities. 
                                      
\section{Baryon mass}
\subsection{Tree approximation}
The classical soliton mass is obtained by inserting the hedgehog ansatz
into the lagrangian (\ref{stanlag}). For convenience, we will give this
quantity in terms of the longitudinal and transversal metric appearing
in (\ref{metr}),
\be
n_{ab}^2=b_L(r) \hat r_a  \hat r_b +b_T(r)(\delta_{ab}-\hat r_a \hat r_b) 
\;,
\ee
which allows to express the mass and most of the other quantities in the
subsequent sections in a very compact way without reference to the
specific terms contained in the lagrangian. For our choice, these
functions are given by
\begin{eqnarray}
b_L &=& 1+2c_4^a \frac{s^2}{r^2} -2c_4^s(F^{\prime
2}+\frac{2s^2}{r^2})-2c_4^k(1-c) +c_6\frac{s^4}{r^4}  \\
b_T &=& 1+c_4^a(F^{\prime
2}+\frac{s^2}{r^2}) -2c_4^s(F^{\prime
2}+\frac{2s^2}{r^2})-2c_4^k(1-c) +c_6\frac{F^{\prime 2} s^2}{r^2} \nonumber
\;,
\end{eqnarray}
where abbreviations $s=sin(F), \;c=cos(F)$ were used.

Variation of the classical soliton mass 
\be
M_0 =\frac{f_{\pi}^2}{3} \int d^3\!r \,\,\left[F^{\prime 2} b_L+
\frac{2s^2}{r^2}b_T \right]
\label{classma}
\ee
with respect to the chiral angle yields the stability condition 
\begin{eqnarray}
\frac{1}{r^2}(r^2 F^{\prime} b_L)^{\prime} &=& \frac{2sc}{r^2} b_T +m_{\pi}^2 s
\nonumber \\
&& -c_4^k \,s (F^{\prime
2}+\frac{2s^2}{r^2}) -2c_4^m\, s\,(1-c)
\:,
\end{eqnarray}
an ordinary nonlinear differential equation, which has to be solved numerically
subject to the boundary conditions $F(0)=\pi, \; F(\infty)=0$ which
guarantee a solution carrying baryon number $B=1$. Clearly the soliton
mass is of ${\cal O}(N_C)$.

It was the fact that the classical mass (\ref{classma}) always came out
much too large in soliton models with realistic parameters which
finally led to the investigation of loop corrections \cite{ch87} 
\cite{zawime}-\cite{mou} which
in the end turned out to considerably lower 
\cite{mouk} \cite{ho} \cite{mou} the numerical 
values for this quantity.

Similar to the soliton mass, we obtain for the moment of inertia
\be
\Theta=\frac{2f_{\pi}^2}{3} \int d^3\!r \,\,s^2 b_T \;,
\ee
which according to (\ref{ndel}) determines the nucleon-$\Delta$ split
\be
\Delta=\frac{3}{2\Theta}
\:,
\ee
a nonadiabatic quantity of ${\cal O}(N_C^{-1})$ related to the angular
rotation. As already mentioned, we do not report about loop corrections to nonadiabatic
quantities although they are by no means less important for the simple
reason that such calculations would become tremendously complicated.

Another nonadiabatic quantity related to the mass is the neutron proton
split which vanishes for the lagrangian  (\ref{stanlag})  because there
exists no term that distinguishes between states of different isospin
$3$-component. In fact, it can be shown that such a terms must not
appear before $ChO \; 8$ \cite{wall}. The $ChO \; 8$ term is isolated
by local approximation from the standard symmetry breaker for
vector mesons and is related to $\varrho \omega$ mixing \cite{symmbr} 
\cite{edal},
\begin{eqnarray}
L^{\varrho\omega} &=& \frac{i g_{\omega}N_C \mu_{\varrho\omega}^2}{16
g_{\varrho}
m_{\varrho}^2 m_{\omega}^2} \int d^3\!r \,\, \partial^{\nu} B^{\mu} <(U\tau_3+
\tau_3 U^{\dagger})[\alpha_{\mu},\,\alpha_{\nu}]> \nonumber \\
&\stackrel{tree}{=} & (\Theta M_{np}) D_{3a}\Omega_a^R
\;.
\end{eqnarray}
Here, $B^{\mu} $ is again the familiar baryon current. With experimental
values for $g_{\omega}=3.3, \; g_{\varrho}=2.8, \; m_{\omega}=782
MeV, \; m_{\varrho}=768 MeV$ and for the $\varrho \omega$ mixing
parameter $\mu_{\varrho\omega}^2=(-4.5 \pm 0.6) \cdot 10^{-3}GeV^2 $
\cite{baal} \cite{mn90} quite a reasonable result is obtained for the $np$ split
\be
M_{np}=\frac{g_{\omega}N_C \mu_{\varrho\omega}^2}{3 \Theta
g_{\varrho}
m_{\varrho}^2 m_{\omega}^2} \int d^3\!r \,\,
B^0(\frac{2F^{\prime}\,s}{r}+\frac{s^2\,c}{r^2}) \,,\quad B^0=-\frac{F^{\prime}
s^2}{2\pi^2 r^2} 
\;,
\ee
compare table \ref{nonares} .
\subsection{Loop corrections and scale dependence of the soliton mass}
For the $1$ loop corrections adiabatic fluctuations around the soliton
background according to the parametrization (\ref{ansat}) are introduced and
conveniently decomposed into longitudinal and transversal components
\be
\mbox{\boldmath$ \eta $}=\eta_L \mbox{\boldmath$ \hat{ r} $} + 
\mbox{\boldmath$ \eta $}_T
\;.
\ee
From the lagrangian expanded to quadratic order in the fluctuations
\begin{eqnarray}
L &\stackrel{\eta^2}{=}& \frac{1}{2}\int d^3\!r\,\,\left\lgroup 
\dot{\mbox{\boldmath$ \eta$}}^2 
-\partial_i \mbox{\boldmath$ \eta$} \partial_i \mbox{\boldmath$ 
\eta$} + \frac{4c}{r} \eta_L \mbox{\boldmath$  \nabla  \eta$}_T \right.
\nonumber \\
&& \qquad + \left. \frac{2(c^2-s^2)}{r^2}\eta_L^2
-(F^{\prime 2}+\frac{2s^2}{r^2}) \mbox{\boldmath$\eta$}_T^2
+\cdots \right\rgroup \nonumber \\
&=& \frac{1}{2}\int d^3\!r\,\, \left\lgroup \dot \eta_a n_{ab}^2 \dot \eta_b 
+\eta_a h_{ab}^2 \eta_b  \right\rgroup
\end{eqnarray}
the operators  $n_{ab}^2$ and $h_{ab}^2$ which determine the e.o.m 
(\ref{fleom}) may
be read off. For simplicity, we  give only the terms generated by the
N$\ell \sigma$ model explictly.

Technically, it is convenient to write the fluctuations in terms of vector
spherical harmonics   
\be
\mbox{\boldmath$ \eta$}(\mbox{\boldmath$ r$}, t)=\sum_{L \ell} f_{L\ell}(r)
\mbox{\boldmath $Y$}_{L \ell M}(\mbox{\boldmath$ \hat{r}$}) e^{-i\omega\, t}
\;,
\ee
which decouple  the e.o.m into electric $\ell=L\pm 1$ and magnetic $\ell=L$ modes.
This procedure is standard and the full differential equations for the radial
functions $f_{L \ell}(r)$ are given elsewhere  \cite{mou} \cite{waeck} \cite{mape}.
The
challenge is now to solve these coupled channel equations for the phaseshifts 
with high enough accuracy which implies large phonon spins ($L_{max} \simeq 
100$) and large momenta ($p_{max} \simeq 25 m_{\pi}$) such that the sum
\be
\delta(p)=\sum_{Lc}(2L+1) \delta_L^c(p)
\ee
converges and the asymptotic behaviour can be reliably extracted. For this
purpose we use the so called variable phase method \cite{calo} which allows to solve for
all desired phonon spins simultaneously. The result is plotted in
fig. \ref{deltsub} where the asymptotic behaviour  has already been subtracted
(a similar  picture appears in \cite{mou} ). Because of the zero modes and in
accordance with Levinson's theorem this subtracted phaseshift starts close to
$6\pi$ and falls off rapidly on a momentum scale of $2 m_{\pi}$.
\vspace*{7.5cm}
\begin{figure}[h]
\begin{center} \parbox{8.2cm}{\caption{\label{deltsub} Scattering 
phaseshift from model $A$ with asymptotic behaviour
subtracted.}}
\end{center}
\end{figure}
It is essentially this function which enters the expression (\ref{cas2}) for
the Casimir energy. The asymptotical constants $a_0,\; a_1,\; a_2$ are  known
analytically for the N$\ell \sigma$ model which fact served as a test for the
program. Numerical values for the full lagrangian are close to those  given by
Moussallam \cite{mou}.

Next, we are going to investigate the dependence of the soliton mass in
$1$-loop approximation (\ref{cas2}, \ref{solmat}) on the chiral 
scale $\mu$. In close
analogy to the $1+1$ dimensional kink \cite{raja} we may expect scale
independence only to zeroth order in the large expansion parameter of
the theory, which in our case is $N_C$ (an ${\cal O}(N_C^{-1})$ scale
dependence must in principle be compensated by 2-loop contributions).

In order to accomplish this task we have to choose the scale, say
$\mu=m_{\varrho}$, such that the lagrangian is dominantly ${\cal
O}(N_C)$ and then we may look at the scale dependence of $M(\mu)$ in
the vicinity of $\mu \simeq m_{\varrho}$. Because we do not know the
$LECs$ of $ChO \ge 6$ and their scale dependence, we use the pure
$4$th order lagrangian $A$.

Before we discuss the numerical results, we will show analytically 
that the mass is
indeed scale independent to ${\cal O}(N_C^0)$ in
the vicinity of $\mu \simeq m_{\varrho}$ provided all $ChOs$ higher than
those already contained in the starting lagrangian are negligible. 

For this purpose we notice that the $LECs$ of $ChO \; 4$ scale
according to
\be
\label{sclec}
\ell^r_i(\mu)=\ell_i^r(m_{\varrho})-\frac{\gamma_i}{32\pi^2}
\ell n(\frac{\mu^2}{m_{\varrho}^2}) \;, \;\; \ell_i^r(m_{\varrho})={\cal
O}(N_C) \;.
\ee
In the vicinity of $\mu \simeq m_{\varrho}$ the soliton mass
(\ref{solmat}) behaves like
\begin{eqnarray}
M_0(\ell_i^r(\mu)) &\stackrel{\mu \simeq m_{\varrho}}{\simeq}&
M_0(\ell_i^r(m_{\varrho}))+\frac{1}{32\pi^2}
\ell n(\frac{\mu^2}{m_{\varrho}^2})  \sum_{i=1}^4 \gamma_i \int d^3\!r\,\,{\cal L}_i
\nonumber \\
&=& M_0(\ell_i^r(m_{\varrho}))+\frac{1}{2\pi} [-\frac{m_{\pi}^2
a_1^{(2)}}{4}+\frac{a_2^{(4)}}{2}] 
\ell n(\frac{\mu^2}{m_{\varrho}^2}) 
\label{scaldep}
\;.
\end{eqnarray}
Because the stability condition for the $N_C$ soliton is calculated
from $M_0(\ell_i^r(m_{\varrho}))$  there is no contribution from the
$\mu$ dependence of the chiral angle.
The low energy constants $\ell_i^r(\mu)$ enter the e.o.m for the
fluctuations and consequently also the phaseshifts $\delta(p)$ as well as
the asymptotical constants $a_0, \; a_1, \; a_2$ as ratios $\ell_i^r(\mu)
/ f^2$ which are all of ${\cal O}(N_C^0)$. From (\ref{sclec}) it is
then immediately clear that the $\mu$ dependence in all these
quantities, $E_{cas}^0$ included, does not appear until ${\cal
O}(N_C^{-1})$. 
In $ChO \; 4$ the remaining terms in (\ref{scaldep})   just
compensate for the scale dependence of the counter terms in
(\ref{cas2}, \ref{solmat}) 
($a_0$ is at least $ChO \; 2$) which is then scale independent
to ${\cal O}(N_C^0)$.

However the constants $m^4  a_0, \; m^2  a_1, \; a_2$
contain all chiral orders, and the higher orders ($ChO \ge 6$) are not
compensated for. Therefore, the expression (\ref{cas2}, \ref{solmat})
cannot be strictly
scale invariant even at ${\cal O}(N_C^0)$. Invariance could only be
restored if higher $ChO$ terms in the lagrangian were switched on upon
leaving $\mu=m_{\varrho}$ where they were assumed to be zero. This
implies that the scale dependence calculated numerically does not only
comprise of small ${\cal O}(N_C^{-1})$ effects but also measures the
magnitude of higher $ChO$ terms not accounted for through
the usage of an effective Skyrme parameter. All the more it comes as a
surprise that the soliton mass in tree $+$ $1$ loop depicted in
fig. \ref{solmasc} ($a$) (solid line) turns out to be almost scale
independent over a very large region of $\mu$. The scale dependence of
the tree contribution (dashed line) is nicely compensated for by that 
of the $1$-loop piece. At small scales $\mu \le 550MeV$ the onset of
scale dependence is rapid till the soliton is destroyed at $\mu \simeq
420MeV$ by the increasing $ChO \;4$ symmetric term 
($c_4^s= 0.017 m_{\pi}^{-2}$,\cite{pt85}). For comparison, we also 
evaluated a model with $e=7.24$ at
$\mu=770MeV$, this choice corresponds to the value implied by the numerical
size of the $LEC$s at scale $\mu=m_{\varrho}$ (table 2.3).  

\hspace{.75cm} (a) \hfill
\vspace*{7.0cm}

\hspace{.75cm} (b) \hfill
\vspace*{7.0cm}
\begin{figure}[h]
\begin{center} \parbox{8.2cm}{\caption{\label{solmasc} Scale dependence of the soliton mass for models which at scale 
$\mu=770 MeV$ 
correspond to ($a$) $e=4.25$ 
($g_{\omega}=0.0$)  and ($b$) $e=7.24$ 
($g_{\omega}=0.0$). Dashed lines: tree values, solid lines:
tree $+$ $1$-loop.}}
\end{center}
\end{figure}
\vfill
\newpage
Any notion of scale independence 
ceases to exist in this case (fig. \ref{solmasc} ($b$) ) which strongly supports the
conjecture that the usage of effective parameters might emulate missing higher
chiral order terms.  

\section{Scalar properties}

The scalar field $\varsigma$ which couples to the quark mass matrix is
introduced into the $SU(2)$ lagrangian (\ref{stanlag}) by the replacement
$m^2 \to m^2(1+\varsigma)$ 
\be
L(\varepsilon)=L-\varepsilon f_{\pi}^2 \int d^3\!r\,\, \varsigma
[m_{\pi}^2(1-u)+2c_4^mu(1-u) -c_4^k
\mbox{\boldmath$ \alpha$}_{\mu} \mbox{\boldmath$ \alpha $}^{\mu} u] \; .
\ee
The special choice
\be
\label{scalfield}
\varsigma=i_0(\sqrt{t}r) =
\frac{sinh(\sqrt{t}r)}{\sqrt{t}r} , \quad t=(p-p^{\prime})^2 \;
\ee
where $p$ and $p^{\prime}$ denote the nucleon four momenta leads to
\be
\label{stabi}
M(\varepsilon) = M+\varepsilon \sigma(t) , \quad \sigma(t)
=\left. \frac{\partial M(\varepsilon)}{\partial \varepsilon} \right|_{\varepsilon =0}
\;,
\ee
the scalar formfactor $\sigma(t)$ in the time like region.

According to the $\sigma$-term update \cite{gals91} the $\sigma$-term should lie
around $\sigma=\sigma(0) \simeq 45MeV$. The interesting quantity for
the extrapolation of the $\sigma$-term from the Cheng-Dashen point is
$\sigma(2m_\pi^2)-\sigma(0) $ which is estimated in \cite{gals91b} to be $15.2
\pm 0.4 MeV$. This value is connected with an extraordinarily large
scalar radius $<r^2>_{\sigma} =6\sigma^{\prime}(t) / \sigma =1.6 fm^2$.

\subsection{Tree approximation}
The scalar formfactor in tree approximation is immediately obtained
from 
\be
\label{sigt}
\sigma(t)  \stackrel{tree}{ =}  f_{\pi}^2 \int d^3\!r\,\, i_0(\sqrt{t}r) [m_{\pi}^2(1-c)+
2c_4^m c(1-c)+c_4^k
(F^{\prime 2}+2\frac{s^2}{r^2}) c] \; .
\ee
There are eventually relativistic corrections \cite{ji} which however neither
affect the $\sigma$-term nor the scalar radius
\begin{eqnarray}
\label{scalrad}
\sigma & \stackrel{tree}{=} & f_{\pi}^2 \int d^3\!r\,\,  [m_{\pi}^2(1-c)+
2c_4^m c(1-c)+c_4^k
(F^{\prime 2}+2\frac{s^2}{r^2}) c]  \\ 
<r^2>_{\sigma} & \stackrel{tree}{=} & \frac{f_{\pi}^2}{\sigma} \int d^3\!r\,\, r^2
 [m_{\pi}^2(1-c)+
2c_4^m c(1-c)+c_4^k
(F^{\prime 2}+2\frac{s^2}{r^2}) c] \nonumber \;.
\end{eqnarray}
The main contribution to the $\sigma$-term is supplied by the familiar
pion mass term of $ChO \, 2$ (first term), the $ChO\, 4$ terms
contribute only about $10 MeV$.

In the chiral limit $m_{\pi} \to 0$ the scalar square radius
$<r^2>_{\sigma}$ diverges as is noticed when the asymptotic behaviour of
the chiral angle
\be
F \stackrel{r \to \infty}{\to} \frac{3g_A}{8 \pi
f_{\pi}^2}  \frac{1+m_{\pi}r}{r^2} e^{-m_{\pi}r}
\ee
is inserted into (\ref{scalrad}) . In particular, we obtain in this
limit 
\be
6\sigma^{\prime}(0)=<r^2>_{\sigma} \sigma \stackrel{m_{\pi}
\to 0}{=} \frac{f_{\pi}^2 m_{\pi}^2}{2} \int d^3\!r\,\,  r^2 F^2 
 \stackrel{m_{\pi} \to 0}{=} \frac{45 g_A^2
m_{\pi}}{128 \pi f_{\pi}^2} \;,
\ee
the standard result of $HBChPT$
with external nucleons \cite{gassa} \cite{bkkm}. This is due to 
the $\Delta$ states
degenerate with the nucleon in soliton models at leading order $N_C$
\cite{cb} .

In general, soliton models, provided realistic parameters are used,
overestimate the $\sigma$-term and underestimate the scalar radius in
tree approximation, the latter deficiency being caused by the former.
Eventually, loop corrections are able to cure both shortcomings.

\subsection{Loop corrections to the scalar formfactor}

Here we will discuss the calculation of the Casimir energy in the
presence of the external scalar field (\ref{scalfield}). The hedgehog
remains solution and the stability condition obtained by variation of
the tree expression (\ref{stabi}) with respect to the chiral angle $F$
picks up contributions from $\sigma(t)$ (\ref{sigt}) .

From the terms quadratic in the fluctuations
\begin{eqnarray}
L(\varepsilon)&\stackrel{\eta^2}{=}&L-\frac{\varepsilon}{2}  \int d^3\!r\,\,  
i_0(\sqrt{t}r)
\left\lgroup -2c_4^k c \dot{\mbox{\boldmath$ \eta$}}^2 
+2c_4^k c (\partial_i \mbox{\boldmath$ \eta$}_T \partial_i \mbox{\boldmath$ 
\eta$}_T + \partial_i \eta_L \partial_i \eta_L) \right. \nonumber \\
&+& \left. 4c_4^k \frac{2c^2-s^2}{r^2} \eta_L r \mbox{\boldmath$  \nabla  \eta$}_T +
m_{\pi}^2 c \mbox{\boldmath$ \eta$}^2 -2c_4^k F^{\prime}s {(\eta_L^2)}^{\prime}+
4c_4^k\frac{c(c^2-3s^2)}{r}\eta_L^2 \right. \nonumber \\
&-& \left. c_4^k c(F^{\prime 2}+2\frac{s^2}{r^2})
(\eta_L^2+ 3\mbox{\boldmath$ \eta$}_T^2) +2c_4^m(c(2c-1)\mbox{\boldmath$ \eta$}^2-2s^2
\eta_L^2) \right\rgroup \;
\label{fleoms}
\end{eqnarray}
the e.o.m for magnetic and electric modes may be generated using the
standard procedure. 

From eq. (\ref{fleoms}) it is noticed that in the vacuum sector or
equivalently far away from the soliton centre ($F\to 0,
s\to  0, c \to 1$)  the $\sigma$-term does not
vanish. This implies that $h_0^2(\varepsilon)$ does in general not
correspond to a free Klein-Gordon equation. However, for $t=0 \quad
(i_0=1)$ and far away from the soliton centre, the sole effect of the
terms in (\ref{fleoms}) stemming from the external scalar field is to
alter the asymptotical pion mass
\be
\label{asmass}
m_{\pi}^2 \to m_a^2=
\frac{m_{\pi}^2(1+\varepsilon)+2c_4^m \varepsilon}{1+2c_4^k \varepsilon}
\ee
such that $h_0^2(\varepsilon)$ corresponds again to a free Klein-Gordon
equation, but with the mass (\ref{asmass}) . This change related to the
vacuum
$\sigma$-term \cite{galeu} is easily taken into account by just using the mass
(\ref{asmass}) in the expression for the Casimir energy (\ref{cas2}) .
Note here, that the renormalized $LECs$ do not depend on the external
field which simplifies the situation considerably.

The external scalar field (\ref{scalfield}) is rotationally
invariant and, for $t=0$, translationally invariant, too. Therefore all
zero modes must be recovered in the presence of the external scalar
field with $t=0$ at zero energy. This serves as a crucial test that
we have solved the stability condition for $F$ and the e.o.m for the
fluctuations correctly.

For $t \neq 0$, i.e. for the scalar radius ($i_0(\sqrt{t}r) \simeq 1+
r^2 t/6 $) the terms proportional to the external scalar field
(\ref{scalfield}) lead to a confining potential (oscillator potential
in the case of the scalar radius) and we are not able to use the
phaseshift formula (\ref{shift}) without manipulation. In coordinate
space the expression $\int d^3\!r\,\, \langle h(\varepsilon)- 
h_0(\varepsilon)\rangle $
collects contributions only from a region $r \le R$
where the soliton profile is unequal zero. We integrate the phase shifts $\delta(\varepsilon)$ and
$\delta_0(\varepsilon)$ of $h^2(\varepsilon)$ and $h_0^2(\varepsilon)$,
respectively, to a radius $R$ where the asymptotical mass becomes
\be
m_a^2=\frac{m_{\pi}^2(1+\varepsilon  i_0(\sqrt{t}R))
+2c_4^m \varepsilon i_0(\sqrt{t}R)}{1+2c_4^k \varepsilon
i_0(\sqrt{t}R)} \;.
\ee
The difference $\delta(\varepsilon)-\delta_0(\varepsilon)$ entering 
the phase shift formula $(\ref{shift})$
should then lead to a Casimir energy independent of $R$, provided $R$
was chosen large enough.
\section{Axial properties}
The most general form of the nucleon axial current in the Breit frame
$p^{\mu}=(E, -\frac{1}{2}\mbox{\boldmath$ q $}),\; p^{\mu \prime}=(E,
\frac{1}{2} \mbox{\boldmath$ q $}) , \;
E=\sqrt{M^2+\frac{1}{4}q^2}$ (no energy transfer
$q^{\mu}=(0,\mbox{\boldmath$   q$})$ ) is
given by
\begin{eqnarray}
\langle N^{\prime}(\frac{1}{2}\mbox{\boldmath$ q$})|A_0^a
|N(-\frac{1}{2} \mbox{\boldmath$ q $})
\rangle  &=& \langle N^{\prime}|\frac{E}{2M^2}G_T(q^2)
(\mbox{\boldmath$ \sigma q$}) \frac{\tau_a}{2}|N\rangle \nonumber \\
\label{deffor}
\langle N^{\prime}(\frac{1}{2}\mbox{\boldmath$ q $})| A_i^a
|N(-\frac{1}{2} \mbox{\boldmath$ q $})
\rangle  &=& \langle N^{\prime}|\left\lgroup\frac{E}{M}G_A(q^2)
(\sigma_i -\hat
q_i(\mbox{\boldmath$ \hat{ q} \sigma$})) \right. \\
&&\left. +(G_A(q^2)-\frac{q^2}{4M^2}G_P(q^2))\hat q_i( 
\mbox{\boldmath$ \hat{q} \sigma$})
\right\rgroup \frac{\tau_a}{2}|N\rangle  \nonumber \;.
\end{eqnarray}
The time component $A^a_0$ and the corresponding pseudotensor
formfactor $G_T(q^2)$ are related to that part of the axial current
which has positive $G$-parity (so-called second class current). In
soliton models the time component of the axial current in tree
approximation 
\begin{eqnarray}
A_0^a & \stackrel{tree}{=}& -f_{\pi}^2 D_{ap} (b_T+c_4^f \Box) \, sc \, 
(\mbox{\boldmath$ {\hat
r} \times \Omega^R$} )_p \nonumber \\
&=& \frac{f_{\pi}^2}{\Theta} (b_T+c_4^f \Box) sc \;
\varepsilon_{piq} \hat
r_i \frac{1}{2 }\{ D_{ap} ,R_q \}
\label{nonad}
\end{eqnarray}
is proportional to the angular velocity $\mbox{\boldmath$  \Omega$}^R$ and to
the right angular momentum $\mbox{\boldmath$  R$}$, respectively. Because after
hermitean ordering the operator 
\be
 \langle N^{\prime}| \varepsilon_{piq} \frac{1}{2 }\{ D_{ap} ,R_q \} |N\rangle =
\frac{i}{2} \langle N^{\prime}|\mbox{\boldmath$  R$}^2 D_{ai} - D_{ai} \mbox{\boldmath$  R$}^2 
|N\rangle =0
\ee
vanishes between nucleon states there is no contribution to the axial
pseudo tensor formfactor \cite{meiss}. There may well be nonvanishing loop
corrections; however, loop corrections to nonadiabatic quantities
(as (\ref{nonad})) become very much involved and are not treated in
this paper. 

The ${\cal{O}}(N_C)$ spatial components of the axial current appear in
soliton models as $A^a_i= D_{ap} \tilde A^p_i$ where the Euler angle
dependent $D$-function transforms the intrinsic $\tilde A^p_i$ to the
lab. system. For the nucleon matrix element we may use
$D_{ap}=-\frac{1}{3} \tau_a \sigma_p$ and the decomposition
\be
\tilde A^p_i=A_L \hat r_i \hat r_p+A_T (\delta_{ip}-\hat r_i \hat r_p)
\label{decomp}
\ee
into longitudinal and transversal parts to obtain
\begin{eqnarray}
\langle N^{\prime}| \int d^3\!r\,\, e^{i\mbox{\boldmath$  q r$}} A_i^a |N\rangle 
&=& \langle N^{\prime}|\int d^3\!r\,\, e^{i\mbox{\boldmath$  q  r$}} D_{ap} 
\tilde A_i^p |N\rangle \nonumber \\
&=& -\frac{2}{3}\langle N^{\prime}| \int d^3\!r\,\, e^{i\mbox{\boldmath$  q  r$}} 
 [ A_L \hat r_i \hat r_p \nonumber \\ 
&&+ A_T (\delta_{ip}-\hat r_i \hat r_p) ] \sigma_p \frac{\tau_a}{2} |N\rangle
\nonumber \\
&=& -\frac{2}{9}\langle N^{\prime}| \int d^3\!r\,\, \left[ j_0(qr)
(A_L+2A_T)\sigma_i \right. \nonumber \\
&& \left. + j_2(qr) (A_L-A_T)(\sigma_i -3\hat q_i( 
\mbox{\boldmath$ \hat{q} \sigma$})) \right] \frac{\tau_a}{2} |N\rangle
\;. \quad
\label{decfor}
\end{eqnarray}
The decomposition (\ref{decomp}) is evident for the tree approximation
(see (\ref{trap}) below) and holds still for the $1$-loop part when
adiabatic fluctuations depending on $\mbox{\boldmath$  r$}$ and momentum $\mbox{\boldmath$  p$}$ are
inserted and the momentum is integrated out. Then $\tilde A^p_i$ is
again a function of only $\mbox{\boldmath$  r$}$ and the decomposition
(\ref{decomp}) is possible.

Comparing (\ref{deffor}) and (\ref{decfor}) the axial formfactor $G_A$
and the induced formfactor $G_P$ can be read off as
$$
\frac{E}{M}G_A(q^2) = -\frac{2}{9}  \int d^3\!r\,\, [ (j_0(qr) + 
j_2(qr)) A_L +(2j_0(qr) - j_2(qr))A_T ]
$$
\begin{eqnarray}
\label{axfor}
G_A(q^2)-\frac{q^2}{4M^2}G_P(q^2)\! &=& \!
-\frac{2}{9}  \int d^3\!r\,\, 
\left[ (j_0(qr) - 2j_2(qr)) A_L \right. \quad\quad \\
&& \quad\qquad \left. +2(j_0(qr) + j_2(qr))A_T \right] \nonumber \;.
\end{eqnarray}
Because $G_P$ is not very well known experimentally we will 
concentrate on $G_A$, in particular on the axial vector
coupling constant
\be
g_A \equiv G_A(0)= -\frac{2}{9}  \int d^3\!r\,\, ( A_L +2A_T ) =
-\frac{2}{9}  \int d^3\!r\,\, \tilde A_i^i \;.
\ee
Relativistic corrections \cite{ji} may be accounted for through the
replacement 
\be
G_A(q^2) \to \frac{M}{E}G_A(\frac{M^2q^2}{E^2}).
\ee
For small momentum transfers these corrections are of minor importance;
$g_A$ is not affected at all and the axial square radius experiences
only a slight increase by $3/4M^2 = 0.03 fm^2$ (compare
(\ref{gart})). 

\subsection{Tree approximation}
In tree approximation, the spatial components of the intrinsic axial
current 
\be
\tilde A_i^p  \stackrel{tree}{=} f_{\pi}^2 \left[ (b_L+c_4^f \Box)
F^{\prime} \hat r_i \hat
r_p + (b_T+c_4^f \Box) \frac{sc}{r}(\delta_{ip}-\hat r_i \hat
r_p) \right]
\label{trap}
\ee
are of the form (\ref{decomp}) as already mentioned and the axial
formfactor is obtained as
\be
\label{axfort}
\frac{E}{M}G_A(q^2) \stackrel{tree}{=} -\frac{2f_{\pi}^2}{9} 
\int d^3\!r\,\, [ (j_0 + 
j_2) (b_L+c_4^f q^2)
F^{\prime} +(2j_0 - j_2)(b_T+c_4^f q^2) \frac{sc}{r} ]
\;.
\ee
For completeness, we give the corresponding formulae for the axial
vector coupling constant and the axial square radius
\begin{eqnarray}
g_A & \stackrel{tree}{=} & -\frac{2f_{\pi}^2}{9} 
\int d^3\!r\,\, ( b_L
F^{\prime} +2 b_T \frac{sc}{r} )
\\
<r^2>_A  & \stackrel{tree}{=} & \frac{3}{4M^2} -\frac{2f_{\pi}^2}{15g_A} 
\int d^3\!r\,\, r^2 ( b_L F^{\prime} +4 b_T \frac{sc}{r}) \nonumber \\
&& + c_4^f\frac{4f_{\pi}^2}{3g_A} 
\int d^3\!r\,\, (F^{\prime} +2 \frac{sc}{r})
\;.
\label{gart}
\end{eqnarray}
It should be mentioned that the main contribution to the axial radius
is due to the non-minimal coupling (last term in (\ref{gart})) .
In tree approximation the axial coupling constant turns out too small
in chiral soliton models provided reasonable parameters are used and
this deficiency can not be cured through inclusion of vector mesons.
Therefore this quantity is considered to be one important candidate for
which loop corrections may prove to be essential.

\subsection{Loop corrections to the axial formfactor}
Here we want to consider the loop corrections to the axial formfactor
(\ref{axfort}). For this purpose we have to discuss the Casimir energy
in the presence of an external axial field $a^a_i$ ,
\be
L(\varepsilon) = L+ \varepsilon \int d^3\!r\,\, a_i^a A_i^a =L+ \varepsilon \int d^3\!r\,\, \tilde a_i^p \tilde A_i^p,
\label{axla}
\ee
$$
A^a_i= D_{ap} \tilde A^p_i, \; a^a_i= D_{ap} \tilde a^p_i
$$
which finally leads to the axial formfactor $g_A(q^2) \equiv
E/M G_A(q^2)$; 
\be
M(\varepsilon) =M +\varepsilon g_A, \quad g_A(q^2)=\left.\frac{\partial
M(\varepsilon)}{\partial \varepsilon} \right|_{\varepsilon =0}
\;.
\label{axma}
\ee
According to (\ref{decomp}) and (\ref{axfor}) the external axial field
has to be chosen as
\begin{eqnarray}
\tilde a_i^p &=&
 \frac{2}{9} [ (j_0+j_2) \hat r_i \hat r_p+
(j_0-\frac{1}{2} j_2)(\delta_{ip}-\hat r_i \hat r_p) ] \nonumber \\
a_i^a &=&
 \frac{2}{9} [ (j_0+j_2) D_{ap} \hat r_i \hat r_p+
(j_0-\frac{1}{2} j_2)(D_{ai}-D_{ap}\hat r_i \hat r_p) ]
\;.
\end{eqnarray}
This simplifies for the axial vector coupling constant $g_A=g_A(0)$ to
\be
\tilde a_i^p =  \frac{2}{9} \delta_{ip} , \quad
a_i^a =  \frac{2}{9} D_{ai} \;.
\ee
We wrote down  here the axial currents and fields in the lab frame (without
tilde)  just to
display their correct axial transformation properties although we do not
need them in the following: the whole calculation is performed in the
intrinsic system using the standard procedure.      

If we expand the lagrangian (\ref{axla}) in the presence of the
external axial field to linear order in the fluctations, we notice that again
the hedgehog ansatz solves the static e.o.m .
The stability condition is obtained by variation of (\ref{axma}) with
the axial formfactor in tree approximation (\ref{axfort}) which yields
a $q$-dependent chiral angle and makes the linear term in the
fluctuations vanish. 

In close correspondence  with the longitudinal and transversal
operators appearing in the external axial field we can distinguish two
pieces in the lagrangian quadratic in the fluctuations
\begin{eqnarray}
L(\varepsilon) & \stackrel{\eta^2}{=}& L - \frac{2\varepsilon}{9} \left\lbrace 
\int d^3\!r\,\, (j_0+j_2) 
[ F^{\prime} \mbox{\boldmath$ \eta$}^2_T+ \cdots ]
\right. \nonumber \\
& +& \left. \int d^3\!r\,\, (j_0-\frac{1}{2}j_2)
\left[
\frac{2s}{r} \eta _L r \mbox{\boldmath$ \nabla \eta$}_T + \frac{2sc}{r}
(2\eta _L^2+ \mbox{\boldmath$ \eta$}_T^2)  + \cdots \right] \right\rbrace
\label{fleomax}
\end{eqnarray}
from which the e.o.m for the fluctuations may be generated (for the
sake of simplicity we have listed only terms produced by the
N$\ell \sigma$ model). From eq.(\ref{fleomax}) it is clear that the axial 
current matrix element is
zero in the vacuum sector $F\to 0$. Therefore
$h_0^2 (\varepsilon)=h_0^2$  corresponds to a free
Klein-Gordon equation with pion mass $m_{\pi}$ and we do not have 
the difficulties discussed in
the preceding section for the $\sigma$-term.

For the $1$ loop calculation of $g_A$ there exists an alternative method
concerning the evaluation of the integral (\ref{axla}) which we shall describe
briefly. The divergence of the axial current is related to the symmetry
breaking terms in the lagrangian (\ref{stanlag})
\begin{eqnarray}
\partial^{\mu} A_{\mu}^a &=& \dot A_0^a + \partial^i A_i^a \nonumber \\
&=& m_{\pi}^2 D_{ap} \left[f_{\pi}^2 \,s\,\hat r_p +f_{\pi}(c\,\hat r_p \eta_L
+ \eta_{p T}) -\frac{s}{2} \hat r_p \mbox{\boldmath$ \eta$}^2 \right] +\cdots
\; ,
\label{alterga}
\end{eqnarray}
where we expanded the standard pion mass term up to quadratic order in the
fluctuations (similarly for the other symmetry breakers not listed here). We  may
now use eq.(\ref{alterga}) at order $\eta^2$ to simplify the integral in
(\ref{axla}) by partial integration,
\begin{eqnarray}
L(\varepsilon) &\stackrel{\eta^2}{=}& L +\frac{2\varepsilon}{9} \int d^3\!r \,\, \partial^i x_p \tilde A_i^p
\nonumber  \\
&=&  L -\frac{2\varepsilon}{9} \int d^3\!r \,\,  x_p \partial^i \tilde A_i^p
\nonumber  \\
&=&  L +\frac{m_{\pi}^2\varepsilon}{9} \int d^3\!r \,\, r\,s \,
\mbox{\boldmath$ \eta$}^2 + \cdots  \;,
\label{altercal}
\end{eqnarray}
where we have checked that the surface terms vanish. Because we neglect the
contribution from $\dot A_0^a$ which is ${\cal O}(N_C^{-1})$ the loop
corrections to $g_A$ calculated according to (\ref{fleomax}) and (\ref{altercal})
, respectively, differ just by ${\cal O}(N_C^{-1})$. 

The disturbing result is now that this difference turns out to be large in our
soliton models ($.45$ for parameter set $A$) which suggests that the 
${\cal O}(N_C^{-1})$ contribution to $g_A$  might be large, too. The
consequence  of a large ${\cal O}(N_C^{-1})$ contribution to $g_A$ and it's
connection to current algebra will be discussed in the following subsection.

\subsection{Current algebra and $1/N_C$-expansion}
From the time component of the axial current in the infinite momentum
frame and the chiral charge commutator
\be
\label{cicu}
[Q_5^a,Q_5^b]=i \varepsilon_{abc} L_c
\;,
\ee
Kirchbach and Riska 
\cite{kr91} derived a model independent version 
of the Adler - Weisberger sum rule
\be
\label{kir}
g_A^2 =1+R
\;.
\ee
In \cite{kr91}, the "$1$", produced by the commutation 
relations, is of ${\cal O}(N_C^0)$ and $R$ is always positive. 

From (\ref{kir}), it is immediately obvious that the $1/N_C$ expansion
cannot converge reasonably for both, $g_A^2$ and $R$, because otherwise
we would expect the ${\cal O}(N_C^0)$ contributions to both of these
quantities to be small compared to the experimental values of $1.56$
and $.56$ respectively, which violates (\ref{kir}) at ${\cal O}(N_C^0)$!
Obviously, the $1/N_C$ expansion can converge rapidly, if it is to
converge at all, only for one of the two quantities. From our soliton
model calculations we conjecture that this is the case for $R$ and not
for $g_A^2$: For example, 
the model $A$ in tree $+$ $1$-loop provides the ${\cal O}(N_C^2)$ and
${\cal O}(N_C)$ contributions $g_A^2=R=.83-.41=.42$. Clearly, this seems
to converge to the experimental value of $R$ rather than to that of
$g_A^2$. Interestingly, such a result (negative ${\cal O}(N_C^0)$ contribution
to $g_A$) was already found in \cite{zawime}, although we disagree with
the procedure used there. If we add the current algebra "$1$" of ${\cal O}(N_C^0)$ from
eq.(\ref{kir}) by hand, thereby assuming that $g_A$ calculated up to ${\cal
O}(1/N_C)$ exhausts relation ($\ref{kir}$), we obtain the quite satisfactory result
$g_A=.91-.25+.54=1.20$ at the expense of a large ${\cal O}(N_C^{-1})$
contribution to $g_A$. The suggestion that there might be a sizeable ${\cal
O}(1/N_C)$ piece contained in $g_A$ is supported by the alternative
calculation in the preceeding subsection, which yields a large difference precisely in this order. However,  
an explicit calculation of this piece, which involves tree (with two angular
velocities), nonadiabatic $1$-loop (one angular velocity) and adiabatic
2-loop contributions
seems forbiddingly complicated.
Using instead the ad hoc addition of the CA " $1$",
we may estimate the contributions to $g_A$ for the other
parameter sets, too. 
The individual $1/N_C$ contributions to $g_A$ for the various models are listed
in the table \ref{axres}. It is noticed that the $1/N_C$ piece is large and positive
throughout and increases the tree $+$ $1$ loop value towards the experimental
datum, except for the case of parameter combination $D \; (e=3.75,\; g_{\omega}=0)$, designed to yield
acceptable numbers for $g_A$ already at tree $+$ $1$ loop level at the expense of
all other
quantities, which turns out to considerably overestimate $g_A$ and should
therefore be discarded.

For the axial radius we may employ the same procedure. Instead of (\ref{kir}),
one has \cite{affr73}
\be
g_A^2<r^2>_A =<r^2>^V_1+X
\ee
with the isovector square radius   
$<r^2>^V_1=.58fm^2$ which determines the nucleon-$\Delta$ transition radius $<r^2>_A^*
$. 
(Note that, in the end, we will be interested in the quantity $g_A<r^2>_A $ 
which is the
slope of the formfactor). 
The isovector square radius related to the isovector charge square radius and 
magnetic moment 
\be
<\!r^2\!>^V_1=<\!r^2\!>^V_E-\frac{3}{M^2}(\mu^V-\frac{1}{2})
\ee
appears due to the CA commutation relations and is ${\cal O}(N_C^0)$. 

From the experimental datum $<r^2>_A=.42fm^2$ we again find a small value $[g_A
<r^2>_A]^{(1+0)}=.19fm^2$ (superscripts here denote $N_C$ orders) in tree $+$ $1$ loop and a large positive $1/N_C$
correction 
$[g_A<r^2>_A]^{(-1)}=.34fm^2$. Again for the reasonable sets $(A)-(C)$ the
pattern
is repeated and the tree $+$ $1$  loop result enhanced toward the experimental
value $g_A<r^2>_A=.53fm^2$ although in this case there is an overestimation for 
the first three models  
and model $(D)$ is not widely off the mark. Altogether, for the axial radius,
numbers are less conclusive in
view of the large error of the axial radius of ${+.18 \atop -.08} fm^2$.
\begin{table}[h]
\begin{center} \parbox{9.5cm}{\caption{\label{axres} 
Comparative listing of tree, $1$-loop and estimated
$1/N_C$ piece of $g_A$ and $g_A <\! r^2\!>_A$ for the four
parameter  combinations $A$, $B$, $C$ and $D$.}}
\begin{tabular}{|c|c|c|c|c|}
\hline
& $A$ & $B$ & $C$ & $D$\\
\hline
$g_A^{(1)}$ & .91 & .96 & 1.0 & 1.14 \\
\hline
$g_A^{(0)}$ & $-$.25 & $-$.15 & $-$.11 & .14  \\
\hline
$g_A^{(-1)}$ & .54 & .48 & .44 & .34  \\
\hline
$g_A$ & 1.20 & 1.29 & 1.36 & 1.62 \\
\hline
$[g_A<\!r^2\!>_A]^{(1)}$ $[fm^2]$ & .41 & .42 & .41 & .62  \\
\hline
$[g_A<\!r^2\!>_A]^{(0)}$ $[fm^2]$ & $-$.13 & $-$.07 & .11 & $-$.16 \\
\hline
$[g_A<\!r^2\!>_A]^{(-1)}$ $[fm^2]$ & .38 & .35 & .28 & .28 \\
\hline
$[g_A<\!r^2\!>_A]$ $[fm^2]$ & .62 & .65 & .76 & .69 \\
\hline
$<\!r^2\!>_A$ $[fm^2]$ & .55 & .54 & .59 & .46 \\
\hline
\end{tabular}
\end{center} 
\end{table}

This unusual scenario (negative ${\cal O}(N_C^0)$ and large
positive ${\cal O}(N_C^{-1})$ contribution) for axial quantities 
is in striking contrast to all other
models (nonrelativistic quark model, bag models, NJL model) which find a
positive ${\cal O}(N_C^0)$ and no ${\cal O}(N_C^{-1})$ correction. This is very
serious because not only the experimental value of $g_A$ itself but also the
individual contributions of different $N_C$ orders are fixed in principle and
if the soliton approach is indeed correct, all others are wrong and vice versa.  
Interestingly, the same numbers for the quantum correction at ${\cal
O}(N_C^{-1})$ come out of a calculation which assumes that Skyrme type
models should give a positive
remainder $R$ and imposes $SU(4)$ current algebra for the spatial
components of the axial current. This calculation is presented in some
detail in appendix A.

As an addendum, we have calculated the scale dependence of $g_A$ starting from
set $A$. The result, depicted in fig. \ref{scalga}
shows the scale independence to be not worse than for the mass considering the
fact that the tree value (dashed line) of $g_A$ is much more sensitive
to the change of
scale than the soliton mass. Due to the too large symmetric $4$th order term
which incurs numerical difficulties, we were unable to calculate the
correction at scales below $600 MeV$. 
\vspace*{7cm} 
\begin{figure}[h]
\begin{center} \parbox{8.2cm}{\caption{ \label{scalga}
Scale dependence of $g_A$ for a model which at scale $\mu=770 MeV$ 
corresponds to $e=4.25$ 
($g_{\omega}=0.0$). Dashed line: Tree value, solid line:
tree $+$ $1$-loop.}}
\end{center}
\end{figure}

\section{Electromagnetic formfactors}
The nucleon matrix element of the electromagnetic current in the Breit
frame 
\begin{eqnarray}
\label{emfor}
\langle N^{\prime}(\frac{1}{2}\mbox{\boldmath$ q $})|J_0(0)
|N(-\frac{1}{2} \mbox{\boldmath$ q $})
\rangle  &=& \langle N^{\prime}|G_E^V(q^2)\tau_3 +G_E^S(q^2)
|N\rangle \\
\langle N^{\prime}(\frac{1}{2}\mbox{\boldmath$ q $})|J_i(0)
|N(-\frac{1}{2} \mbox{\boldmath$ q $})
\rangle  &=& \langle N^{\prime}|\left[\frac{G_M^V(q^2)}{2M}\tau_3
+\frac{G_M^S(q^2)}{2M} \right] i(\mbox{\boldmath$  \sigma \times q$})_i
|N\rangle \nonumber
\end{eqnarray}
fixes the isoscalar and isovector formfactors which are linear
combinations of the proton and neutron formfactors
\be
G_{E,M}^S=\frac{1}{2}(G_{E,M}^p+G_{E,M}^n) \; , \; \;
G_{E,M}^V=\frac{1}{2}(G_{E,M}^p-G_{E,M}^n) 
\ee
normalized to the isoscalar and isovector charges and magnetic moments
\be
G_E^S(0)=\frac{1}{2} , \;\; G_E^V(0)=\frac{1}{2} ,\; \; G_M^S(0)=\mu^S ,\; \;
G_M^V(0)=\mu^V \; .
\ee  
The corresponding radii
\be
<r^2>_{E,M}^{S,V} =-\frac{6}{G_{E,M}^{S,V}(q^2)}
\left. \frac{dG_{E,M}^{S,V}(q^2)}{dq^2}\right|_{q^2=0} 
\ee
are related to the slopes of these formfactors.

According to the charge operator $Q=\frac{1}{2}(\tau_3+1/N_C)$
which guarantees integer charges of ${\cal{O}}(N_C^0)$ for mesons and
baryons the electromagnetic current in soliton models 
\be
J_{\mu} = V_{\mu}^3 +\frac{1}{2}V_{\mu}^0 =D_{3p} \tilde V_{\mu}^p
+\frac{1}{2}V_{\mu}^0 
\ee
decomposes into an isovector part represented by the third component of
the vector current and an isoscalar part related to the baryon current.
For the nucleon matrix element we may again use
$D_{3p}=-\frac{1}{3} \tau_3 \sigma_p$ to obtain
\be
\langle N^{\prime}| \int d^3\!r\,\, e^{i\mbox{\boldmath$  q  r$}} J_{\mu} |N\rangle 
= \langle N^{\prime}|\int d^3\!r\,\, e^{i\mbox{\boldmath$  q r$}}
\left[-\frac{1}{3}\tau_3 \sigma_p  
\tilde V_\mu^p+ \frac{1}{2}V_{\mu}^0 \right] |N\rangle \;.
\ee
Comparison with (\ref{emfor}) gives the electromagnetic formfactors
\begin{eqnarray}
&&G_E^V(q^2) = -\frac{1}{3}  \int d^3\!r\,\, j_0(qr) \sigma_p \tilde V_0^p
\nonumber \\
&&G_E^S(q^2) = \frac{1}{2}  \int d^3\!r\,\, j_0(qr) V_0^0 
\nonumber
\\
&&\frac{G_M^V(q^2)}{2M} = \frac{1}{6}  \int d^3\!r\,\, \frac{j_1(qr)}{qr}
\varepsilon_{pij} x_j \tilde V_i^p 
\nonumber \\
&&\frac{G_M^S(q^2)}{2M} = \frac{1}{4}  \int d^3\!r\,\, \frac{j_1(qr)}{qr}
(\mbox{\boldmath$  \sigma \times r$} )_i V_i^0 \;.
\label{emfor2}
\end{eqnarray}
Relativistic corrections \cite{ji} may be taken into account by the
replacements 
\be
\label{relcor}
G_E^{S,V}(q^2) \to G_E^{S,V}(\frac{M^2q^2}{E^2}) \;, \;\;
G_M^{S,V}(q^2) \to
\frac{M^2}{E^2}G_M^{S,V}(\frac{M^2q^2}{E^2}) \;.
\ee
Again, for small momentum transfers these corrections are of minor
importance; charges and magnetic moments as well as the electric radii
are not affected, the magnetic square radii collect a small
contribution $3/2M^2 = 0.06 fm^2$.
\subsection{Tree approximation}
In tree approximation we obtain for the intrinsic isovector current
\begin{eqnarray}
\tilde V_0^p  &\stackrel{tree}{=}& \frac{f_{\pi}^2}{\Theta} (b_T+c_4^f \Box)
s^2 \{ R_p-\hat r_p (\mbox{\boldmath$\hat{r} R$}) \} \nonumber \\
\tilde V_i^p  &\stackrel{tree}{=}& f_{\pi}^2 (b_T+c_4^f \Box)
\frac{s^2}{r} \varepsilon_{pij} \hat r_j
\label{vecu}
\end{eqnarray}
and similarly for the isoscalar current
\begin{eqnarray}
V_0^0 & \stackrel{tree}{=} & (1-\frac{1}{m_\omega^2} \Box) B_0 ,\;\;
B_0=-\frac{1}{4 \pi^2} \frac{2F^{\prime}s^2}{r^2} , 
\nonumber \\
V_i^0 & \stackrel{tree}{=} & (1-\frac{1}{m_\omega^2} \Box) B_i , \;\; 
B_i=\frac{B_0}{\Theta}(\mbox{\boldmath$ r\times R$})_i
\label{iscu}
\; ,
\end{eqnarray}
where the nonminimal $ChO \; 6$ coupling through the $\omega$-meson,
essential for the isoscalar radius, was taken into account.
With the currents (\ref{vecu}, \ref{iscu}) the formfactors
(\ref{emfor2}) read in tree
approximation ($R_p=-\sigma_p/2$ for nucleon states)
\begin{eqnarray}
G_E^V(q^2) = \frac{f_{\pi}^2}{3\Theta} \int d^3\!r\,\, j_0(qr)
(b_T+c_4^f q^2) s^2 \quad\quad & & G_E^V(0)=\frac{1}{2} \nonumber \\ 
G_E^S(q^2) = \frac{1}{2}(1-\frac{q^2}{m_\omega^2}) \int d^3\!r\,\, j_0(qr)
B_0 \quad \qquad&&G_E^S(0)=\frac{1}{2}  \\
G_M^V(q^2) = 2M \frac{f_{\pi}^2}{9}  \int d^3\!r\,\, \frac{3j_1(qr)}{qr}
(b_T+c_4^f q^2) s^2  &&G_M^V(0)=\mu^V=\frac{M}{3} \Theta \nonumber \\
G_M^S(q^2) = \frac{M}{6 \Theta}(1-\frac{q^2}{m_\omega^2}) 
\int d^3\!r\,\, \frac{3j_1(qr)}{qr} r^2 B_0 \quad &&G_M^S(0)=\mu^S=\frac{M}{6
\Theta} <r^2>_B \;. \nonumber
\end{eqnarray}
The corresponding radii are
\begin{eqnarray}
\label{radii}
<r^2>_{E}^{V} &=& <r^2>_{\Theta} - c_4^f \frac{4f_{\pi}^2}{\Theta}
\int d^3\!r\,\, s^2 \; , \quad <r^2>_{\Theta} =
\frac{2f_{\pi}^2}{3\Theta} \int d^3\!r\,\, r^2 b_T s^2 \nonumber \\
<r^2>_{E}^{S} &=& <r^2>_{B}+\frac{6}{m_{\omega}^2} \;, \qquad \qquad \qquad 
<r^n>_{B}=\int d^3\!r\,\, r^n B_0 \nonumber \\
<r^2>_{M}^{V} &=& \frac{3}{2M^2}+ \frac{3}{5} <r^2>_{\Theta} 
 - c_4^f \frac{4f_{\pi}^2}{\Theta} \int d^3\!r\,\, s^2\nonumber \\
<r^2>_{M}^{S} &=& \frac{3}{2M^2}+\frac{3}{5} \frac{<r^4>_B}{<r^2>_B}
+\frac{6}{m_{\omega}^2} \;.
\end{eqnarray}
where the relativistic correction due to (\ref{relcor}) was included.
The nonminimal couplings through the $\varrho$-meson in $ChO \, 4$ and
through the $\omega$-meson in $ChO \, 6$ are essential and lead to
a considerable enhancement of all these radii.

In the following we will calculate loop corrections to some of these
quantities. As is noticed from (\ref{vecu}, \ref{iscu}) the adiabatic
ones containing no angular velocity repectively no angular momentum
for which the method to calculate the Casimir energy outlined above applies 
are the isovector magnetic formfactor of ${\cal{O}}(N_C^1)$ and the 
isoscalar electric formfactor of ${\cal{O}}(N_C^0)$.

\subsection{Loop corrections to the isovector magnetic formfactor}
For the calculation of the Casimir energy in the presence of an
external isovector field $v_i^a$
\be
L(\varepsilon) = L+ \varepsilon \int d^3\!r\,\, v_i^a V_i^a =L+ \varepsilon \int d^3\!r\,\, \tilde v_i^p \tilde V_i^p,
\label{vecla}
\ee
$$
V^a_i= D_{ap} \tilde V^p_i, \quad v^a_i= D_{ap} \tilde v^p_i
$$
we proceed in very much the same way as for the axial field of the
previous section. The external field which leads to the magnetic
formfactor $g_M^V(q^2) \equiv
G_M^V(q^2)/2M $ in the form
\be
M(\varepsilon) =M +\varepsilon g_M^V, \;\; g_M^V(q^2)=\left. \frac{\partial
M(\varepsilon)}{\partial \varepsilon} \right|_{\varepsilon =0}
\;
\ee
follows from (\ref{emfor2}) and (\ref{vecla}):
\be
\tilde v^p_i=\frac{3}{2}\frac{j_1(qr)}{qr}
\varepsilon_{ip\ell} x_{\ell} , \quad
v^3_i=\frac{3}{2}\frac{j_1(qr)}{qr} D_{3p}\varepsilon_{ip\ell} x_{\ell}
\label{extv}
.
\ee
The isovector field $v_i^a$ in the lab. frame has the correct
transformation property. Again the hedgehog solves the static e.o.m.
and makes the term linear in the fluctuations vanish. From the term
quadratic in the fluctuations, 
\begin{eqnarray}
L(\varepsilon) &\stackrel{\eta^2}{=}& L + \frac{\varepsilon}{18} \int d^3\!r\,\, 
\frac{3j_1}{qr} 
[ (c^2- s^2) ( 2\eta _L^2+\mbox{\boldmath$ \eta$} _T^2 ) 
+ 2r\,c\, \eta _L \mbox{\boldmath$\nabla \eta$}_T  
+\cdots ] \nonumber \\
&&
\end{eqnarray}
the e.o.m. for the fluctuations are set up. Here, apparently the
additional terms due to the interaction do not vanish in the vacuum
sector such that $h_0^2 (\varepsilon)=h_0^2+\varepsilon w_0$ with 
an asymptotic potential  $w_0 \neq 0$.
However, for the Casimir energy we have to evaluate  $
\int d^3\!r\, \langle \tilde h(\varepsilon) - h_0(\varepsilon) \rangle \simeq 
\int d^3\!r\, \langle \tilde h(\varepsilon) - h_0-\varepsilon w_0/2h_0
 \rangle = \int d^3\!r\,\, \langle \tilde h(\varepsilon) - h_0
\rangle$.  The term
which contains the asymptotic potential turns out to vanish because of 
$\int d^3\!r\,\, \langle \varepsilon w_0/2h_0\rangle =0$, which
implies that the vacuum matrix
element of the isovector current is zero, as it should be. For
practical reasons, we must subtract the corresponding term which is
also zero in the soliton sector from $\tilde h^2 (\varepsilon)$ in
order to obtain stable phase
shifts.

In the presence of the vector field (\ref{extv}) which is neither
rotationally nor translationally invariant (even at $q=0$) the e.o.m.
for the fluctuations unfortunately cannot be checked by the zero modes. 

\subsection{Loop corrections to the isoscalar formfactor}
The external isoscalar field $v_0^0=\varepsilon j_0(qr)$,
\be
L(\varepsilon) = L- \frac{\varepsilon}{2} \int d^3\!r\,\, v_0^0 V_0^0 =L 
-\frac{\varepsilon}{2} (1-\frac{q^2}{m_{\omega}^2}) 
\int d^3\!r\,\, v_0^0 B_0
\ee
immediately leads to the isoscalar electric formfactor 
\be
M(\varepsilon) =M +\varepsilon G_E^S, \; G_E^S(q^2)=\left. \frac{\partial
M(\varepsilon)}{\partial \varepsilon} \right|_{\varepsilon =0} \;
.
\ee
Of course, the hedgehog is again solution and the e.o.m. for the
fluctuations are obtained from 
\be
L(\varepsilon) \stackrel{\eta^2}{=} L - \frac{\varepsilon}{2}
(1-\frac{q^2}{m_{\omega}^2}) \int d^3\!r\,\, j_0^{\prime} 
\left[ \frac{sc}{r^2} ( 2 \eta _L^2+\mbox{\boldmath$ \eta$} _T^2) +\frac{2s}{r}
\eta _L \mbox{\boldmath$  \nabla 
\eta$}_T \right] \;,
\label{fleomiso}
\ee
where we have used partial integration. It is noticed that for $q=0$
the loop contributions in (\ref{fleomiso}) vanish identically. This is
of course to be expected because the isoscalar charge
$G_E^S(0)=1/2$ should not be subject to quantum corrections.
Apart from this (trivial) test we checked the e.o.m. by the rotational
zero mode, which remains at zero energy because the external field is
rotationally invariant.

\section{Polarizabilities}
The electromagnetic polarizabilities of the nucleon have been calculated in 
various approaches,
amongst them nonrelativistic quark- \cite{damapro}-\cite{schle}, MIT bag-
\cite{hebe} \cite{schae} and chiral quark models \cite{wewe} as well as
$BChPT$ \cite{gassa} \cite{bkkm} \cite{be1} \cite{be2}.
The static polarizability is a measure for how easily an electric (magnetic)
dipole moment may be induced by a constant external electric field
$\mbox{\boldmath$ \varepsilon$}$ (magnetic field $\mbox{\boldmath$b$}$).
In the lagrangian this leads to additional terms
\be
\label{3.4.1}
L(\mbox{\boldmath$\varepsilon$},\mbox{\boldmath$b$})=L+\frac{1}{2}(4\pi\alpha^b) 
\mbox{\boldmath$\varepsilon$}^2+ 
\frac{1}{2}(4\pi\beta^b) \mbox{\boldmath$b$}^2  
\ee
Experimentally there
is information about the dynamic polarizabilities obtained from Compton
scattering \cite{f90}
\cite{r91}, which have to be corrected for relativistic and
retardation effects caused by the finite size of the nucleon
\cite{ehue}-\cite{s91} and about
the static electric polarizability of the neutron from neutron nucleus
scattering \cite{s91} . There is no direct measurement of the static
polarizabilities of the proton. The experimental results for the neutron
and proton electric polarizabilities $\alpha^n=12.0 \pm 3.5 \cdot
10^{-4} fm^3$ and $\alpha^p=7.0 \pm 3.5 \cdot
10^{-4} fm^3$  are arranged into the linear combinations
\begin{eqnarray}
  \alpha &=& \frac{1}{2}(\alpha^n+\alpha^p)=9.5 \pm 5.0 \cdot
10^{-4} fm^3 \nonumber \\
{\scriptstyle \triangle}\! \alpha &=& \alpha^n-\alpha^p =5.0 \pm 5.0 \cdot
10^{-4} fm^3
\end{eqnarray}
which turn out to have different $N_C$ behaviour in chiral soliton
models. In these models the adiabatic quantity $  \alpha$ is leading
${\cal{O}}(N_C^1)$ in tree approximation (section 3.5.1) and quantum
corrections of ${\cal{O}}(N_C^0)$ (section 3.5.2) may be important. The
non-adiabatic quantity ${\scriptstyle \triangle} \alpha$ comes with an angular velocity
in tree approximation \cite{bbc92} and gives rise to a ${\cal{O}}(N_C^{-1})$
splitting of neutron and proton electric polarizabilities (section
3.5.3). For the magnetic polarizabilities this is quite analogous, only
the experimental information is much poorer there. All previous attempts to attack the problem in the framework of chiral
soliton models \cite{n84} \cite{ch87} \cite{sw89} \cite{sm92}
\cite{sch92} rely on the adiabatic
tree approximation. Trivially there is no splitting of the neutron's
and proton's electric polarizabilities $({\scriptstyle \triangle}
\alpha = {\scriptstyle \triangle} \beta  = 0)$.

More severely the
numerical values obtained, in purely pseudoscalar models $  \alpha \simeq
18 - 30 \cdot 10^{-4} fm^3$ depending on the parameters used, come out
much too large. Nonminimal couplings to the photon do not help here,
they are of negligible influence (section 3.5.1). This finding is at
variance with the seemingly much smaller results $  \alpha = 10.4 -
14.8 \cdot 10^{-4} fm^3$ obtained by adding vector mesons explicitly
\cite{sw89} which include nonminimal couplings naturally. The reason for
this discrepancy is due to $\varrho_0$ and $\omega_0$ components in the
vector meson soliton profiles induced by the electric field $\mbox{\boldmath$
\varepsilon$}$
which lead to additional ${\cal{O}}(N_C)$ contributions to the electric
polarizability and which are simply neglected in the approach mentioned
\cite{sw89}. The situation is easily understood and the missing terms can
be traced by looking at the local approximation which converts the
vector meson into a purly pseudoscalar model.
Taking these contributions properly into account, vector
meson models would end up with large electric polarizabilities
comparable to those of the corresponding pseudoscalar models. 

For the magnetic polarizabilities the situation is even worse.
Numerical values obtained so far range from $  \beta = -7 -
(+8.5) \cdot 10^{-4} fm^3$ \cite{ch87} \cite{sw90} \cite{sm92}. 
Up to now there does not even exist a correct
and complete calculation of the leading ${\cal{O}}(N_C)$ contribution
in the simplest Skyrme model.

\subsection{Electric polarizability in tree approximation}

The isovector photon field $v^3_0 = - \mbox{\boldmath$  \varepsilon r$}$ leads to
a contribution in the
lagrangian quadratic in the electric field $\mbox{\boldmath$ \varepsilon$}$ which in tree
approximation is ${\cal O}(N^1_C)$
\begin{eqnarray}
\label{3.4.3}
L_{\gamma \gamma} &= & \frac{f^2_\pi e^2}{2} \int d^3\!r\,\, (\mbox{\boldmath$
\varepsilon$}
\mbox{\boldmath$  r$})^2 s^2 b_T D_{3a} D_{3b} (\delta_{ab} - \hat r_a \hat r_b)
\nonumber \\
& + & f^2_\pi (2c^e_4-c^f_4) e^2 \mbox{\boldmath$ \varepsilon$}^2 \int d^3\!r\,\,  
s^2  D_{3a} D_{3b} (\delta_{ab} - \hat r_a \hat r_b) 
\end{eqnarray}
(We use $\frac{e^2}{4\pi}=\frac{1}{137}$ throughout).
There is no such contribution for the isoscalar photon field and most
importantly there is no ${\cal O}(N_C^1)$ term at all which is linear
in the electric field as e.g. in vector meson models which would induce
adiabatic soliton components driven by $\mbox{\boldmath$ \varepsilon$}$. 
There is, however,
such a term in ${\cal O}(N_C^0)$, nonadiabatic, which contains an
angular velocity and which consequently splits neutron and proton
polarizabilities. This term is treated in subsection 3.4.3. Concluding,
in leading ${\cal O}(N_C^1)$ eq. (\ref{3.4.3}) represents the complete
contribution to the electric polarizability
\be
\label{3.4.4}
 \alpha = \frac{f_{\pi}^2 e^2}{18 \pi} \int d^3\!r\,\, r^2 s^2 b_T +
\frac{f_{\pi}^2 e^2}{3 \pi} (2c^e_4 - c^f_4) \int d^3\!r\,\, s^2  \; .
\ee
Here we used that for nucleon states $D_{3a} D_{3b} = \frac{1}{3}
\delta_{ab}$. Alternatively we could have averaged over the
orientations of the electric field $\mbox{\boldmath$ \varepsilon$}$ which leads to the same
result. For the calculation of the quantum corrections in the next
subsection we will have to do both simultaneously, taking nucleon
matrix elements and averaging over the $\mbox{\boldmath$ \varepsilon$}$-orientations.

The second term in (\ref{3.4.4}) represents the non-minimal couplings.
If the chiral scale is chosen such that the $LECs$ $\ell^r_6 \simeq 2
\ell^r_5 \simeq - 1/8g^2$ are exhausted by the
$\varrho$-meson
resonance \cite{edar} then the contribution from the non-minimal couplings
$c^f_4 \simeq 2 c^e_4$ vanishes exactly. The Gasser Leutwyler $LECs$ at
$\mu=m_\varrho$ lead to an unsignificant enhancement of the electric
polarizability due to the non-minimal couplings. Thus, in contrast to
the common belief, neither non-minimal couplings nor the explicit
inclusion of vector mesons is able to cure the problem with the much
too large electric polarizabilities obtained in tree approximation from
soliton models. 

Such a conclusion is of course valid only if the
way to calculate polarizabilities as sketched above is correct, which has
recently been doubted \cite{poldo1} \cite{poldo2}. The explicit
statement was that the terms quadratic in the external electric or
magnetic fields (seagull terms) do not contribute at all. Since seagull
terms are in fact the only contribution at leading order $N_C$, this
claim, if true,
would make quite a difference, and we shall in the following discuss it
in some detail.

A first hint that not all is well with the statement voiced in 
\cite{poldo1} \cite{poldo2}  may be derived from the fact that the
tree expression for the electric polarizability (\ref{3.4.4}) evaluated
in the chiral limit (where an analytical treatment akin to the case of
the $\sigma$- term is possible) agrees with the result obtained in
$HBChPT$ \cite{bkkm}  \cite{cb}:
\be
\label{alchpt}
  \alpha \stackrel{m_{\pi} \to 0}{\rightarrow} \frac{2
f_{\pi}^2 e^2}{9} \int_0^{\infty} d r\,\, r^4 F^2 =\frac{e^2}{4\pi} \frac{5g_A^2}{32 \pi f_{\pi}^2
m_{\pi}}
\;. 
\ee
In order to put our concerns on a more solid footing, we insert the
ansatz (\ref{ansat}) for the time dependent matrix $U$ into the lagrangian
coupled to the external field  $\tilde v_0^a$ (intrinsic system). 
It's relevant terms then read
\begin{eqnarray}
L(\varepsilon) \;\;= -M_0&+&\frac{1}{2} \int d^3\!r\,\, (\dot \eta_a n_{ab}^2 \dot \eta_b -
\eta_a h_{ab}^2 \eta_b) \nonumber \\
&-& \sqrt{\Theta} \int d^3\!r\,\, (\Omega_e^R+e \tilde v_0^e)z_a^e n_{ab}^2 \dot
\eta_b \nonumber \\
&+& \frac{1}{2} \Theta \int d^3\!r\,\, (\Omega_e^R+e \tilde v_0^e)z_a^e
n_{ab}^2 z_b^f (\Omega_f^R+e \tilde v_0^f)  \nonumber \\
&+& \cdots
\label{seal}
\end{eqnarray}
with the moment of inertia $\Theta$ and $z_a^e$ the normalized rotational zero
mode (\ref{constr}),
\be
z_a^e=\frac{f_{\pi}}{\sqrt{\Theta}} s \; \varepsilon_{aei} \hat r_i \;.
\ee
The last term explicitly written down in (\ref{seal})
contains the rotational energy as well as the seagull contribution. 
The collective rotation (translations are unimportant in the present context)
together with the fluctuations $\eta$ gives
rise to a redundancy which shows up in the conjugate
momenta 
\be
\pi_a =\frac{\delta L}{\delta \dot \eta_a} = n_{ab}^2 \left[ \dot
\eta_b - \sqrt{\Theta}  z_b^f (\Omega_f^R+e \tilde v^f_0)
\right]
\label{conmo}
\ee
$$
R_e = - \frac{\partial L}{\partial \Omega_e} = \sqrt{\Theta} \int d^3\!r\,\,
z_a^e n_{ab}^2 \dot \eta_b -\Theta \int d^3\!r\,\, z_a^e n_{ab}^2 z_b^f
(\Omega_f^R+e \tilde v^f_0)
$$ 
making them linearly dependent (so called primary constraint \cite{dir}):
\be
R_e=\sqrt{\Theta} \int d^3\!r\,\, z_a^e \pi_a \;.
\label{pric}
\ee
Due to this fact, the naive Legendre transformation (used in
\cite{poldo1} \cite{poldo2}  ) to the hamiltonian 
\be
H=M+\frac{1}{2} \int d^3\!r\,\, \left[ \pi_a n_{ab}^{-2} \pi_b +\eta_a
h_{ab}^2  \eta_b \right] +e \sqrt{\Theta} \int d^3\!r\,\,\tilde v^f_0 z_a^f \pi_a
\label{hamil}
\ee
is not well behaved: The collective coordinates have been removed
completely, and gone are the seagull term as well as the soliton's
rotational energy.

In order to keep the collective coordinates, one has to impose
secondary constraints \cite{dir} conjugate to the primary ones
(\ref{pric}). There is an obvious choice for these: the fluctuations should be
orthogonal to the zero modes, which have already been accounted for
through the collective variables 
\begin{eqnarray}
\int d^3\!r\,\, z_a^c n_{ab}^2 \eta_b &=& 0 \;, \nonumber \\
\int d^3\!r\,\, z_a^c n_{ab}^2 \dot \eta_b &=& \int d^3\!r\,\, z_a^c 
\bar \pi_a=0 ,
\;\;
\bar \pi_a \equiv n_{ab}^2 \dot \eta_b 
\;.
\end{eqnarray}
These constraints have to be added with multipliers to the lagrangian
and, going through Dirac's procedure \cite{dir}, one finally obtains the
proper hamiltonian. However, the same objective may be achieved faster
by decomposing the conjugate momenta (\ref{conmo}) 
\be
\pi_a =\pi_a^{col} + \bar \pi_a ,\quad \pi_a^{col}=-\sqrt{\Theta}n_{ab}^2 z_b^f
(\Omega_f^R+e \tilde v^f_0) 
\ee
into collective and fluctuational parts and inserting them into the
hamiltonian (\ref{hamil}) . It is noticed that the collective motion
decouples completely from the fluctuations
\begin{eqnarray}
H &=& H_0 +H_{\pi\pi} +H_{\gamma} +H_{\gamma\gamma} \\
H_0&=&M_0 + \frac{\mbox{\boldmath$R$}^2}{2 \Theta} \nonumber \\
H_{\pi\pi}&=&\frac{1}{2} \int d^3\!r\,\, (\bar \pi_a n_{ab}^{-2} \bar \pi_b +
\eta_a h_{ab}^2 \eta_b) \nonumber \\
H_{\gamma} &=& e R_e \int d^3\!r\,\, z_a^e n_{ab}^{2} z_b^f  \tilde v^f_0
\nonumber  \\
H_{\gamma\gamma}&=& \frac{e^2}{2} \Theta \left[ \left( \int 
d^3\!r\,\, z_a^e
n_{ab}^{2} z_b^f \tilde v^f_0 \right)^2-\int d^3\!r\,\, z_a^e
n_{ab}^{2} z_b^f \tilde v^e_0 \tilde v^f_0 \right]  \nonumber 
\; .
\end{eqnarray}
The rotational energy $\mbox{\boldmath$R$}^2 /2\Theta $ is recovered as well
as the seagull term $H_{\gamma\gamma}$. So far we have not specialized
on the particular photon field $\tilde v^a_0$. For a spatially constant
potential $\tilde v^a_0=const.$, $H_{\gamma\gamma}$ does indeed vanish,
as is to be expected since $\tilde v^a_0=const.$ corresponds to a
global rotation. For the constant electric field $\tilde v^a_0=
-\mbox{\boldmath$ \varepsilon r$}D_{3a}$ 
needed to calculate the electric polarizability the first
integral in $H_{\gamma\gamma}$ vanishes for parity reasons and we end
up with $H_{\gamma}=0$ and
\be
H_{\gamma \gamma} =  -\frac{f^2_\pi e^2}{2} \int d^3\!r\,\, 
(\mbox{\boldmath$\varepsilon r$})^2 s^2b_T D_{3a} D_{3b} (\delta_{ab} - 
\hat r_a \hat r_b)
\;,
\ee
the familiar soliton model result, compare eq.(\ref{3.4.3}). Therefore
we definitely do not agree
with the statement that the seagull terms should vanish and are
confident that the result (\ref{3.4.4}) for the tree contribution to the
electric polarizabiliy in soliton models is indeed correct. Similar
conclusions were independently drawn by Scoccola and Cohen \cite{scoc}

In the next section we will provide the method to calculate the quantum
corrections to this interesting quantity in order to learn whether
these $1$-loop contributions are eventually able to reduce the too
large tree level values.

\subsection{Loop corrections to the electric polarizability}

The calculation of loop corrections to the electric polarizability is very
much involved, concerning both, the theoretical input and the
computational effort. Although the output is only one number we decided
to undergo this pain, because this quantity  persistently comes out too
large in tree regardless of the model considered and therefore may
serve as a crucial test of  the loop expansion.

In the end we want to calculate our baryon property from the
Casimir energy
\be
\label{3.4.6}
M(\varepsilon^2) = M -2\pi \varepsilon^2   \alpha,  \quad  
\alpha =-\frac{1}{2\pi}\left.  \frac{\partial
M(\varepsilon^2)}{\partial \varepsilon^2}
\right|_{\varepsilon^2 = 0}
\ee
in the familiar way. For this purpose the photon field $v^3_0 = -
\mbox{\boldmath$  \varepsilon  r$} $ has to be coupled to the lagrangian (via the
covariant derivatives) which then ought to be expanded up to quadratic
order in the adiabatic fluctuations. Of course if we consider a certain
fixed direction of the electric field the soliton gets deformed under
its influence and our whole concept, which is based on the
spherical hedgehog solution would fail. Luckily we are interested only
in the static electric polarizability which, defined as the coefficient
of $\mbox{\boldmath$ \varepsilon$}^2$ (\ref{3.4.1}), is independent of the direction
of $\mbox{\boldmath$ \varepsilon$}$ and 
we are allowed to average over these directions. In addition we may use
the relation $D_{3a} D_{3b} = \frac{1}{3} \delta_{ab}$ for nucleon
states. With these two allowed manipulations we are able to show that
the hedgehog still solves the e.o.m. in the presence of the electric
field $\mbox{\boldmath$ \varepsilon$}$ and makes the term linear in the fluctuation
vanish if
the stability condition for the chiral angle is calculated by variation
of the tree contribution (\ref{3.4.4}) contained in (\ref{3.4.6}). So 
far everything is quite
similar to the quantities calculated in the previous sections. The
complications arise with the terms quadratic in the fluctuations, or
more precisely, with the e.o.m. for the fluctuations.

The lagrangian expanded to second order in the fluctuations
\begin{eqnarray}
\label{3.4.5}
L(\varepsilon) & = & L + L_\gamma + L_{\gamma \gamma} \nonumber \\
L_\gamma & \stackrel{\eta^2}{=} & e D_{3 a} \int d^3\!r\,\,
(\mbox{\boldmath$ \varepsilon r$}) [ (\mbox{\boldmath$ \eta
\times {\dot \eta}$})_a-2(1-c) \eta_L (
\mbox{\boldmath$ \hat{r} \times \dot{\eta}$})_a + \cdots ] \nonumber \\
L_{\gamma \gamma} & \stackrel{\eta^2}{=} & \frac{1}{2} e^2
D_{3 a}  D_{3b} \int d^3\!r\, (\mbox{\boldmath$ \varepsilon r$})^2
\left[ \mbox{\boldmath$ \eta$}^2 \delta_{ab}-\eta_a \eta_b
-s^2(2\eta_L^2 +\mbox{\boldmath$\eta$}_T^2)(\delta_{ab}-\hat r_a \hat
r_b) \right. \nonumber \\
&& \; \; \; \left. \quad \quad \quad + (1-c) \eta_L (\hat r_a \eta_{b T} 
+\eta_{a T} \hat r_b) + \cdots \right]
\end{eqnarray}
contains a piece linear in the electric field with one time derivative
and a piece quadratic in the electric field (for simplicity again only
the N$\ell \sigma$ contributions are listed). The second term can be
simplified by using
$D_{3a} D_{3b} =  \frac{1}{3} \delta_{ab}$ for nucleon states and
averaging over the electric field orientations 
$(\mbox{\boldmath$ \varepsilon r$})^2 \to \frac{1}{3}
\varepsilon^2 r^2$. 

Naively one might think that the first term
vanishes by averaging over the electric field orientations but this is
not the case for the e.o.m. as we will demonstrate in the following.

The lagrangian (\ref{3.4.5}) introduces additional terms $\sim
\varepsilon$ and $\sim \varepsilon^2$ into the e.o.m.
which may be written formally
\be \label{liti}
h^2 \eta + 2 i \varepsilon w_{\gamma} \dot \eta - \varepsilon^2
w_{\gamma \gamma} \eta = - n^2(\varepsilon) \ddot \eta
\ee
with hermitean differential operators $w_{\gamma}$ and
$w_{\gamma\gamma}$ 
\begin{eqnarray}
(w_\gamma)_{ab} &=& ie D_{3p} \mbox{\boldmath$ {\hat \varepsilon}
r$} \left[ \varepsilon_{pab} - (1-c) \hat r_q (\hat r_a
\varepsilon_{pqb}-\varepsilon_{pqa} \hat r_b) + \cdots \right]
\\
(w_{\gamma\gamma})_{ab} &=& e^2 D_{3p}D_{3q} (\mbox{\boldmath$ {\hat 
\varepsilon }
r$})^2 \left[\delta_{pq} \delta_{ab}-\delta_{ap}\delta_{bq}
-s^2(\delta_{pq}-\hat r_p \hat r_q)(\delta_{ab}+\hat r_a \hat r_b)
\right. \nonumber \\
&& \qquad \qquad \qquad \left. +(1-c) (\hat r_a \hat r_p \delta_{qb} +\delta_{ap} \hat r_q
\hat r_b -2 \hat r_a \hat r_p \hat r_q \hat r_b) + \cdots \right]
\nonumber \;.  
\label{potpol}
\end{eqnarray}
Differentiations appear for the higher chiral order terms omitted in
(\ref{3.4.5}). For $w_{\gamma}$ alone we obtain 9 terms of different
isospin structure from the lagrangian (\ref{3.4.1}).The single time
derivative in eq.(\ref{liti}) is exactly of the form discussed in
eqs.(\ref{cave1}, \ref{cave2}) and consequently
\be
\label{trlog}
\frac{i}{2} \int d^4 x \langle \ell n \Omega \rangle 
 = - \frac{T}{2}\int d^3\!r\,\, \langle \sqrt{\tilde h^2-
\varepsilon^2(\tilde w_{\gamma\gamma}- \tilde w_{\gamma}^2)} \rangle 
\ee
there is no term linear in the electric field $\varepsilon$. Thus
considering the trace log (\ref{trlog}) we may equally well solve the e.o.m. 
\be
h^2 \eta - \varepsilon^2(w_{\gamma \gamma} -w_{\gamma} n^{-2}
w_{\gamma}) \eta = - n^2 \ddot \eta
\ee
instead of (\ref{liti}). This is of the standard form with an
additional potential, simple only in the case of the N$\ell \sigma$ model ($n
\equiv 1$)
\begin{eqnarray}
(w_{\gamma} n^{-2}
w_{\gamma})_{ab} &=& e^2 D_{3p}D_{3q} (\mbox{\boldmath$ {\hat \varepsilon}
r$})^2 \left[\delta_{pq} \delta_{ab}-\delta_{ap}\delta_{bq}
-s^2 \hat r_i \hat r_j \varepsilon_{pia}\varepsilon_{qjb} \right. 
\nonumber \\
&& \qquad \qquad \qquad \left. -s^2\delta_{pq} \hat r_a \hat r_b+(1-c) (\hat r_a \hat r_p 
\delta_{qb} +\delta_{ap} \hat r_q
\hat r_b \right. \nonumber \\
&& \qquad \qquad \qquad \left. -(1-c)^2 \hat r_a \hat r_p \hat r_q \hat r_b) + \cdots \right] \;.
\label{quadpot}
\end{eqnarray} 
In general we had to work out, according to (\ref{potpol}) and the
longitudinal and transversal metric, $2\times9\times9$ terms.
From (\ref{potpol}) and (\ref{quadpot}) it is recognised that in
contrast to the individual potentials $w_{\gamma\gamma}$ and 
$w_{\gamma} n^{-2} w_{\gamma}$ their difference is well behaved for $r
\to \infty$ ($F \to 0$).

Finally we may simplify the potentials using again $D_{3p} D_{3q} =
\frac{1}{3} \delta_{pq}$ and averaging
over $\varepsilon$-field orientations to obtain
\begin{eqnarray}
(w_{\gamma\gamma})_{ab} &=& \frac{e^2}{9}r^2 \left[2\delta_{ab}
-2s^2(\delta_{ab}+\hat r_a \hat r_b)
+\cdots \right] \nonumber \\
(w_{\gamma} n^{-2}
w_{\gamma})_{ab} &=& \frac{e^2}{9}r^2 \left[2\delta_{ab}
-s^2(\delta_{ab} + \hat r_a \hat r_b)
+ \cdots \right] \;.
\end{eqnarray}
With $w_{\gamma\gamma}$ alone we would end up with an infinite loop
correction to the polarizability of the vacuum which of course should
be zero.
This is how we noticed that we need the first term in
(\ref{3.4.5}) linear in the electric field in order to get a reasonable
finite result for the quantum correction to the electric polarizability
in the vacuum as well as in the soliton sector.

\subsection{Neutron-proton split of the electric polarizability}

In this subsection we present a derivation of the neutron-proton split 
${\scriptstyle \triangle} 
\alpha$ of the electric polarizability in chiral soliton models.
A similar approach for a chiral soliton model with quarks was published
by Broniowski et al. \cite{bbc92}.

The quantity under investigation must
contain an angular velocity in order to split the values for neutron
and proton and is therefore of lower order $N_C$ compared to the tree
and $1$-loop contributions to the electric polarizability discussed in
the previous sections.

We consider the term linear in the electric field (non-minimal
couplings do not contribute here)
\be
\label{3.4.11}
L_\gamma = e \int d^3\!r\,\, (\mbox{\boldmath$  \varepsilon r$}) (D_{3a} \tilde V^a_0 +
\frac{1}{2} B_0)
\ee
which provides the driving terms for static 
fluctuations $\mbox{\boldmath$ \eta$}^S$ and $\mbox{\boldmath$
\eta$}^V$ induced by the isoscalar and isovector electric
field (the terms linear in the fluctuations do not vanish)\footnote{This was
first noticed by B. Schwesinger}. To
guarantee these fluctuations to be orthogonal on the zero modes $z^c_a$
we have to implement the constraints
\be
\label{3.4.12}
\int d^3\!r\,\, z^c_a n_{ab}^2 \eta^{S,V}_b = 0 \quad \mbox{for all} \; \; c
\ee
simply by adding them with Lagrange multipliers
$\lambda^S_c$ and $\lambda^V_c$ to the lagrangian. The relevant zero
modes are the three infinitesimal translations
\be
z^c_a = \frac{f_ \pi}{\sqrt{M_0}} \left[ F^{\prime} \hat r_a \hat r_c +
\frac{s}{r} ( \delta_{ac} - \hat r_a\hat r_c) \right] \; ,
\ee
normalized appropriately (\ref{constr}) with $M_0$ being the classical soliton mass
(\ref{classma}). Mathematically  fluctuations unquestionably have to be defined in
the space orthogonal to the one spanned by the collective coordinates
(Dirac constraints), physically the constraints (\ref{3.4.12}) make sure that
simple translations of the system may not contribute to the electric
polarizability. 

With these preparations we are ready to solve the e.o.m.  
\be
\label{3.4.14}
h_{ab}^2 \eta^S_b = \frac{e}{2} (\mbox{\boldmath$  \varepsilon  r$}) \left. \frac{\delta
B_0}{\delta \eta_a} \right|_{\mbox{\boldmath$ \eta$} = 0} + \lambda^S_c n_{ab}^2 z^c_b
\ee
\be
h_{ab}^2 \eta^V_b = e (\mbox{\boldmath$ \varepsilon r$}) D_{3p} \left. \frac{\delta
\tilde V_0^p}{\delta \eta_a} \right|_{\mbox{\boldmath$ \eta $}= 0} + 
\lambda^V_c n_{ab}^2 z^c_b 
\ee
for the  components induced by the electric 
field, which reinserted into the lagrangian
\begin{eqnarray}
L_{\gamma \gamma} & = & \int d^3\!r\,\, (\eta^S_a + \eta^V_a) \left[-\frac{1}{2}
h_{ab}^2 (\eta^S_b + \eta^V_b) + e (\mbox{\boldmath$  \varepsilon  r$}) \left. \left( D_{3p} \frac{\delta
\tilde V_0^p}{\delta \eta_a} + \frac{1}{2} \frac{\delta B_0}{\delta
\eta_a} \right)  \right|_{\mbox{\boldmath$ \eta$} = 0} \right] \nonumber \\
& = & \frac{e}{2} \int d^3\!r\,\,  (\mbox{\boldmath$ \varepsilon r$}) (\eta^S_a + \eta^V_a)
\left. \left( D_{3p} \frac{\delta
\tilde V_0^p}{\delta \eta_a} + \frac{1}{2} \frac{\delta B_0}{\delta
\eta_a} \right)  \right|_{\mbox{\boldmath$ \eta$} = 0}
 \nonumber \\
& \equiv & L^{SS}_{\gamma \gamma} + L^{SV}_{\gamma \gamma} +
L^{VV}_{\gamma\gamma}
\label{3.4.16} 
\end{eqnarray}
lead to terms proportional to $\mbox{\boldmath$ \varepsilon$}^2$ which contribute to the
electric polarizability. By looking at the Legendre transformation we
made sure that this naive insertion into the lagrangian is here
as an exception of the rule correct.

All three terms in (\ref{3.4.16}) are ${\cal O} (N^{-1}_c)$, $L^{SS}_{\gamma
\gamma}$ and $L^{VV}_{\gamma \gamma}$ contribute to $  \alpha$ and
only $L^{SV}_{\gamma \gamma}$ contains an angular velocity which splits
neutron and proton electric polarizabilities. In contrast to $ 
\alpha$ where we have many other contributions in ${\cal O}(N^{-1}_C)$
(e.g. 2-loops) the leading contribution to ${\scriptstyle \triangle} \alpha$ is contained
in (\ref{3.4.16})
\begin{eqnarray}
L^{SV}_{\gamma \gamma} & = &  e  D_{3p} \int d^3\!r\,\, (\mbox{\boldmath$
\varepsilon  r$ })
\left. \frac{\delta
\tilde V^p_0}{\delta \eta_a} \right|_{\mbox{\boldmath$ \eta$} = 0} \eta^S_a \nonumber
\\
& =&  \frac{e}{2} \int d^3\!r\,\, (\mbox{\boldmath$ \varepsilon r$}) \left. \frac{\delta B_0}{\delta
\eta_a}  \right|_{\mbox{\boldmath$  \eta$}= 0} \eta^V_a
\;.
\label{3.4.17}
\end{eqnarray}
Although we have calculated both isoscalar and isovector induced
components and checked our calculation using the equivalence of the two
expressions in (\ref{3.4.17}) which follows from the e.o.m., we shall
present here only the evaluation via the first one using the isoscalar
component in some detail.

\begin{figure}[h]
\vspace*{7.5cm}
\begin{center} \parbox{8.2cm}{\caption{\label{polrad} Radial functions 
$u(r)$ and $v(r)$ of the induced
isoscalar electric dipole mode plotted over a range of $3 fm$. 
The node of $v(r)$ is due
to orthogonality on the translation.}}
\end{center}
\end{figure}

With $B_0$ expanded linearly in the fluctations, the e.o.m. for the induced
isoscalar component follows from (\ref{3.4.14}):
\be
\label{3.4.18}
h_{ab}^2 \eta^S_b=\frac{e}{4\pi^2 f_{\pi}} \left[
\frac{F^{\prime}s}{r}(\varepsilon_a-\hat r_a(\mbox{\boldmath$ \hat{r} 
\varepsilon$})) +
\frac{s^2}{r^2}\hat r_a(\mbox{\boldmath$ \hat{r} \varepsilon$}) \right] 
+\lambda^S_c n_{ab}^2 z^c_b
\;
.
\ee
The Lagrange multipliers $\lambda_c^S=\frac{1}{2} e
\varepsilon_c/\sqrt{M_0}$, determined by multiplying (\ref{3.4.18}) 
with $z_a^d$ from the
left are inserted into the e.o.m. 
\begin{eqnarray}
h_{ab}^2 \eta^S_b &=& \frac{e}{2 f_{\pi}} \left[
(\frac{F^{\prime}s}{2\pi^2 r} +
\frac{f_{\pi}^2}{M_0}\frac{s}{r}b_T)(\varepsilon_a-\hat r_a( 
\mbox{\boldmath$ \hat{r} \varepsilon$}))
\right. \nonumber  \\
&&\qquad + \left.
(\frac{s^2}{2\pi^2 r^2} +
\frac{f_{\pi}^2}{M_0} F^{\prime}b_L) \hat r_a( 
\mbox{\boldmath$ \hat{r} \varepsilon$}) \right]
\;.
\label{3.4.19}
\end{eqnarray}
From this equation it is noticed that the electric field induces in this
case a pure $E1$ mode
\be
\mbox{\boldmath$ \eta$}^S=-\frac{e}{4\pi^2 f_{\pi}^2} \left[u(r)
\mbox{\boldmath$ \hat{r} (\hat{r} \varepsilon$})  +v(r)(\mbox{\boldmath$
\varepsilon$}-  
\mbox{\boldmath$ \hat{r} (\hat{r} \varepsilon$})) \right] 
\ee
and it is evident that it is the translational zero mode which enters the
constraints (\ref{3.4.12}) .
The radial functions $u(r)$ and $v(r)$, subject to boundary conditions
$u^{\prime}(0)=v^{\prime}(0)=0$ , $u(\infty)=v(\infty)=0$, are depicted in fig.
\ref{polrad}; $v(r)$ has a node which is caused by orthogonality on
the translational zero mode.

The $E1$ mode $\eta^S$ yields after averaging over the electric field
directions (first expression in (\ref{3.4.17}) )
\begin{eqnarray}
L^{SV}_{\gamma \gamma} &=& \frac{1}{2} (4\pi\Theta {\scriptstyle \triangle}\! \alpha)
D_{3a} \Omega^R_a \mbox{\boldmath$ \varepsilon$}^2 =-\frac{1}{2} 
(4\pi{\scriptstyle \triangle}\! \alpha) L_3
\mbox{\boldmath$ \varepsilon$}^2 \nonumber \\
{\scriptstyle \triangle}\! \alpha &=& \frac{2 e^2}{9 \pi^2  f_\pi \Theta}
\int_0^{\infty}  dr\,\,
r^3 \left[sc \; b_T u +(c_4^a-4c^s_4 +c_6 {F^{\prime}}^2) \frac{s^3}{r^2}(c u
-v) \right. \nonumber \\
&& \quad \quad \quad \left. -c^k_4 s^3 u +(c_4^a-2c^s_4 +c_6\frac{s^2}{r^2} )
F^{\prime}s^2 u^{\prime} \right]
\;.
\label{3.4.21}
\end{eqnarray}
Here we used that the isospin $L_3=-\Theta D_{3a} \Omega^R_a$ is
related to the angular momentum $\Omega^R_a$.

After presenting the mechanism which splits neutron and proton
electric polarizabilities with a positive ${\scriptstyle \triangle} \alpha$ of
${\cal O}(N_C^{-1})$ 
we will give an analytical result in the chiral
limit $m_{\pi} \to 0$.  It turns out that  ${\scriptstyle \triangle} \alpha$
diverges in this limit such that the integral in (\ref{3.4.21}) becomes
dominated by the asymptotic region for which we are able to solve the
differential equations (\ref{3.4.19}) analytically
\be
u \stackrel{r \to \infty} =  \frac{3\pi f_\pi g_A}{8 M_0 r}
(2+m_{\pi}r) e^{-m_{\pi}r} \, , \quad
v \stackrel{r \to \infty} =  \frac{3\pi f_\pi g_A}{8 M_0 r}
e^{-m_{\pi}r}
\;.
\ee
With these solutions we obtain
\be
{\scriptstyle \triangle}\! \alpha \stackrel{m_{\pi} \to 0}{\to}
 \frac{2 e^2}{9 \pi^2 f_\pi \Theta} \int_0^{\infty} d r\,\,
r^3 F u = \frac{e^2}{4 \pi} \frac{g_A^2}{6\pi f_{\pi}^2 m_{\pi}}
\frac{\Delta}{M_0}
\;, 
\ee
where $\Delta$ represents the nucleon-$\Delta$ split. Physical
values for $g_A$, $f_{\pi}$, $M_0$, $\Delta$ and $m_{\pi}$
result in ${\scriptstyle \triangle} \alpha=12.5 \cdot 10^{-4} fm^3$, which is too
large. However, the ratio
\be
\frac{{\scriptstyle \triangle}\! \alpha}{  \alpha}
\stackrel{m_{\pi} \to 0}{\to} \frac{16}{15}
\frac{\Delta}{M_0} \simeq \frac{1}{3}
\ee
turns out to be not unreasonable.
Numerical values for the exact expression (\ref{3.4.21}) are given in
the results chapter.

\subsection{Magnetic polarizability}

Finally we will present a complete calculation of the static magnetic
polarizability $\beta$ to leading order ${\cal O}(N_C^{1})$. All
previous attempts \cite{ch87}\cite{sw90}\cite{sm92} 
miss an important part which may be noticed by the
violation of the $HBChPT$ result $\beta
\stackrel{m_{\pi} \to 0} = \alpha/10$ \cite{bkkm} \cite{be1}. In the course of our
derivation we will recover this relation although unfortunately
in the end we will loose it again for reasons which will become
obvious. Consequently the magnetic polarizability is the only quantity where
the $HBChPT$ result in the chiral limit is not contained in our tree
approximation. 

There is agreement on the contribution due to the seagull term (see
discussion in section 3.5.1) which is the term quadratic in the magnetic
field $\mbox{\boldmath$b$}$ when the isovector photon field 
$v_i^3 = \frac{1}{2}(\mbox{\boldmath$ r\times b$})_i$ is inserted
into the lagrangian
\begin{eqnarray}
\label{lgg}
L_{\gamma \gamma} &= & - \frac{f^2_\pi e^2}{120}
D_{3a} D_{3b} \int d^3\!r\,r^2 s^2
\left[ (b_T-c^a_4 \frac{s^2}{r^2}) (6 \mbox{\boldmath$b$}^2 \delta_{ab}
+2b_a b_b) \right.  \\
& & \left. +(c^a_4-2c_4^s) \frac{s^2}{r^2} (\mbox{\boldmath$b$}^2 \delta_{ab}
+7b_a b_b) \right] - \frac{2f^2_\pi e^2}{3} (2c^e_4-c^f_4)  \int d^3\!r\,\,  
s^2  \mbox{\boldmath$b$}^2 \; . \nonumber
\end{eqnarray}
Using $D_{3a} D_{3b} = \frac{1}{3} \delta_{ab}$ for nucleon states we
obtain the direct ${\cal O}(N_C^{1})$ contribution
\begin{eqnarray}
\label{betadir}
\beta_{dir} & = & -\frac{f_{\pi}^2 e^2}{36 \pi} \int d^3\!r\,\, r^2
s^2 \left[ b_T - \frac{1}{2}(c^a_4-2c_4^s)\frac{s^2}{r^2} \right] \nonumber \\
&& -\frac{f_{\pi}^2 e^2}{3 \pi}(2c^e_4 - c^f_4) \int d^3\!r\,\, s^2 
\end{eqnarray}
to the magnetic polarizability. The last term again represents the
non-minimal couplings which vanish for $c_4^f=2c_4^e$. In the chiral
limit we obtain $\beta_{dir}\stackrel{m_{\pi} \to 0} = -\alpha/2$
(compare (\ref{alchpt})) which is not the desired $HBChPT$ result
indicating a missing contribution.

In analogy to (\ref{3.4.11}) we consider the term linear in the
magnetic field (non-minimal couplings do not contribute)
\be
\label{lgamma}
L_\gamma = \frac{e}{2} \int d^3\!r\,\, 
(\mbox{\boldmath$ r\times b$})_i (D_{3a} \tilde V^a_i +\frac{1}{2} B_i).
\ee
Because the magnetic field couples via the spatial components of the
vector current, the isovector part is leading ${\cal O}(N_C^{1})$ in
contrast to the electric field case in (\ref{3.4.11}) (but similar
to the coupling of the electric field in vector meson lagrangians) and
will therefore give a contribution to the magnetic polarizability in
leading order. As in the electric case of the previous subsection the
isoscalar part of ${\cal O}(N_C^{-1})$ splits the polarizabilities of
neutron and proton. In the following we concentrate on the leading
${\cal O}(N_C^{1})$ term.

Previous attempts consider this term with the spherical hedgehog
inserted
\be
L_\gamma = -\frac{3e}{2M} \mu^V D_{3p} b_p \, ,
\ee
in second order perturbation theory which leads to the
standard paramagnetic $\Delta$ contribution
\be
\label{para}
\beta_{para} = \frac{e^2}{4 \pi M^2} \frac{(\mu^V)^2}{M_\Delta - M_N}
\; . 
\ee
This procedure overestimates the effect of (\ref{lgamma}) considerably
(see table \ref{betares}) because it neglects the soliton's response
to the external magnetic field. Although $\beta_{para}$ has the correct  
${\cal O}(N_C^{1})$ it remains finite in the chiral limit and cannot
add to the relation $\beta \stackrel{m_{\pi} \to 0} = -\alpha/2$.
Therefore it is not the desired term we are looking for.

The magnetic field in (\ref{lgamma}) gives rise to deformations of the
soliton. Because we are able to make the magnetic field arbitrarily
small, it suffices to calculate the linear response via the vector
current in (\ref{lgamma}) which provides the driving term for the
static fluctuation
\begin{eqnarray}
\label{betaind}
h_{ab}^2 \eta^V_b&=&\frac{e}{2} (\mbox{\boldmath$ r\times b $})_i
D_{3p} \left. \frac{\delta
\tilde V_i^p}{\delta \eta_a} \right|_{\mbox{\boldmath$ \eta $}= 0} + 
\lambda^V_c n_{ab}^2 z^c_b  \\
&=&- \frac{e f_{\pi}}{2} D_{3p} \left[ 2sc \hat r_a (b_p- \hat r_p 
(\mbox{\boldmath$ \hat{r} b$})) - 2s \hat r_p
(b_a- \hat r_a (\mbox{\boldmath$ \hat{r} b$})) 
+ \cdots \right] \nonumber \\
&&+\lambda^V_c n_{ab}^2
z^c_b \;. \nonumber 
\end{eqnarray}
The N$\ell \sigma$ model terms are listed explicitly and the zero mode
constraints (compare previous subsection) are added for later
convenience. The solution reinserted into the lagrangian leads
\begin{eqnarray}
\label{lvv}
L^{VV}_{\gamma \gamma} & = & \int d^3\!r\,\, \eta^V_a \left[-\frac{1}{2}
h_{ab}^2 \eta^V_b + \frac{e}{2} (\mbox{\boldmath$ r\times b $})_i
D_{3p} \left. \frac{\delta
\tilde V_i^p}{\delta \eta_a} \right|_{\mbox{\boldmath$ \eta$} = 0} 
\right] \\
& = &- \frac{e f_{\pi}}{4} D_{3p} \int d^3\!r\,\,
\left[ 2sc \eta_L (b_p- \hat r_p (\mbox{\boldmath$ \hat{r} b$})) 
- 2s \hat r_p (\mbox{\boldmath$
b \eta_T$}) + \cdots \right] \nonumber \\
& = & \frac{1}{2} (4 \pi \beta_{ind}) \mbox{\boldmath$ b $}^2 \nonumber
\end{eqnarray}
to a term proportional to $\mbox{\boldmath$ b $}^2$. Lets forget about
the constraint for a moment and solve (\ref{betaind}) analytically in
the asymptotical region. The result
\be
\label{evv}
\eta^V_a \stackrel{r \to \infty} = \frac{f_\pi e}{2} (\frac{3g_A}
{8\pi f_\pi^2}) e^{-m_\pi r} D_{3p} (\hat r_p b_a - \hat r_a b_p)
\ee
inserted into (\ref{lvv}) leads after some simple integrations to
$\beta \stackrel{m_{\pi} \to 0} = -3\alpha/5$ which is exactly the
$HBChPT$ result $\beta = \beta_{dir} + \beta_{ind} \stackrel{m_{\pi} 
\to 0} = \alpha/10$ as promised. However the full equation
(\ref{betaind}) cannot be solved without the constraint because the
rotational zero mode is solution to the homogeneous equation. More
explicitly this is seen by multiplying (\ref{betaind}) with $z_a^d$
from the left
\be
\label{lamu}
\lambda^V_c = -\frac{e}{2} \sqrt{\Theta} D_{3p}
\epsilon_{cpq} b_q \; ,
\ee
which determines the lagrangian mulipiers $\lambda^V_c \neq 0$.
It is worth noting that (\ref{lamu}) does not only hold for the
N$\ell \sigma$ model but also for the full lagrangian.
Apart from the mathematical necessity of the constraints in eq.
(\ref{betaind}) they are also needed for physical reasons: we do not
want simple rotations of the soliton contribute to the magnetic
polarizability. When the lagrangian multipliers (\ref{lamu}) are
inserted into the e.o.m. (\ref{betaind})
\begin{eqnarray}
\label{betaind2}
h_{ab}^2 \eta^V_b&=&- \frac{e f_{\pi}}{2} D_{3p} 
\left[ 2sc \hat r_a (b_p- \hat r_p (\mbox{\boldmath$ \hat{r} b$}))
 - s(\delta_{ap} (\mbox{\boldmath$ \hat{r} b$}) 
+ b_a \hat r_p- 2 \hat r_a \hat r_p (\mbox{\boldmath$ \hat{r} b$}) ) 
\right. \nonumber \\
&&+ \left. \cdots \right] \; ,  
\end{eqnarray}
it is noticed that the magnetic field induces a monopole and quadrupole
deformation
\begin{eqnarray}
\label{betafluc}
\eta_L^V &=& -\frac{e}{6f_\pi} D_{3p} \left[
2f(r) b_p + u(r) (b_p-3 \hat{r}_p 
(\mbox{\boldmath$ \hat{r} b $})) \right]  \\
\eta_{aT}^V &=& \frac{e}{4f_\pi}  D_{3p} v(r)
\left[ \delta_{ap} (\mbox{\boldmath$ \hat{r} b $}) +\hat{r}_p b_a
-2 \hat{r}_a \hat{r}_p (\mbox{\boldmath$ \hat{r} b $}) \right]
\; , \nonumber
\end{eqnarray}
while the magnetic dipole is projected completely from the e.o.m.
(\ref{betaind}) by the constraint. The radial
functions subject to the boundary conditions $f(0)=u(0)=v(0)=0$ and
\be   
\label{betafluca}
f(r) \stackrel{r \to \infty} = u(r) \stackrel{r \to \infty} {=} 
v(r) \stackrel{r \to \infty} {=} \frac{3g_A}{8\pi} e^{-m_\pi r} 
\ee
(compare (\ref{evv})) are depicted in fig. \ref{betarad}. 
Because of the constraint we now obtain
$\beta_{ind} \stackrel{m_{\pi} \to 0} = 9\alpha/20$ and $\beta
\stackrel{m_{\pi} \to 0} = -\alpha/20$ in the chiral limit in variance
with the $HBChPT$ result \cite{bkkm} \cite{be1}. We could not decide
whether this discrepancy is located in $HBChPT$ which does not know zero
modes, or is caused by the soliton model's
rigid rotator approach which rotates the profile rigidly for
all distances even for $r \to \infty$. 

\begin{figure}[h]
\vspace*{7.5cm}
\begin{center} \parbox{8.2cm}{\caption{\label{betarad} Radial functions 
$f(r)$ of the monopole deformation and $u(r)$ and $v(r)$ of the 
quadrupole deformation induced by a constant magnetic
field (full lines). For comparison the asymptotical function
relevant in the chiral limit is also plotted (dashed
line).}} 
\end{center}
\end{figure}

Table \ref{betares} lists the complete leading ${\cal O}(N_C^{1})$
contributions for the parameter sets $A$, $B$, and $C$. For comparison we
also give the perturbation theory result $\beta_{para}$ (\ref{para})
which overestimates $\beta_{ind}$ considerably and may at best be
considered a rough estimate. From this table it is noticed that the
total magnetic polarizability comes out small in magnitude for all
three parameter sets due to a strong cancellation of the direct and
induced contributions. Due to this cancellation in leading  
${\cal O}(N_C^{1})$, $1$-loop corrections of ${\cal O}(N_C^{0})$ may play
a decisive role for this quantity.  

However loop corrections for the magnetic polarizabilities are not
presented in this paper for the following reasons: even if we were
able to calculate the $1$-loop contributions to $\beta_{dir}$
(\ref{betadir}), where we expect a drastic reduction of the absolute
tree value similar to the electric polarizability in subsection 3.5.2,
we are not yet in a position to calcuate loop corrections to a deformed
soliton necessary for $\beta_{ind}$. Because both loop corrections are
of the same ${\cal O}(N_C^{0})$, it does not make sense to consider
only one of them alone, still it is likely that the loop corrections
will shift the magnetic polarizability in table \ref{betares} to small
positive values.

Because experimental information on static polarizabilities is scarce, 
it would be appropriate to calculate directly the Compton scattering
amplitudes within chiral soliton models in order to avoid the model
dependent and uncertain extraction of the static polarizabilities from
these data. In principle this should be possible.

\begin{table}[h]
\begin{center} \parbox{8.3cm}{\caption{\label{betares}
Direct and induced contributions to the magnetic polarizabilities are
compared for the three
parameter  combinations $A$, $B$, $C$. The perturbation theory result
$\beta_{para}$ overestimates $\beta_{ind}$ by a large margin. All 
polarizabilities are given in units $10^{-4}fm^3$. }}
\begin{tabular}{|c|c|c|c|}
\hline
& $A$ & $B$ & $C$ \\
\hline
$\beta_{dir}$  & -8.6 & -9.4 & -10.7  \\
\hline
$\beta_{ind}$ (E0)  & 2.7 & 2.5 & 2.0  \\
$\beta_{ind}$ (E2)  & 5.4 & 5.9 & 6.8 \\
$\beta_{ind}$ & 8.1 & 8.4 & 8.8   \\
\hline
$\beta_{para}$ & 11.0 & 15.1 & 23.4   \\
\hline
$\beta   $ & -0.5 & -1.0 & -1.9   \\
\hline
\end{tabular}
\end{center} \end{table}

\section{Electromagnetic properties of the $\Delta$ isobar}

To complete our discussion of electromagnetic properties we briefly
adress some of these quantities connected with the $\Delta$ isobar. In
general  the static properties of the $\Delta$ are related to those of
the nucleon by simple geometrical coefficients which follow from the
collective operators involved. For adiabatic quantities this operator
structure remains unaffected by $1$-loop corrections and consequently the
model independent relations survive for these quantities. The
experimental verifications of these relations may be considered as a
crucial test of the validity of the
$1/N_C$ expansion. As an example we discuss the magnetic moments in the
following paragraph. There we also mention the quadrupole moment (which
vanishes for the nucleon) as well as the corresponding transition
moments. 

Another interesting quantity subject to recent measurements is
discussed in the second subsection, namely the electromagnetic ratio of
the $\Delta$ photo-decay amplitudes. All soliton model attempts to
calculate this quantity so far make use of Siegert's theorem at the
finite photon momentum where it is not stictly valid. We show that the
correct treatment reduces the electromagnetic ratio considerably.

\subsection{Magnetic and quadrupole moments}

The operators for the calculation of magnetic and quadrupole moments
are given by
\begin{eqnarray}
\label{emop}
\mu & = & -3 \mu^V D_{33} -2 \mu^S R_3 + {\cal O}(N_C^{-1})
\nonumber \\
Q & = & -\frac{1}{5} <r^2>_{\Theta} ( D_{33} R_3  
-\frac{1}{3} D_{3p} R_p ) 
\; ,
\end{eqnarray}
with the isoscalar and isovector magnetic moments $\mu^S$ and $\mu^V$
and the squared radius $<r^2>_{\Theta}$ of the moment of inertia's
density defined in section 3.4. The adiabatic quantity $\mu^V$ may
contain the $1$-loop corrections described in the section just mentioned.
Evaluating the matrix elements of $\mu$ (\ref{emop}) with respect to
$\Delta$ states or nucleon and $\Delta$ states
\begin{eqnarray}
\label{mamo}
\mu^{\Delta ^{++}} & = & <\Delta ^{++},S_3=\frac{3}{2}|\mu|
\Delta ^{++},S_3=\frac{3}{2}>  =  \frac{9}{5} \mu^V + 3 \mu^S
+ {\cal O}(N_C^{-1})\nonumber \\
\mu^{N \Delta } & = & <\Delta ^{+},S_3=\frac{1}{2}|\mu|
p,S_3=\frac{1}{2}>  =  \sqrt{2} \mu^V + {\cal O}(N_C^{-1})\; , 
\end{eqnarray}
relates the magnetic (transition) moments by simple geometrical
coefficients to the isoscalar and isovector magnetic moments. Upon
using the experimental values $\mu^V =2.35$ and $\mu^S =.44$ we obtain 
$\mu^{\Delta^{++}} =5.55$ and $\mu^{N \Delta} =3.32$.
If accurate experimental information about these magnetic moments were
available we had a check on the $1/N_C$ expansion i.e. we could
estimate the next-to-next to leading order contribution generated by the
vector current. In our model such ${\cal O}(N_C^{-1})$ contributions are
produced in tree approximation by small soliton deformations
proportional to the angular velocity squared (next subsection), by
nonadiabatic $1$-loop processes which come with one angular velocity,
and by $2$-loop diagrams (compare a similar discussion for the axial
current in appendix A). Unfortunately there are no precise experimental
data: for $\mu^{\Delta^{++}}$ the particle data group gives a range of
$3.7-7.5$ nuclear magnetons \cite{pdg94}, for the transition magnetic moment there
exist photoproduction data, but these may not directly be compared to the
matrix element (\ref{mamo}) (see table 3.4).

Many other $\Delta$ observables like rms radii and polarizabilities are
related to those of the nucleon by simple geometrical coefficients just
as in the case of the magnetic moments. We do not list them here
because there are no experimental data available, instead we
concentrate on the electromagnetic ratio of the $\Delta$ photo-decay
amplitudes. For this purpose we add here still the 
quadrupole moments 
\begin{eqnarray}
\label{qumo}
Q^{\Delta ^{++}} & = & <\Delta ^{++},S_3=\frac{3}{2}|Q|
\Delta ^{++},S_3=\frac{3}{2}>  =  -\frac{2}{25} <r^2>_\Theta
\nonumber \\
Q^{N \Delta } & = & <\Delta ^{+},S_3=\frac{1}{2}|Q|
p,S_3=\frac{1}{2}>  =  -\frac{\sqrt{2}}{30} <r^2>_\Theta \;.
\end{eqnarray}
Our favorite models $A$ and $B$ give $Q^{\Delta ^{++}} \simeq -0.06
fm^2$, the transition moments are listed in table 3.3. Note in this
context that the
quadrupole moments are nonadiabatic quantities of ${\cal O}(N_C^{0})$
which we calculate in tree approximation only.

\subsection{$E2/M1$ ratio of the $\Delta$ photo-decay amplitudes}

The desire for precise values of the $\Delta$ photo-decay
amplitudes has spawned numerous theoretical and experimental
investigations in recent years. Much of this effort has been directed
toward the $E2/M1$ ratio of these amplitudes. In soliton models from
the earliest attempts on \cite{an85}\cite{a87}\cite{ww87} down to very
recent publications \cite{Goeke}\cite{awr96} the $E2$ transition was
always calculated via the time component of the vector current $V_0^3$
by employing Siegert's theorem \cite{s37} at the finite photon
momentum, where it is not strictly
valid. We calculate the $E2$ transition directly from the
spatial components of the vector current $V_i^3$, show that indeed for
$q \to 0$ the transition matrix element may be expressed by the
corresponding matrix element of $V_0^3$ (Siegert's theorem), but we
find nonnegligible corrections at the photon point 
leading to a noticeable reduction of the electromagnetic ratio.

For this purpose we have to solve the e.o.m.
\be   
\label{veom}
-\partial^i V_i^a = \partial_i V_i^a  =  \dot V_0^a =
\frac{f_\pi^2}{\Theta^2} s^2 b_T D_{ap}
(\mbox{\boldmath$ \hat{r} \times R$})_p (\mbox{\boldmath$ \hat{r} R$})   
\ee
to ${\cal O}(N_C^{-1})$ because the $E2$ transition is supressed by $1/
N_C^2$ as compared to the $M1$ transition. This leads to small
rotationally induced monopole and quadrupole deformations
\begin{eqnarray}
\label{rott}
\eta_L^\Omega &=& \frac{1}{6f_\pi \Theta^2} \left[
2f(r)\mbox{\boldmath$R$}^2+u(r)
(\mbox{\boldmath$R$}^2-3(\mbox{\boldmath$\hat{r} R$})^2) \right]  \\
\mbox{\boldmath$\eta$}_T^{\Omega} &=& -\frac{1}{2f_\pi \Theta ^2}
v(r)(\mbox{\boldmath$R$}-
\mbox{\boldmath$\hat{r}$} (\mbox{\boldmath$ \hat{r} R$})) 
(\mbox{\boldmath$ \hat{r} R$}) \nonumber
\end{eqnarray}
described in appendix A.1. The system of differential equations for the
radial functions $f(r)$, $u(r)$ and $v(r)$ is identical to the one
discussed for the magnetic polarizabilities (subsection 3.5.4), and
consequently also the radial functions shown in fig. 3.5. With these
components inserted, the vector current obeys (\ref{veom}). For this
reason we will not encounter ambiguous results
calculating the $E2$ transition via the time or the spatial components
of the electromagnetic current, discussed for constituent quark models
in \cite{dg84}.

In order to minimize errors, the so-called "equal-velocity" frame where
the incoming nucleon and the outgoing $\Delta$ have opposite velocities
and which reduces to the Breit frame as the nucleon-$\Delta$ split
vanishes, is chosen as a convenient reference frame \cite{wspr90}. In
this frame the photon momentum is $q_\Delta = (M_\Delta - M_N)/2
\sqrt{M_\Delta M_N} = 296 MeV$. 
We define the electromagnetic ratio by the ratio of the corresponding
helicity amplitudes
\be   
\label{emr1}
\frac{E2}{M1} = \frac{1}{3} \frac{A_{\frac{1}{2}}(E2)}{A_{\frac{1}{2}}(M1)} 
=\frac{1}{3} \frac
{<\Delta ^{+},S_3=\frac{1}{2}|M^{E2}_{\lambda=1}|p,S_3=-\frac{1}{2}>}
{<\Delta ^{+},S_3=\frac{1}{2}|M^{M1}_{\lambda=1}|p,S_3=-\frac{1}{2}>}
\ee
with the multipole operators \cite{fw66}
\begin{eqnarray}
\label{mupo}
M^{M1}_\lambda (q^2) = i \sqrt{6 \pi} \lambda \int d^3\!r\, V^3_i \, j_1(qr) \left[
\mbox{\boldmath$Y$}_{11\lambda}( \mbox{\boldmath$ \hat{r}$} ) \right]_i
\nonumber \\
M^{E2}_\lambda (q^2) = \frac{\sqrt{10 \pi}}{q} \int d^3\!r\, V^3_i \, \left[
\mbox{\boldmath$\nabla$} \times 
(j_2(qr) \mbox{\boldmath$Y$}_{22\lambda}( \mbox{\boldmath$
\hat{r}$} )) \right]_i \,.
\end{eqnarray}
First we are going to evaluate the $M1$
transition operator by inserting the vector current in tree $+$
$1$-loop (which part may be expressed by the isovector magnetic
formfactor discussed in section 3.4) as well as 
the rotationally induced soliton deformations
of ${\cal O}(N_C^{-1})$
\begin{eqnarray}
\label{m1po1}
M^{M1}_\lambda (q^2) &=& -\frac{3 \lambda}{2} \left[ 
\frac{q}{M_N} G_M^V (q^2) D_{3 \lambda} \right. \nonumber \\ 
&& \left. \qquad +2 f_\pi D_{3p} \int d^3\!r\, j_1 \frac{s}{r} \left(
c(\delta_{p \lambda} - \hat{r}_p \hat{r}_\lambda) \eta_L -
\hat{r}_p \eta_{T \lambda}   + \cdots \right) \right] \nonumber \\
&=& -\frac{3 \lambda}{2} \left[ 
\frac{q}{M_N} G_M^V (q^2) D_{3 \lambda} \right. \\
&& \left. \qquad + \frac{1}{45 \Theta^2} \int d^3\!r\, j_1 \frac{s}{r} 
(20cf - 2cu - 3v + \cdots ) 
\frac{1}{2} \{ D_{3\lambda},\mbox{\boldmath${R}$}^2 \} \right. \nonumber \\  
&& \left. \qquad + \frac{1}{15 \Theta^2} \int d^3\!r\, j_1 \frac{s}{r} 
(2cu + 3v + \cdots) L_3 R_\lambda \right] \; . \nonumber   
\end{eqnarray} 
With this multipole operator,
the $M1$ transition matrix element in tree $+$
$1$-loop with soliton deformations included 
\begin{eqnarray}
\label{m1po}
&&<\Delta ^{+},S_3=\frac{1}{2}|M^{M1}_{\lambda=1}|p,S_3=-\frac{1}{2}>
= - \frac{q}{2 \sqrt{2} M_N} G^{N \Delta}_{M1}(q^2)  \\
&&G^{N \Delta}_{M1}(q^2) = \sqrt{2} \left\{ G^{V}_{M}(q^2)
+\frac{M_N}{20q \Theta^2}
\int d^3\!r\, j_1 \frac{s}{r}   
[2c(10f-u)-3v+ \cdots ] \right\} \; . \nonumber
\end{eqnarray}
is determined.

For the $E2$ transition there is no choice: the soliton
deformations (\ref{rott}) must be inserted, otherwise the operator vanishes.
Alternatively, and of course with the same result, we may employ
partial integration and the vector current conservation (\ref{veom})
\begin{eqnarray}
\label{e2po1}
M^{E2}_\lambda (q^2) &=& \frac{1}{iq} \sqrt{\frac{5 \pi}{3}} \int d^3\!r\, \left[
(3j_2-qrj_3) \partial_i V_i^3 - q^2 j_2 x_i V_i^3 \right] Y_{2 \lambda}
\nonumber \\
&=& \frac{1}{iq} \sqrt{\frac{5 \pi}{3}} \int d^3\!r\, \left[ 
(3j_2-qrj_3) \dot V_0^3  \right.  \\ 
&&  \left. \qquad \qquad \qquad + f_\pi q^2 j_2 D_{3p} 
(r F^{\prime} c \mbox{\boldmath$\hat{r}$} \times \mbox{\boldmath$\eta$}
- r s \mbox{\boldmath$\hat{r}$} \times \mbox{\boldmath$\eta$} ^{\prime}
+ \cdots )_p \right] Y_{2 \lambda} \nonumber \\
&=& \sqrt{\frac{5 \pi}{3}} \frac{1}{\Theta} \int d^3\!r\, \left[
f_\pi^2 (3j_2-qrj_3) s^2 b_T - \frac{q^2}{2} j_2  
(r F^{\prime} cv - rsv^{\prime} + \cdots ) \right] \nonumber \\
&& \qquad \qquad \qquad D_{3p} (R_p - \hat{r}_p (\mbox{\boldmath$\hat{r} R$}))
Y_{2 \lambda} \; . \nonumber
\end{eqnarray}
In the last step the soliton deformations (\ref{rott}) (only 
$N\ell\sigma$ contributions are written out in explicit form) are
inserted, and it is then noticed that the entire operator may be
written as a total time derivative which in the end may be replaced by
$iq$ (compare the derivation of Siegert's theorem). The $E2$ transition
matrix element is now
\be
\label{e2po2}
<\Delta ^{+},S_3=\frac{1}{2}|M^{E2}_{\lambda=1}|p,S_3=-\frac{1}{2}>
= - \frac{3q^2}{4 \sqrt{2}} G^{N \Delta}_{E2}(q^2)
\ee
$$
G^{N \Delta}_{E2}(q^2) = -\frac{\sqrt{2}}{9 q^2 \Theta}
\int d^3\!r\, \left[
f_\pi^2 (3j_2-qrj_3) s^2 b_T - \frac{q^2 j_2}{2}  
(r F^{\prime} cv - rsv^{\prime}
+ \cdots ) \right] \, \, . 
$$
It should be remembered here that $G^{N \Delta}_{E2}$ is a nonadiabatic
quantity which we calculate in tree approximation only. There may
possibly be nonadiabatic $1$-loop contributions of the same 
order which to calculate we are not yet in a position.

For $q \to 0$ we obtain $G^{N \Delta}_{E2} \to Q^{N \Delta}$
(\ref{qumo}) which is indeed Siegert's theorem. At the photon point
the corrections (second and third term in (\ref{e2po2}))
are sizeable, numerical values for the parameter combinations $A$, $B$,
and $C$ are compared in table 3.3. The reason for $|G^{N \Delta}_{E2}
-Q^{N \Delta}|$ being larger than expected is connected with the fact
that the induced rotational components (fig. 3.5) fall off quite slowly
which corresponds to a large effective radius of these corrections. 
With (\ref{m1po}, \ref{e2po2}) the electromagnetic ratio (\ref{emr1})
is determined as
\be   
\label{emr2}
\frac{E2}{M1} = \frac{q_\Delta M}{2}
\frac{G^{N \Delta}_{E2}(q_\Delta^2)}{G^{N \Delta}_{M1}(q_\Delta^2)} 
\to \frac{q_\Delta M}{2} \frac{G^{N \Delta}_{E2}(\Delta^2)}{G^{N \Delta}_{M1}(\Delta^2)}
\; .
\ee
In the last step a relativistic correction similar to (3.32) leads to
the replacement $q_\Delta \to \Delta = M_\Delta - M_N =293 MeV$ in the
transition formfactors which
numerically is of negligible influence.
The results are listed in table 3.3 with an absolute value
considerably smaller than in previous soliton calculations mainly for two 
reasons:
\begin{itemize}
\item $G^{N \Delta}_{M1}$ is larger and
close to the experimental value at least for our favored models $A$
and $B$ (accidentally, maodel $A$ hits the experimental value of the
corresponding helicity amplitude \cite{pdg94} exactly) because we included loop corrections
of ${\cal O}(N_C^{0})$ and soliton deformations of ${\cal O}(N_C^{-1})$
\item The consideration of $G^{N \Delta}_{E2}(\Delta^2)$ instead of
$Q^{N \Delta} (\Delta^2)$ (deviation from Siegert's theorem due to
finite $\Delta$) reduces the absolute value of the
electromagnetic ratio further.
\end{itemize}
Our result of $E2/M1 \simeq -1.7\%$ has to be compared with the
experimental value reported by the particle data group
\cite{pdg94} $-1.5 \pm 0.4 \%$ and a recent MAMI experiment 
with
the preliminary result $-2.4 \pm 0.2 \%$ \cite{Beck}.

\begin{table}[h]
\begin{center} \parbox{6cm}{\caption{\label{eomres}
Magnetic and qua\-dru\-pole transition moments
in units of nuclear magnetons and $fm^2$ resp. are
compared for the three
parameter  combinations $A$, $B$, $C$. The corresponding transition
formfactors
are evaluated at the photon point.
}}

\begin{tabular}{|c|c|c|c|}
\hline
& $A$ & $B$ & $C$ \\
\hline
$\mu^{N \Delta}            $  & 4.1  & 4.4  & 4.9   \\
$G^{N \Delta}_{M1} (\Delta^2)   $  & 3.2  & 3.4  & 3.8  \\
\hline
$Q^{N \Delta}              $  & -.034 & -.035 & -.038  \\
$Q^{N \Delta} (\Delta^2)        $  & -.021 & -.022 & -.023  \\
$G^{N \Delta}_{E2} (\Delta^2)   $  & -.015 & -.015 & -.016   \\
\hline
$\frac{E2}{M1} [\%]   $ & -1.7 & -1.6 & -1.5   \\
\hline
\end{tabular}
\end{center} \end{table}
Because the extraction of the electromagnetic ratio from experimental
data is far from being trivial \cite{dmw91}\cite{wal92}, a complete
calculation of the $\gamma N \to \pi N$ reaction, which includes the
background with the $\Delta$ resonance lying in the continuum would be
desirable in order to obtain complex helicity amplitudes for the $N
\Delta$ transition. Although such a calculation should in principle be
feasible within soliton models, the effort would be tremendeous.

\chapter{Results}
In this chapter, we are finally going to present and discuss results obtained
by applying the method outlined above to the quantities introduced in the
preceding paragraphs. As a shorthand, we shall in the following denote 
quantities whose loop
corrections can be treated in adiabatic approximation (and are therefore
calculated in this paper) as adiabatic quantities, whereas the others 
(for which we are
unable to calculate the loop correction) will be summarized
as nonadiabatic ones; only their tree level values will be given. 
For all quantities considered, we evaluate the Lagrangian
(\ref{stanlag}) for the three parameter sets $A$
($e=4.25,\; g_{\omega}=0$), $B$ ($e=4.5,\; g_{\omega}=1.0$) and $C$
($e=5.845 ,\; g_{\omega}=2.2$) introduced in section 2.2 . From the remarks
in that section, it should be clear that we cannot expect model $C$ to
result in sensible numbers since the local approximation with
vector meson
parameters overestimates the effect from the $\omega$ meson
and neglects important contributions of higher chiral
orders.  However, sets $A$ and $B$, designed to simulate these missing
pieces by introducing 'effective' parameter values should give an
accurate picture of the abilities (and defects) of the soliton approach
with respect to the quantities under investigation. Since amongst these
three, set $B$ comes
closest to the original vector meson model, we consider this to be the
most realistic one. 

Before entering the discussion of magnitude and sign of the $1$ loop 
corrections it is worthwhile to recall the salient features of the tree
approximation within the three parameter sets $A$, $B$ and $C$, at least for
the adiabatic quantities:
\begin{itemize}
\item The {\bf soliton mass} has been a constant source of embarrassment for the 
past decade, since it came out too large by a factor of more than $3/2$
almost regardless of parameter combinations used or meson species included in
the model. This is also obvious from tables \ref{ares}, \ref{bres} and
\ref{cres}, where
the tree level mass is shown to be in the $1500-1630 MeV$ range for the three
models.
\item With values between $54$ and $67 MeV$ against experimental $45MeV$, the 
$\mbox{\boldmath$ \sigma$} $ {\bf term} is also 
overestimated in tree for all parameter
combinations under consideration here. This is not a common feature of all
Skyrme type models and depends on our usage  of the full $ChO \; 4$ lagrangian
including symmetry breakers. The effect is much less pronounced than in case of
the mass, especially if the comparably large error margin of the experimental
datum is taken into account. Nevertheless, a sizeable reduction due to $1$ loop corrections would
be necessary. Partly due to the overestimated $\sigma$ term, the scalar radius
comes out consistently much too low in tree, its value of around $1fm$ as
opposed to experimental $1.6fm$ being rather insensitive
to
the parameters used.
\item The {\bf axial vector constant} is a weak point of Skyrme type
models, since the tree contribution doggedly refuses to grow appreciably beyond
$1.0$ (our values being $.91-1.03$) as long as reasonable parameters are 
considered, where reasonable means that the other quantities should not be
completely off the mark. Likewise, the quantity $g_A <r^2>_A$ is underpredicted
in all cases, numerical values ranging from $.36fm^2$ to $.42fm^2$ whereas
experiment points to $.53fm^2$.
\item  Concerning the {\bf electromagnetic formfactors}, it has long been recognized
that magnetic moments generally come out too small, most notably the (nonadiabatic) 
isoscalar one, which is wrong by a factor of $2$, but also the isovector moment
for which our parameters give $1.6-1.8$ nuclear magnetons in contrast to
the experimental finding of $2.35$. Similarly, for some time, conventional 
wisdom had it that
e.m. radii were seriously underpredicted in purely pionic models. However,
inclusion of nonminimal couplings with coupling constants chosen in accordance
with the logic presented  in section 2.2, as well as taking into account
relativistic corrections serve to remedy this problem already at tree level,
and for
our parameter sets, actually the converse  is true: isoscalar electric as well
as
isovector magnetic radius are slightly overestimated with values of
$.61-.67fm^2$ for the former and $.77-.72 fm^2$ for the latter case to
be compared to experimental data of $.59fm^2$ and $.73fm^2$, respectively. 
\item The {\bf electric polarizability} is consistently too large in tree
without  regard of the model considered; the discussion in the corresponding
section of the previous chapter has revealed  that this would also hold in
models containing vector mesons explicitly, which were for some time believed
to cure that deficiency. Actual numbers are in the $18-22 \cdot 10^{-4} fm^3$ 
bracket for our models; however, the amount by which the polarizability is
overestimated is difficult to quantify in view of the large error bar,
the experimental value being $9.5 \pm 5 \cdot 10^{-4} fm^3$. 
\end{itemize}

Turning now to the loop corrections, inspection of tables \ref{ares},
\ref{bres} and \ref{cres} provides a rough classification in terms of the
sensitivity to the stabilisation mechanism: quantities with a loop contribution
depending sensitively on
the choice of parameters
comprise $\sigma$ term, $g_A$, axial radius end electric polarizability. All
others are rather insensitive to the details of the lagrangian. We shall
postpone discussion of the first kind of properties and deal firstly with the
latter species. Here, the most pronounced results are the following:
\begin{itemize}
\item As had been shown previously (\cite{mou} \cite{ho}) the 
{\bf soliton mass} is considerably reduced towards its experimental value once
loop corrections are taken into account. Every parameter set considered
here results in a
tree $+$ $1$ loop mass of around $950$ $MeV$. 
\item No less importantly, the {\bf scalar radius} is considerably enhanced
towards the experimental datum, with tree $+$ $1$ loop results in the range of
$1.3-1.5fm^2$.
\item The same is true for the {\bf isovector magnetic moment}, which shows an
enhancement to $2.24-3.13$. The latter number ($3.13$ for set $C$) actually 
reverses the 
tree level underprediction into a considerable overestimation, but both the
other models are satisfactory.
\end{itemize}
In all these cases, the quantum correction has the right sign and a magnitude
of slightly more than $1/3$ relative to the tree level value.This serves as an 
impressive
justification of the conjecture that loop corrections might play a crucial
role in the understanding of baryon properties in soliton models. 

Less marked changes
occur in the e.m. radii, where the correction is around $1/6$ of tree, yet has
the right sign in all cases except $C$, where the too small tree magnetic
radius
is further reduced. Since the tree level values were already rather good, it is
not really justified to speak of an improvement here; however, the fact that
loop contributions  are tiny where one would expect them to be, and generally
have, once again, the right sign supports the hypothesis that quantum
corections, properly taken into account, could eventually cure the defects of
tree level evaluated Skyrme type models without worsening things with respect
to quantities 
where tree already gives acceptable results.

So far, there has been little to choose between sets $A$, $B$ and $C$ with the
exception of magnetic properties in case $C$. This radically changes if one
considers the first kind of quantities mentioned above, namely $\sigma$ 
term, $g_A$, axial radius and electric polarizability.

Dealing first with the $\mbox{\boldmath$ \sigma$}$ term, 
this happens to be corrected in the
right direction and with the right magnitude for the first two parameter sets
$A$ and $B$; actually, set $A$ turns the too large tree value into a too small
tree $+$ $1$ loop one ($32 MeV$), whereas set $B$ is just about right ($44MeV$).
However, the startling observation is
that for a strong enough sixth order term (set $C$) the $1$-loop 
contribution changes
sign and increases the already too large tree value to about $80 MeV$.
A similar phenomenon occurs in case of the $\mbox{\boldmath$ g$}_A$ if 
one takes into account set
$D$, and also for the {\bf axial radius}. We do not want to repeat the discussion of
section 3.3 here, but once again we emphasise that the changing sign of the loop
correction serves as a filter to decide which model to discard and which to
keep; sets $C$ and $D$ are clearly ruled out by the numbers.
(Note that in contrast to the situation for the slope of the axial formfactor
$g_A<\!r^2\!>_A$, the $1$ loop correction to the radius itself is not
worrisome, and that $<\!r^2\!>_A$ comes out too large, yet within the error margin
of the experimental datum, if one includes the estimated $1/N_C$ piece).  
The fairly good stability of the tree $+$ $1$ loop value of $g_A$ over
a wide range of scale as shown in fig.\ref{scalga} probably rules
out missing higher chiral orders to
result in a scenario different from the one presented here (e.g.,
positive loop correction for otherwise reasonable parameter sets).
Eventually, one of the success storys of our endeavour is the loop contribution
to the {\bf electric polarizability}, which turns out to bring both, models $A$ and
$B$, to the experimental datum including error bars (tree $+$ $1$ loop values
being $9.8$ and $14.1$ respectively). Like its precursors discussed above, the
correction changes sign as one proceeds from set $B$ to set $C$, inflating the
tree value of $22$ to about $27$, therefore adding further to the list of its
deficiencies. 
We believe this sensitivity to the strength of the sixth order term in the
lagrangian to reflect the problems of the local approximation on
the $\omega $ meson discussed above: A strong sixth order term gives
rise to unphysical features in the scattering phaseshifts, and consequently  
fails to result in sensible predictions for certain quantities. Since
we cannot, for the moment, properly account for the $\omega$ meson, we
have to be content with the philosophy adopted in section 2.2, namely to
use effective values for $e$ and $g_{\omega}$ which then results in
good agreement with experiment.

Summarising, we have shown loop corrections to bring all adiabatic quantities
with the exception of the axial ones, close to their experimental values for
parameter sets $A$ and $B$. In case of $g_A$ and $<\!r^2\!>_A$, we believe
(section 3.3) that their status is exceptional and that the $1/N_C$ expansion must
run into difficulties in their case. We have provided an estimate of the next-
to-next-to-leading order correction to them, which turned out to be large
\begin{table}[p]
\begin{center} \parbox{10cm}{\caption{\label{ares}
Tree and $1$-loop contribution to various
quantities for
parameter set $A$
($e=4.25, g_{\omega}=0$).}}
\begin{tabular}{|c|c|c|c|c|}
\hline
&tree& $1$-loop& $\sum$ & exp.\\
\hline
$M [MeV]$ & 1629 & $-$683 & 946.0 & 939 \\
\hline
$\sigma$ $[MeV]$ & 54 & $-$22 & 32 & 45 $\pm$ 7 \\
\hline
$<\!r^2\!>^S$ $[fm^2]$ & 1.0 & $+$.3 & 1.3 & 1.6 $\pm$ .3 \\
\hline
$g_A$ & .91 & $-$.25 & .66 & 1.26 \\
\hline
$<\!r^2\!>_A$ $[fm^2]$ & .45 & $-$.04 & .41 & .42 ${+.18 \atop -.08}$ \\
\hline
$<\!r^2\!>_E^S$ $[fm^2]$ & .62 & $-$.11 & .51 & .59 \\
\hline
$\mu^V$ & 1.62 & +.62 & 2.24 & 2.35 \\
\hline
$<\!r^2\!>_M^V$ $[fm^2]$ & .77 & $-$.13 & .64 & .73 \\
\hline
$\alpha$ $[10^{-4} fm^3]$ & 17.8 & $-$8.0 & 9.8 & 9.5 $\pm$ 5 \\
\hline
\end{tabular}
\parbox{10cm}{\caption{\label{bres}
Tree and $1$-loop contribution to various
quantities for
parameter set $B$
($e=4.5, g_{\omega}=1.0$).}}
\begin{tabular}{|c|c|c|c|c|}
\hline
&tree& $1$-loop& $\sum$ & exp.\\
\hline
$M [MeV]$ & 1599 & $-$646 & 953 & 939 \\
\hline
$\sigma$ $[MeV]$ & 58 & $-$14 & 44 & 45 $\pm$ 7 \\
\hline
$<\!r^2\!>^S$ $[fm^2]$ & 1.1 & $+$.3 & 1.4 & 1.6 $\pm$ .3 \\
\hline
$g_A$ & .96 & $-$.15 & .81 & 1.26 \\
\hline
$<\!r^2\!>_A$ $[fm^2]$ & .44 & $-$.01 & .43 & .42 ${+.18 \atop -.08}$ \\
\hline
$<\!r^2\!>_E^S$ $[fm^2]$ & .64 & $-$.09 & .55 & .59 \\
\hline
$\mu^V$ & 1.69 & $+$.88 & 2.57 & 2.35 \\
\hline
$<\!r^2\!>_M^V$ $[fm^2]$ & .76 & $-$.17 & .59 & .73 \\
\hline
$\alpha$ $[10^{-4} fm^3]$ & 19.4 & $-$5.3 & 14.1 & 9.5 $\pm$ 5 \\
\hline
\end{tabular}
\parbox{10cm}{\caption{\label{cres}
Tree and $1$-loop contribution to various
quantities for parameter set $C$
($e=5.845, g_{\omega}=2.2$).}}
\begin{tabular}{|c|c|c|c|c|}
\hline
&tree& $1$-loop& $\sum$ & exp.\\
\hline
$M [MeV]$ & 1490 & $-$539 & 951 & 939 \\
\hline
$\sigma$ $[MeV]$ & 67 & $+$13 & 80 & 45 $\pm$ 7 \\
\hline
$<\!r^2\!>^S$ $[fm^2]$ & 1.1 & $+$.4 & 1.5 & 1.6 $\pm$ .3 \\
\hline
$g_A$ & 1.03 & $-$.11 & .92 & 1.26 \\
\hline
$<\!r^2\!>_A$ $[fm^2]$ & .40 & $+$.16 & .56 & .42 ${+.18 \atop -.08}$ \\
\hline
$<\!r^2\!>_E^S$ $[fm^2]$ & .68 & $-$.06 & .62 & .59 \\
\hline
$\mu^V$ & 1.75 & $+$1.38 & 3.13 & 2.35 \\
\hline
$<\!r^2\!>_M^V$ $[fm^2]$ & .72 & $-$.19 & .53 & .73 \\
\hline
$\alpha$ $[10^{-4} fm^3]$ & 21.8 & $+$5.4 & 27.2 & 9.5 $\pm$ 5 \\
\hline
\end{tabular}
\end{center} 
\end{table}
and
remedies the defects that show up including only ${\cal O}(1)$.
Although there is in principle the possibility of sizeable ${\cal O}(N_C^{-1})$ effects
for the other quantities, too, already at this point, we can safely state that
whatever parameter set (if any) might come out of a consideration
including these must lie in the vicinity of $A$ or $B$: $C$-like sets are clearly ruled out from the electric
polarizability alone since they would
require a $1/N_C$ correction of around $-10$ as opposed to $+5$ at ${\cal
O}(1)$.
\begin{table}[h]
\begin{center} \parbox{9.1cm}{\caption{\label{cores}
Comparative listing of tree $+$ $1$ loop values for all
adiabatic quantities considered for the three
parameter  combinations $A$, $B$, $C$. Axial quantities
include estimated $1/N_C$ piece.}}
\begin{tabular}{|c|c|c|c|c|}
\hline
& $A$ & $B$ & $C$ & exp.\\
\hline
$M [MeV]$ & 946 & 953 & 951 & 939 \\
\hline
$\sigma$ $[MeV]$ & 32 & 44 & 80 & 45 $\pm$ 7 \\
\hline
$<\!r^2\!>^S$ $[fm^2]$ & 1.3 & 1.4 & 1.5 & 1.6 $\pm$ .3 \\
\hline
$g_A$ & 1.20 & 1.29 & 1.36 & 1.26 \\
\hline
$<\!r^2\!>_A$ $[fm^2]$ & .55 & .54 & .59 & .42 ${+.18 \atop -.08}$ \\
\hline
$<\!r^2\!>_E^S$ $[fm^2]$ & .51 & .55 & .62 & .59 \\
\hline
$\mu^V$ & 2.24 & 2.57 & 3.13 & 2.35 \\
\hline
$<\!r^2\!>_M^V$ $[fm^2]$ & .64 & .59 & .53 & .73 \\
\hline
$\alpha$ $[10^{-4} fm^3]$ & 9.8 & 14.1 & 27.2 & 9.5 $\pm$ 5 \\
\hline
\end{tabular}
\end{center} 
\end{table} 
Table \ref{nonares} gives tree results for nonadiabatic
quantities. 
\begin{table}[h]
\begin{center} \parbox{8.3cm}{\caption{\label{nonares}
Comparative listing of tree values for all
nonadiabatic quantities considered for the three
parameter  combinations $A$, $B$, $C$.}}
\begin{tabular}{|c|c|c|c|c|}
\hline
& $A$ & $B$ & $C$ & exp.\\
\hline
$\Delta$ $[MeV]$ & 290 & 278 & 265 & 293 \\
\hline
$M_{np}$ $[MeV]$ & 2.0 & 1.7 & 1.4 & 2.05 $\pm$ .30 \\
\hline
$<\!r^2\!>_E^V$ $[fm^2]$ & .86 & .90 & .98 & .85 $\pm$ .03 \\
\hline
$\mu^S$ & .18 & .19 & .21 & .44 \\
\hline
$<\!r^2\!>_M^S$ $[fm^2]$ & .74 & .75 & .77 & .59 $\pm$ .02 \\
\hline
${\scriptstyle \triangle}\alpha$ $[10^{-4} fm^3]$ & 1.7 & 1.8 & 1.9 & 5.0 $\pm$ 5.0 \\
\hline
\end{tabular}
\end{center} \end{table}
Here, one observes that all quantities except two
are rather well reproduced already at this level. The two exceptions
are the isoscalar
magnetic moment and the neutron-proton-split of the electric
polarizability. The
latter is less severe in view of the huge error of the experimental
value, but $\mu_S$ is seriously defective and it seems unlikely that a
next order quantum correction would be able to remedy this discrepancy. In
principle, however, there is the possibility of a quantum correction of the same
order as the tree level in $1/N_C$ since the time derivative contained in the
spatial component of the winding number current expanded  to quadratic order in
the fluctuations may act on a fluctuation instead
of a collective coordinate and twin fluctuations count as $1/N_C$ like the angular
velocity. We are not yet in a position to decide whether this is the root cause
for the too small isoscalar magnetic moment.

Finally, we shall briefly recapitulate the course of this report: In the
beginning we posed the question whether it is possible to renormalize a chiral
lagrangian evaluated in a nontrivial topological sector in analogy to the
methods known as chiral perturbatuion theory for the vacuum sector. The main 
problem is the restriction to a finite number of gradients contained in the
lagrangian, which is not a priori justifiable in the presence of a soliton
background. Upon considering a lagrangian incorporating resonance degrees of
freedom we tried to better understand this issue and concluded that using the
$4$th order chiral lagrangian augmented by a standard $6$th order extension
including nonminimal couplings would be sufficient provided a shift
away from the experimental values of $LEC$s would be allowed for.
Later on, we could add substance to this conclusion by considering the scale dependence
of the tree $+$ $1$-loop soliton mass and the axial coupling constant $g_A$
whose values were shown to be remarkably stable over a wide range of the scale.

Starting from this lagrangian, we devised a method to calculate the self energy
of the soliton in the presence of external fields in adiabatic approximation.
In the course of the actual calculation, we had to clear  several obstacles
only indirectly related to our subject: We
clarified the treatment of zero modes in the presence of external fields,
showed how to properly compute the tree value of the electric polarizability
and calculated the neutron-proton-split of this quantity.
In the end questions remain as to the convergence of  the $1/N_C$ expansion for
axial quantities and to the size of the next-to-next-to-leading order
contributions. However, we once again want to stress that in $7$ out of $9$
cases, the $1$ loop correction has the right sign and magnitude to compensate
for the discrepancy between tree level and experiment, which is truly
remarkable since there are only two parameters ($e$ and $g_\omega$)
involved (which moreover have not been tuned to
achieve some specific result).

The fact that sets $A$ and $B$ designed to simulate the higher chiral order
pieces contained in a vector meson model indeed give much better
results than set $C$ strongly supports models which include vector
mesons explicitly. We decided to use a pseudoscalar lagrangian only
because of the arguments presented in section 2.3 . Nevertheless, we expect
that, concerning $1$-loop corrections, the results for a
lagrangian with explicit $\varrho$ and $\omega$ -mesons would come 
close to those of our model $B$. In particular the problems with the
axial quantities discussed in section 3.3 will probably not be cured
by the introduction of vector mesons.

\section*{Acknowledgements}
The authors would like to express their gratitude for numerous
stimulating discussions with their colleagues G. Holzwarth and B.
Schwesinger. They are particularly indebted to B. Moussallam, who
patiently taught them his method to calculate the Casimir energy in all
details.

\begin{appendix}
\chapter{ $1/N_C$ contribution to $g_A$}
Although the relation (\ref{kir}) is model independent, the positivity of the
remainder $R$ is not a direct implication of the charge algebra. Rather, this
follows either from the experimental value of $g_A$ (which would be a circular
argument in our context) or from the fact that $R$ can
be expressed  as a difference of pion nucleon cross sections which is experimentally
known to be positive. In soliton models, it is hard to see whether $R$
is indeed positive independently of the parameter set chosen.
Provided this were the
case and the $1/N_C$ expansion would converge for $R$, then $g_A$ would
necessarily have to be larger than $1$ as soon as the charge algebra
relation (\ref{cicu}) were fulfilled. We could then  use a different strategy to
obtain an estimate for the $1/N_C$ piece of $g_A$, which in the end is
equivalent to the naive addition of the CA "$1$" to $g_A^2$ used in
section 3.3.3 .Such an estimate is the subject of this appendix. 

Soliton models obey (\ref{cicu}) provided nonadiabatic 
fluctuations which come with a collective angular velocity are taken into 
account \cite{sw94}. Relative to the adiabatic fluctuations employed for the $1$ 
loop contributions, these are suppressed by a factor of $1/N_C$. As a 
consequence current algebra is reflected in $g_A$ not before ${\cal O}(N_C^{-1})
$ which, of course, is also immediately clear from (\ref{kir}).

Quite generally, the (spatially integrated) axial current 
up to ${\cal O}(N_C^{-1})$ may be written as
\be
{\cal A}_i^a= -\frac{3}{2} g_A^{(1+0)}D_{ai} + \frac{1}{2}a^{(-1)}
 \{ D_{ai} , \mbox{\boldmath$R$}^2 \}+b^{(-1)}L_a R_i+c^{(-1)}D_{ai}
\label{ac}
\ee
where the $N_C$ orders are indicated by the superscripts.

In this object, the tree contribution proportional to the angular velocity squared of 
${\cal O}(N_C^{-1})$ due to the soliton's
rotation can be shown to be negligible (subsection A.1). Adiabatic
$2$-loop diagrams
cannot produce the operator structure of (\ref{ac}). There remain the
nonadiabatic $1$ loop processes: We believe that these should contribute sizeably
in ${\cal O}(N_C^{-1})$ to eq. (\ref{ac}) because exactly they 
are responsible for the
restoration of the charge algebra commutation relation (\ref{cicu}).
However, they are too complicated to investigate in detail.  

Therefore, in the second subsection of this appendix, in order to circumvent this
obstacle, we are going to estimate the 
$1/N_C$ contribution
to $g_A$ produced by nonadiabatic $1$ loop processes
by imposing the $SU(4)$ current algebra commutation relations
\be
\label{su4}
[{\cal A}_3^a ,\,\, {\cal A}_3^b ]=i\, \varepsilon_{abc} L_c
\:.
\ee

\section{Contribution due to rotationally induced soliton deformations}

Here we show that the effect of soliton deformations on the axial
current is too weak to explain the relatively large $1/N_C$
contributions needed in order to fulfill the CA commutation relation
(A.2). This does of course not imply that soliton deformations
are small by themselves, in the vector current they are responsible
for sizeable contributions to the magnetic moments as we have
noticed in subsection 3.6.2.
Small deformations due to the soliton's rotation are driven by the term
\be
L_{\Omega}=f_{\pi} \int d^3\!r\,\, \left[ sc (\mbox{\boldmath$ \hat{r} \times
\Omega$}^R)^2 \eta^{\Omega}_L-s(\mbox{\boldmath$\hat{r}\Omega$}^R)
(\mbox{\boldmath$ \Omega$}^R \mbox{\boldmath$ \eta$}^{\Omega}_T) + \cdots \right]
\label{soldef}
\ee
in the lagrangian. The corresponding e.o.m.
\be
\label{soldef2}
h_{ab}^2 \eta^{\Omega}_b=f_{\pi} \left[sc \hat r_a (\mbox{\boldmath$
\hat{r} \times
\Omega$}^R)^2 -s (\Omega_a-\hat r_a (\mbox{\boldmath$\hat{r}\Omega$}^R))
(\mbox{\boldmath$\hat{r} \Omega$}^R) + \cdots \right]
\ee
may again be solved analytically in the asymptotical region
\be
\label{defas}
\mbox{\boldmath$\eta$}^{\Omega}\stackrel{r \to \infty}{=}
\frac{3g_A}{16 \pi f_{\pi}} e^{-m_{\pi}r}\left[ \mbox{\boldmath$\hat{r}$}
(\mbox{\boldmath$\Omega$}^R)^2-\mbox{\boldmath$\Omega$}^R 
(\mbox{\boldmath$ \hat{r}\Omega$}^R) \right] \; ,
\ee
which is in accordance with the result in ref.\cite{dohuma}. The full
equations induce a monopole and a quadrupole deformation
\begin{eqnarray}
\label{defeom}
\eta_L^{\Omega} &=& \frac{1}{6f_\pi} \left[
2f(r)(\mbox{\boldmath$ \Omega$}^R)^2+u(r)
((\mbox{\boldmath$ \Omega$}^R)^2-3(\mbox{\boldmath$\hat{r}
\Omega$}^R)^2) \right]  \\
\mbox{\boldmath$\eta$}_T^{\Omega} &=& -\frac{1}{2f_\pi}v(r)(\mbox{\boldmath$\Omega$}^R-
\mbox{\boldmath$\hat{r}$} (\mbox{\boldmath$ \hat{r}\Omega$}^R)
(\mbox{\boldmath$ \hat{r}\Omega$}^R)) \; ,\nonumber
\end{eqnarray}
and it turns out that the system of differential equations
for the radial functions $f(r)$, $u(r)$ and $v(r)$ is identical to the
ones discussed for the magnetic polarizabilities (subsection 3.5.4) and
the electromagnetic ratio of the photo-decay amplitudes (subsection 3.7.2).
The radial functions are depicted in fig.\ref{betarad}.
These rotationally induced
components may give rise to $1/N_C$ contributions to the spatial axial current
\begin{eqnarray}
{\cal A}_i^a &=&  \int d^3\!r\,\,x_i (\dot{A_0^a}-\partial^{\mu}A_{\mu}^a)
\nonumber \\
            &=& -\frac{3}{2}g_A D_{ai} \nonumber \\
            & & +\frac{f_{\pi}^2}{2 \Theta^2} \int d^3\!r\,\,r\, sc\,b_T
(\frac{1}{2} \{ D_{ai},\mbox{\boldmath${R}$}^2 \} -L_a R_i) \nonumber \\
& & -\frac{m_{\pi}^2}{9 \Theta^2}\int d^3\!r\,\,r\, cf
\frac{1}{2} \{ D_{ai},\mbox{\boldmath${R}$}^2 \}           \nonumber \\
& & -\frac{m_{\pi}^2}{15 \Theta^2} \int d^3\!r\,\,r
(\frac{1}{3} cu+\frac{1}{2} v)
(\frac{1}{2} \{ D_{ai},\mbox{\boldmath${R}$}^2 \} -3L_a R_i)
\:.              
\label{defcu}
\end{eqnarray}
Here we used partial integration and we listed only the contribution of the
familiar symmetry breaking mass term. Equivalently, we could have inserted
(\ref{defeom}) directly into ${\cal A}_i^a$ with the same result.

From (\ref{defcu}), we recover the operator structure (\ref{ac}) and we may read
off the tree level pieces of the coefficients $a^{(-1)}$ and $b^{(-1)}$.
Evaluating (\ref{defcu}) in the chiral limit with the appropriate asymptotic
expressions (\ref{betafluca})
\begin{eqnarray}
{\cal A}_i^a+\frac{3}{2}g_A D_{ai} &\stackrel{m_{\pi} \to 0}{=}&
\frac{3g_A}{2 \Theta^2 m_{\pi}^2}
(\frac{1}{2} \{ D_{ai},\mbox{\boldmath${R^2}$} \} -L_a R_i) \nonumber \\
& & -\frac{g_A}{\Theta^2 m_{\pi}^2}
\frac{1}{2} \{ D_{ai},\mbox{\boldmath${R^2}$} \}           \nonumber \\
& & -\frac{g_A}{2 \Theta^2 m_{\pi}^2} 
(\frac{1}{2} \{ D_{ai},\mbox{\boldmath${R^2}$} \} -3L_a R_i) \nonumber \\
&=& 0 \cdot {\cal O}(m_{\pi}^{-2})             
\end{eqnarray}
the divergent terms cancel exactly which is pleasing because we expect the
coefficients $a^{(-1)}$ and $b^{(-1)}$ to be finite in the chiral limit in
order to fulfill the CA commutation relation(\ref{su4}). According to this
cancellation, the contributions $a^{(-1)}$ and $b^{(-1)}$ remain small 
also for finite pion mass. For example, model $A$ provides the
tree contributions $a^{(-1)}=-.04$ and $b^{(-1)}=-.09$ which are negligible
compared to the values $a^{(-1)}=-b^{(-1)}=.36$ necessary to fulfill the CA
relation (\ref{su4}) (compare also the following subsection).
Therefore we do not expect soliton deformations to
appreciably contribute to this relation.

\section{Contribution due to nonadiabatic loops}

In this section we estimate the $1/N_C$ contribution to the axial
current due to nonadiabatic loops by imposing the CA relation (A.2).
Before doing so, we notice that in a pure
Skyrme model (N$\ell \sigma$ model with $f_\pi$ and Skyrme term with parameter $e$)
the coefficient in the axial current (A.1) scale according to
\be
g_A^1 \sim \frac{1}{e^2} \,, \quad
g_A^0 \sim \frac{1}{f_{\pi}^2 e^2} \,,\quad
a^{(-1)} \sim b^{(-1)} \sim c^{(-1)} \sim \frac{1}{f_{\pi}^2}
\ee
($e=2g$ is ${\cal O}(N_C^{-1/2})$). Thus by choosing a large
Skyrme parameter we can make the tree $+$ $1$ loop contribution small. The
property $g_A \ge 1$ must then almost entirely be achieved
due to the $1/N_C$ contribution.

The CA commutation relation (A.2) requires
\be
b^{(-1)} = - a^{(-1)}\;,\quad
(a^{(-1)})^2-\frac{4}{3}(g_A^{(1+0)}-\frac{2}{3}c^{(-1)})a^{(-1)}+\frac{4}{9}=0
\quad 
\ee
and we may simplify the expression for the axial current (A.1)
\be
{\cal A}_i^a= -\frac{3}{2} g_A^{(1+0)}D_{ai} + a^{(-1)}
 \left[\frac{1}{2} \{ D_{ai} \mbox{\boldmath$R$}^2 \}-L_a R_i-\kappa
D_{ai} \right]
\;,
\ee
where we replaced $c^{(-1)}=-\kappa a^{(-1)}$ according to the scaling
behaviour (A.9). The constant $\kappa$ is parameter independent and 
will be determined immediately.
With (A.11) we may evaluate the weak axial nucleon-nucleon
and nucleon-$\Delta$ transition amplitudes
\begin{eqnarray}
g_A &=& g_A^{(1+0)} + \frac{2}{3}\kappa a^{(-1)}   \nonumber \\
\frac{\sqrt{2}}{3}g_A^* &=& g_A^{(1+0)} -
\frac{3}{2}a^{(-1)}+\frac{2}{3}\kappa a^{(-1)}
\;.
\end{eqnarray}
With these quantities the condition (A.10) may be rewritten as
\begin{eqnarray}
&&g_A^2 -\frac{2}{9}g_A^{*2}=1 \nonumber \\
&&\frac{3}{2}a^{(-1)} = g_A-\frac{\sqrt{2}}{3}g_A^*
\;,
\end{eqnarray}
which is Oehme's relation \cite{oehm} and a direct consequence of the
CA relation (A.2). From this relation it follows that $g_A \ge 1$.
With the coefficient $a^{(-1)}$ inserted into (A.12)
we obtain 
\begin{eqnarray}\label{ga0}
g_A&=&g_A^{(1+0)} + \frac{4}{9} \kappa (g_A-\frac{\sqrt{2}}{3}g_A^*)
\nonumber \\
   &=& g_A^{(1+0)} + \frac{4}{9} \kappa (g_A-\sqrt{g_A^2-1})
\end{eqnarray}
The last step involves the identification $\sqrt{2}/3 g_A^*=+\sqrt{g_A^2-1}$
which is reasonable because of the large $N_C$ requirement $\sqrt{2}/3
g_A^*=g_A$. Since relation \ref{ga0} must hold independently of the
value of $g_A^{(1+0)}$ (c.f. the discussion in the beginning of this
section), it follows that $\kappa \ge 9/4$. The choice $\kappa = 9/4$
is conservative in the sense that it yields,  for any given
$g_A^{(1+0)}$, the smallest possible 
correction $g_A^{-1}$ (fig. \ref{corr}). Therefore, upon using this value for $\kappa$ we
obtain 
\begin{eqnarray}
g_A^2 &=& 1+(g_A^{(1+0)})^2 \nonumber \\ 
\frac{\sqrt{2}}{3}g_A^* &=& g_A^{(1+0)}
\;,
\end{eqnarray}
which was the estimate used in section 3.3.3 .

\begin{figure}[h]
\vspace*{7.5cm}
\begin{center} \parbox{8.2cm}{\caption{\label{corr} Plot of $g_A$
for different values of $g_A^{(1+0)}$.}} 
\end{center}
\end{figure}

From the experimental value $g_A=1.26$ we obtain a large $1/N_C$ contribution
$g_A^{(-1)}=.40$ and a small value $g_A^{(1+0)}=.86$ for tree $+$ $1$ loop. If
we accept the assumption underlying this paragraph, namely
\begin{itemize}
\item $SU(4)$ symmetry of
the spatial axial current
\end{itemize}
then this
seems to be the reason why soliton models in tree ($+$ $1$ loop) approxmation
always underestimate $g_A$ by a large margin.
\end{appendix}
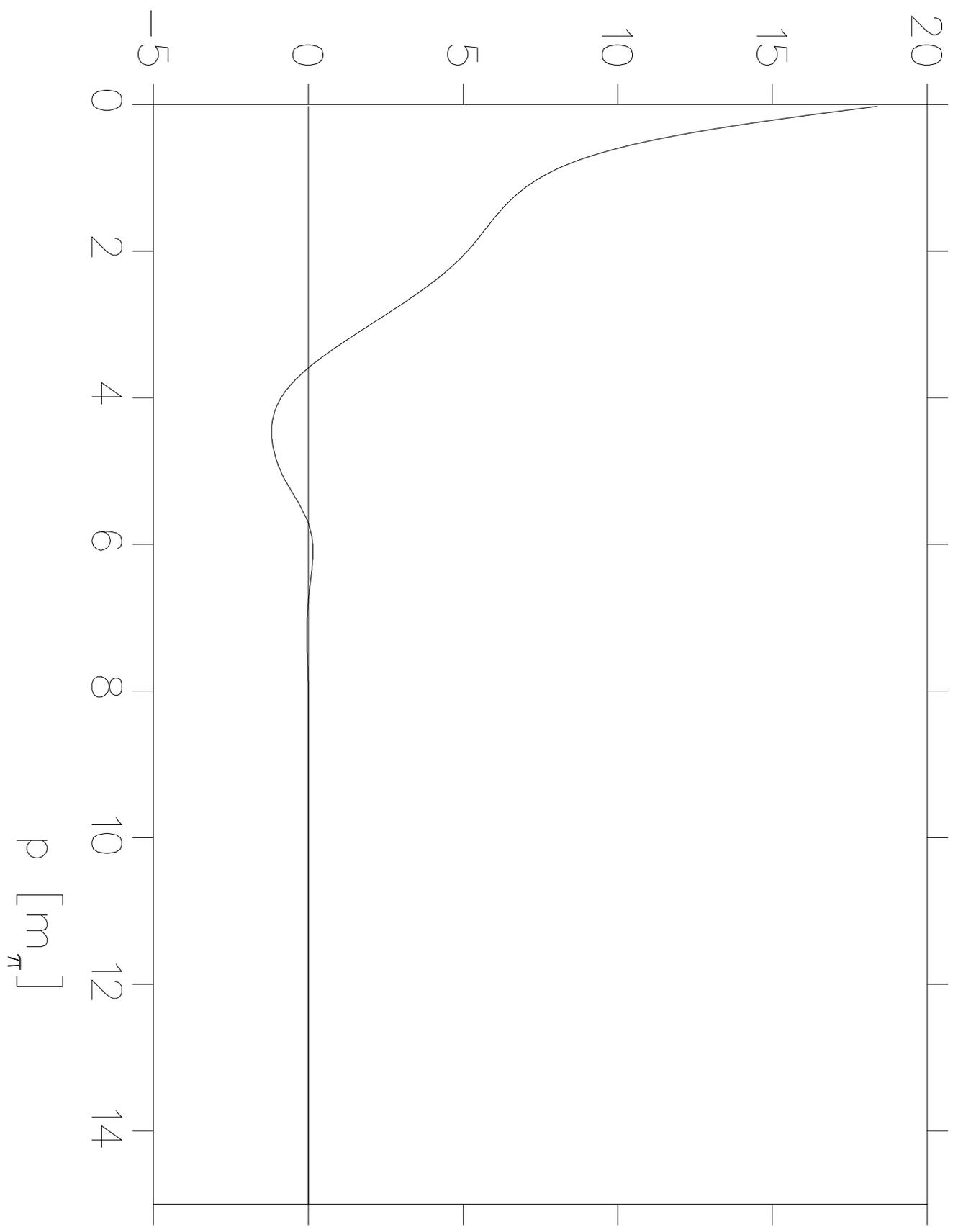
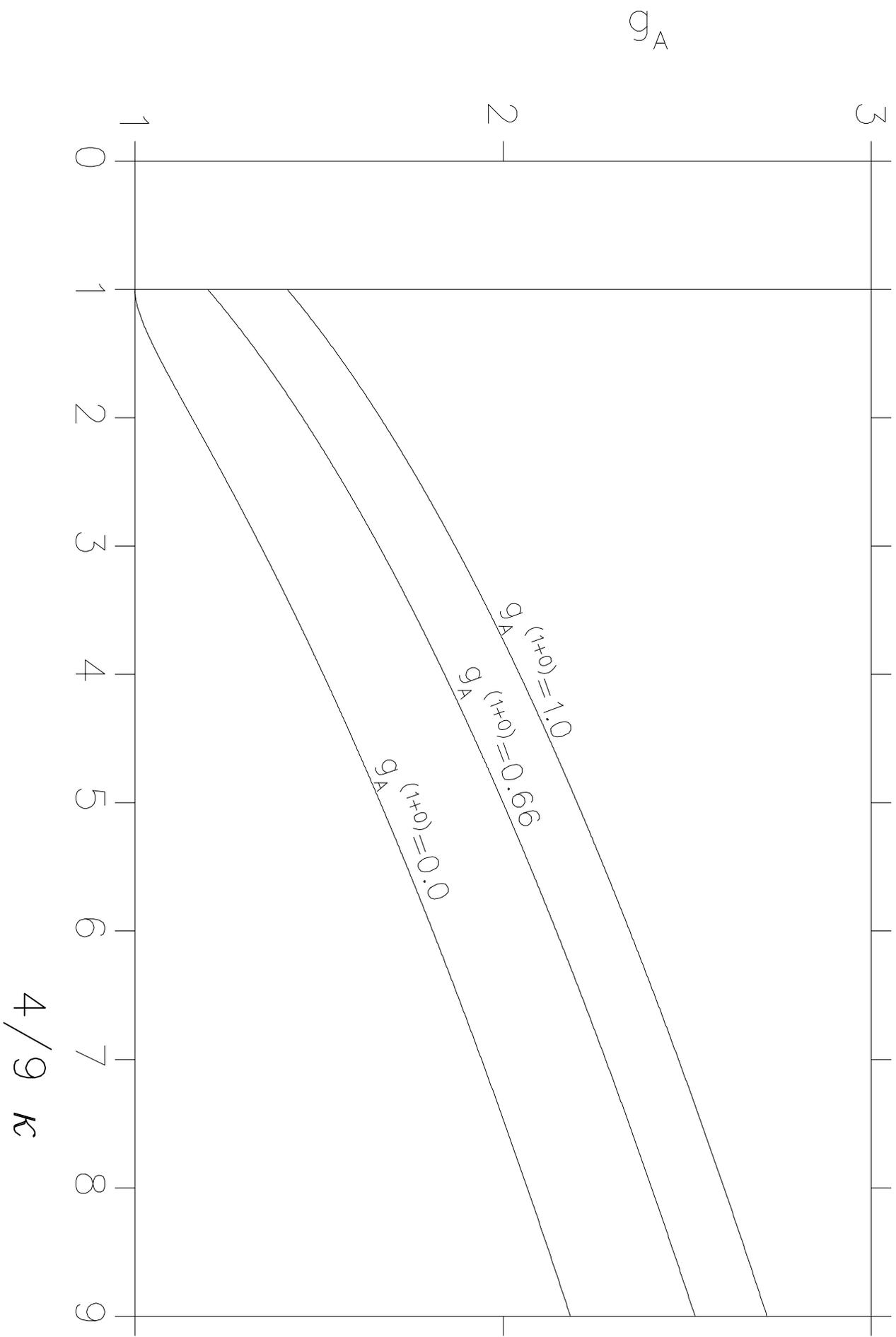
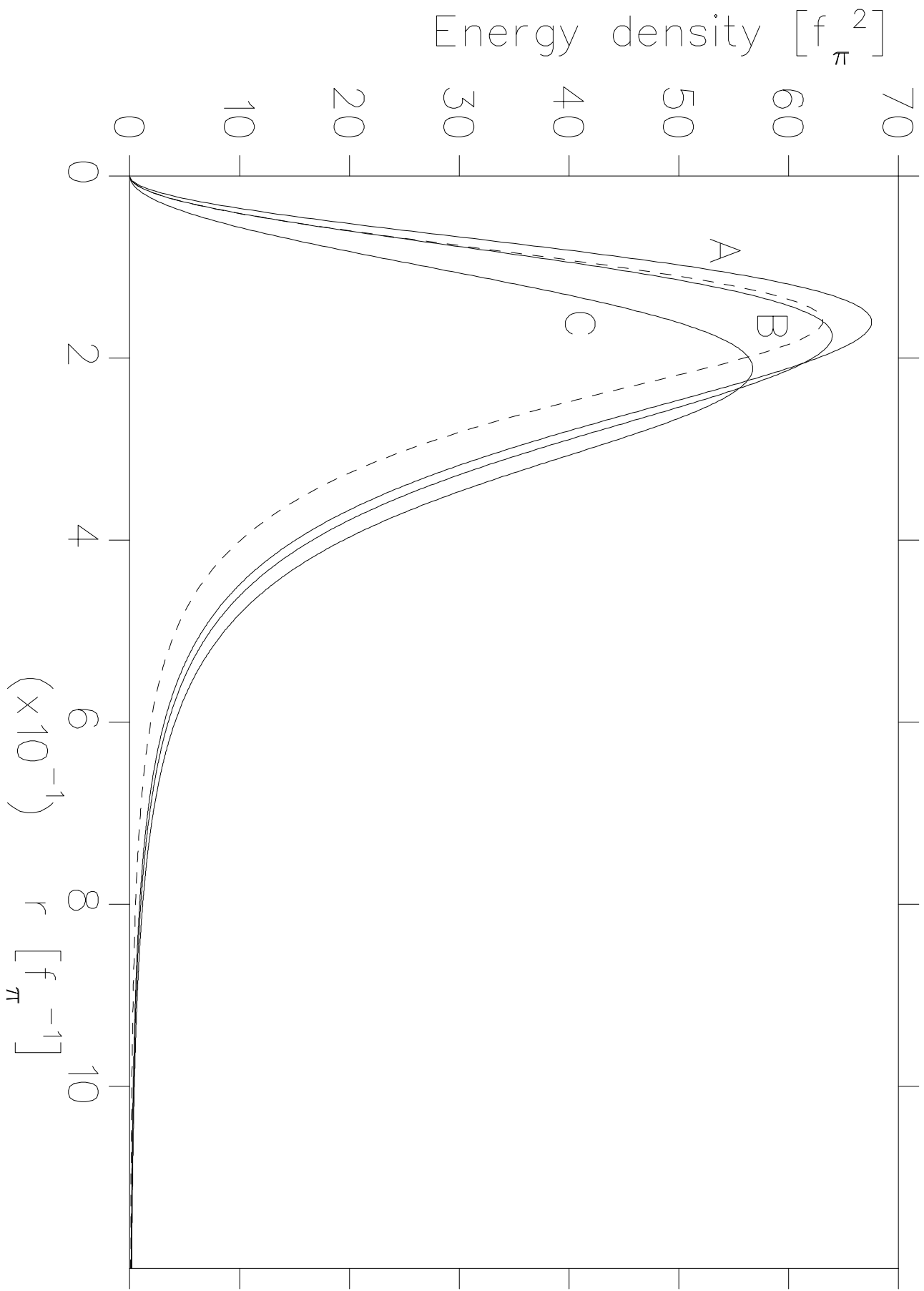
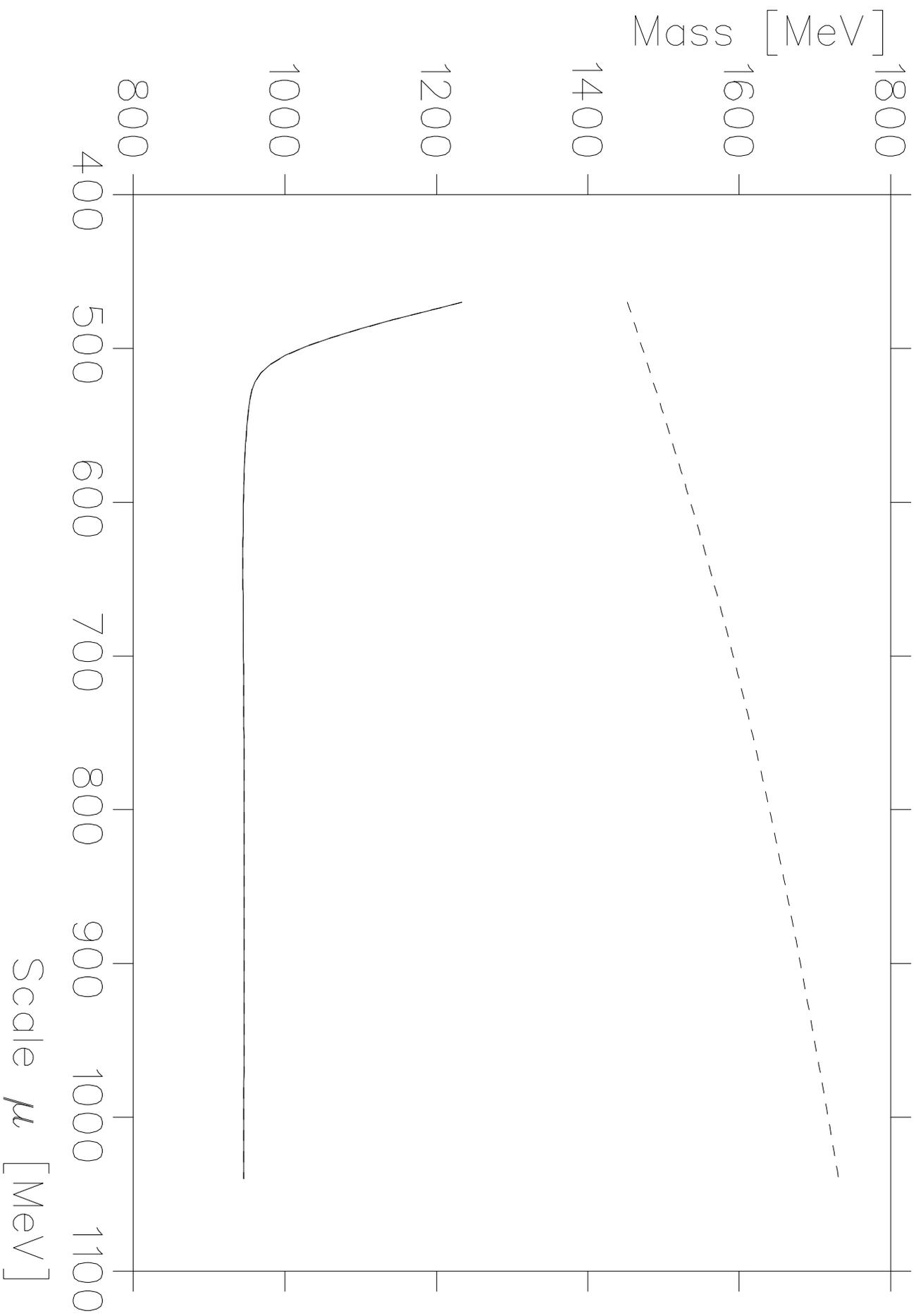
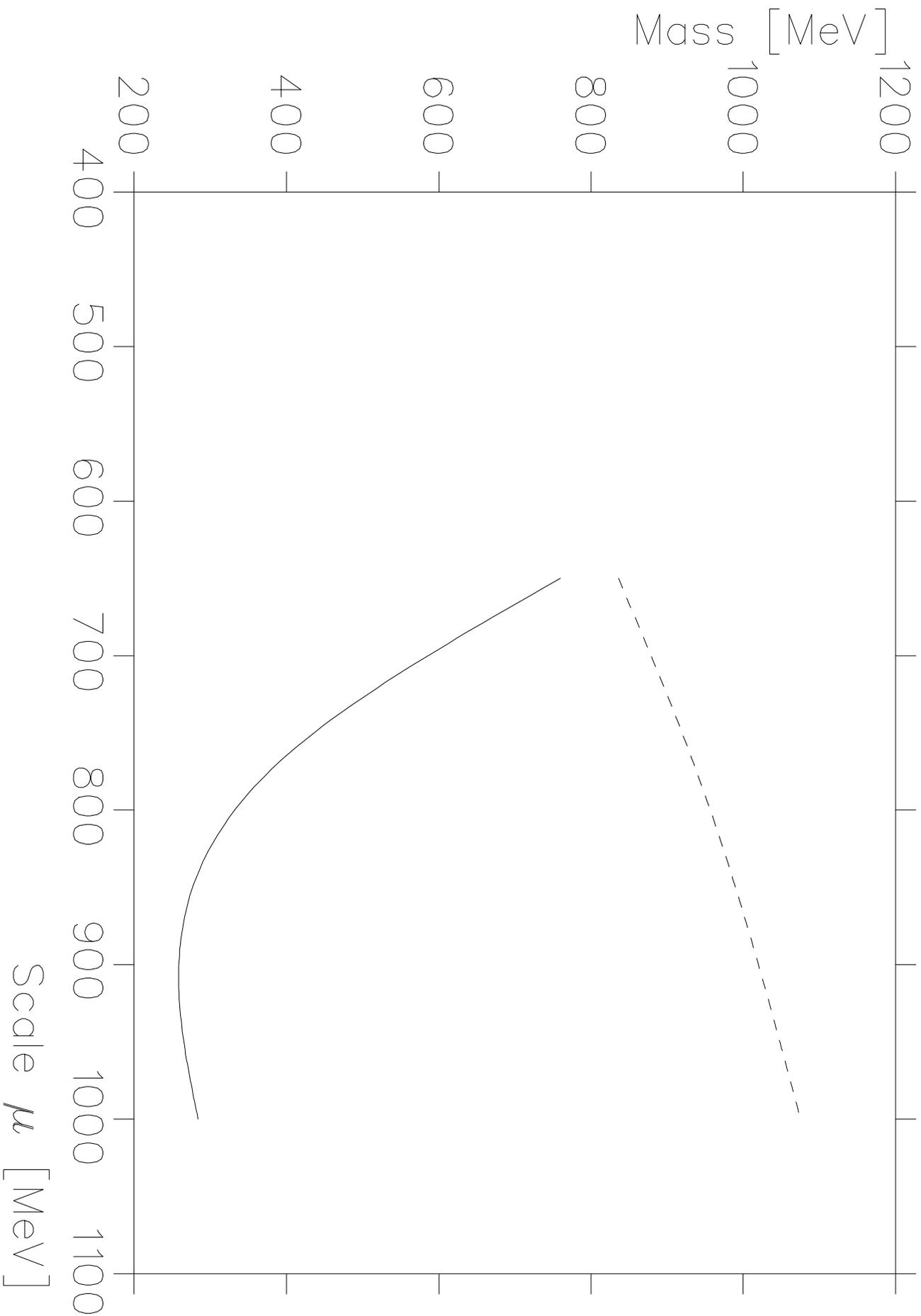
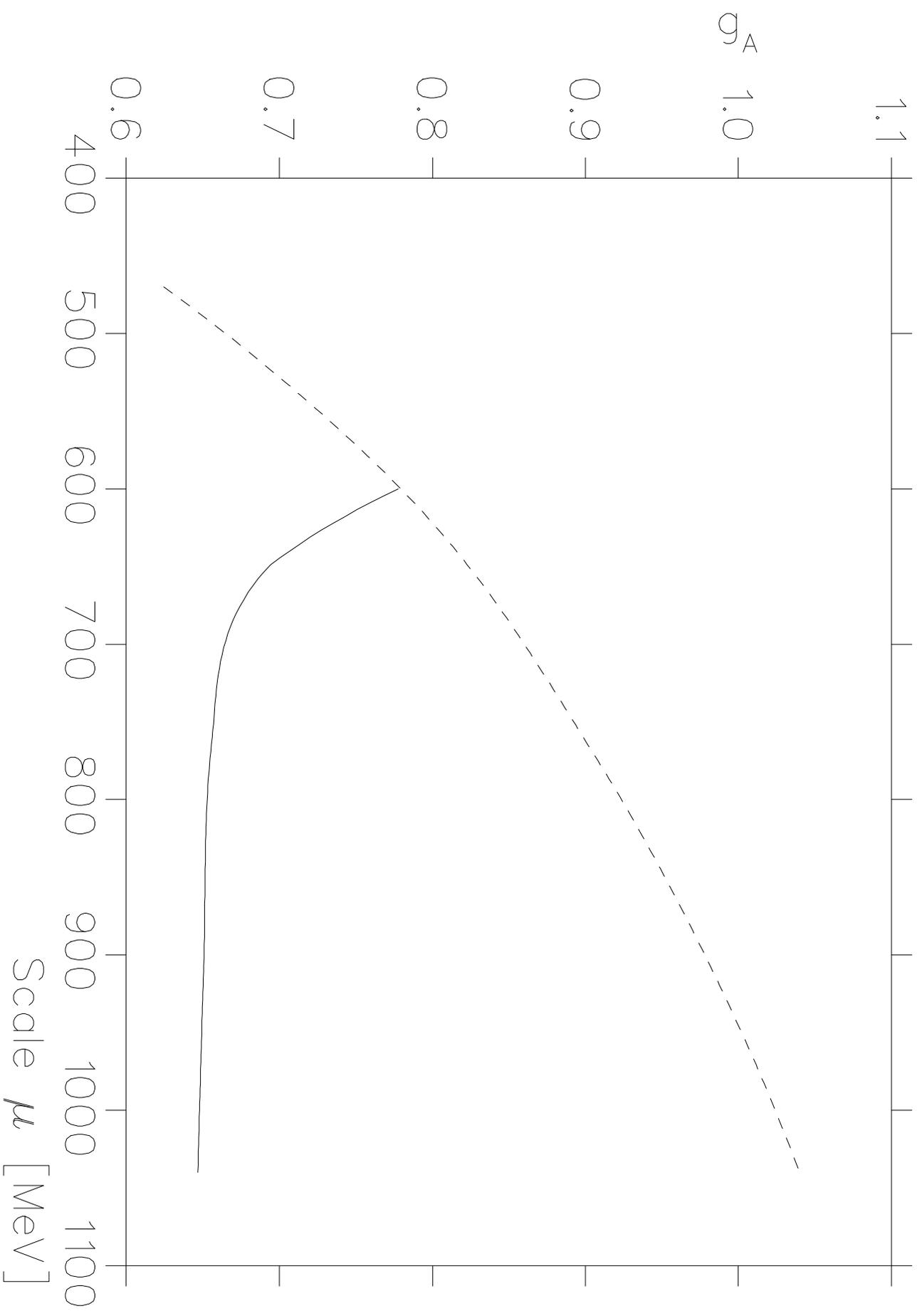
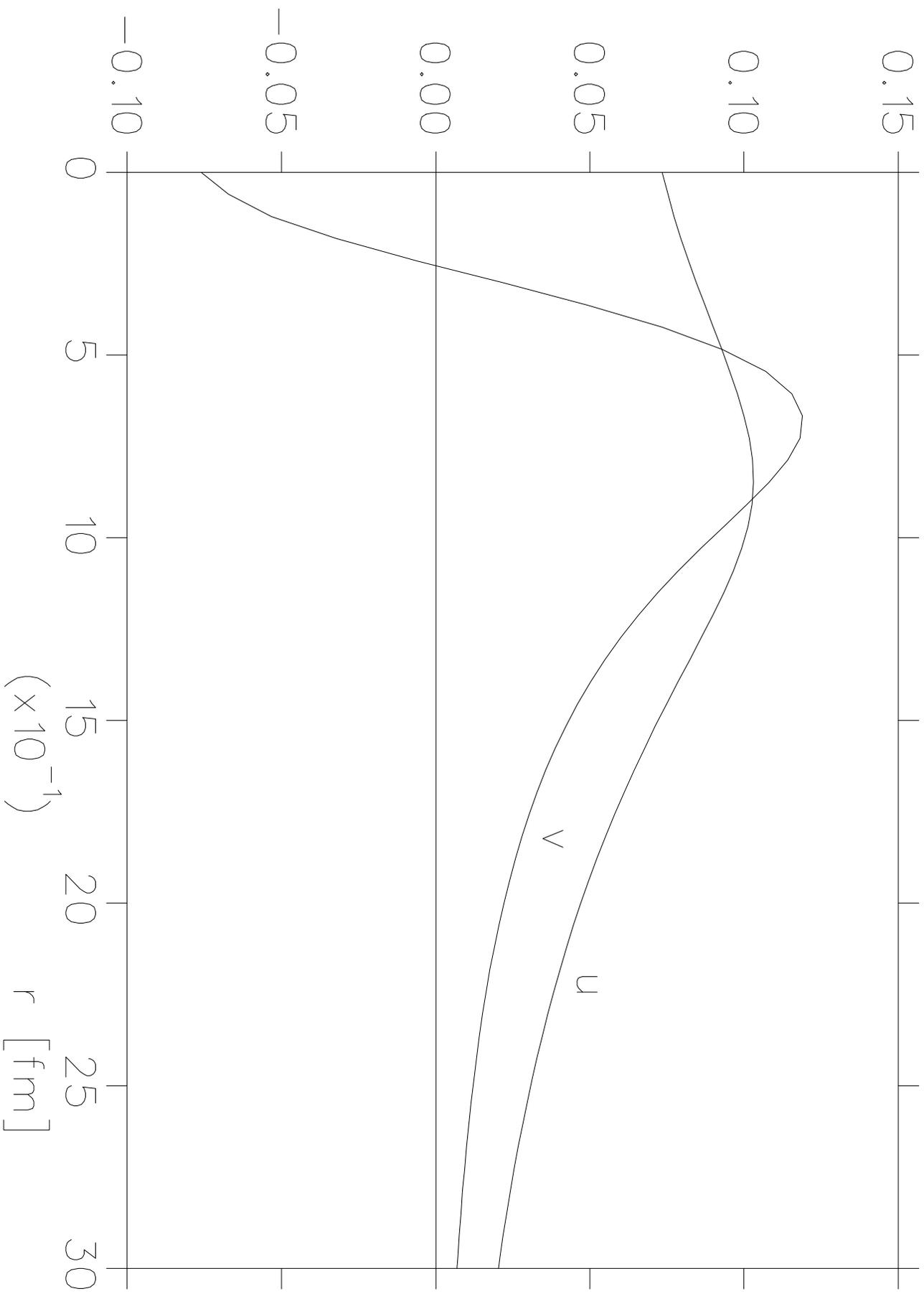
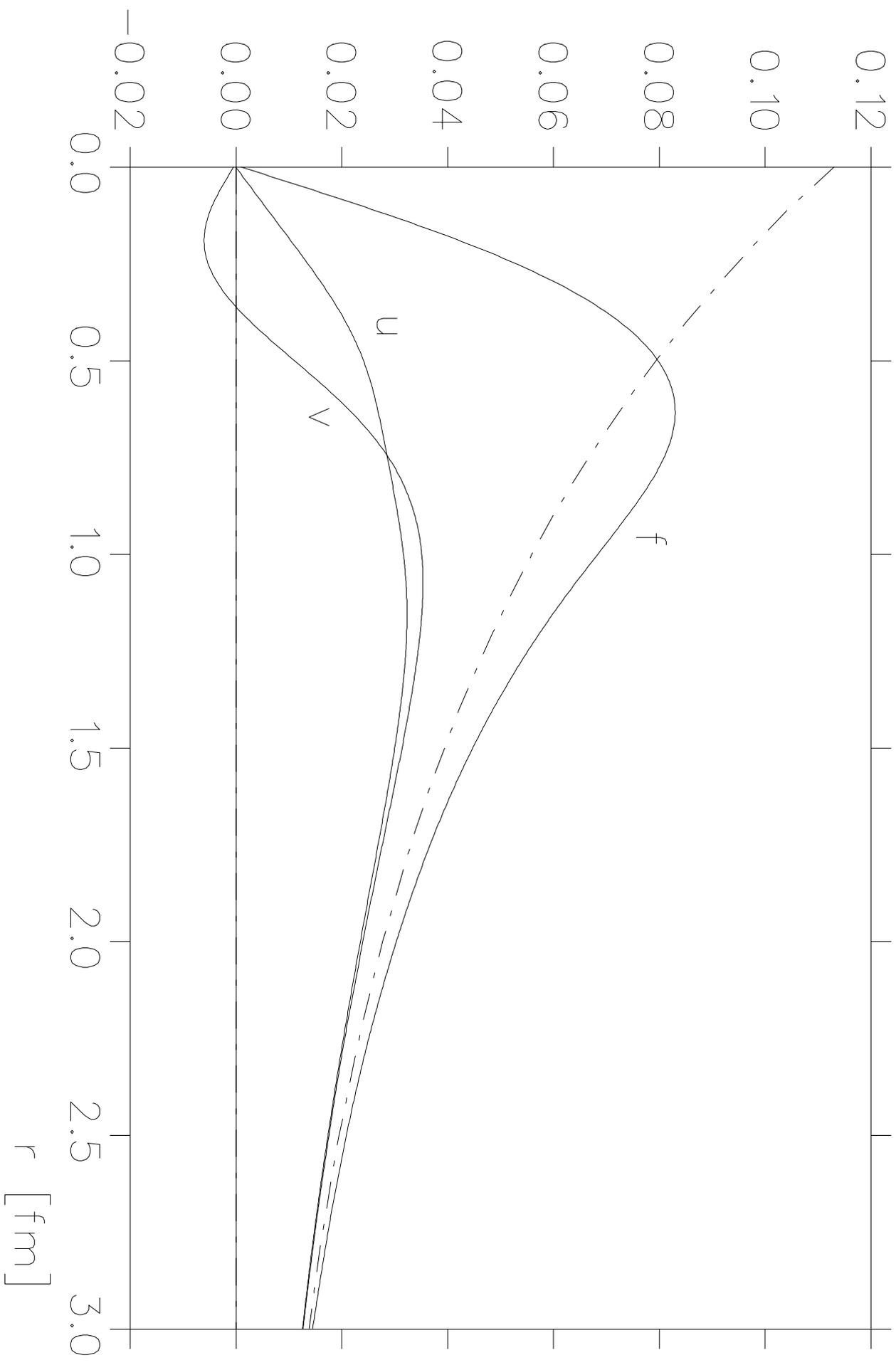
\end{document}